\documentclass[11pt]{article}

\usepackage{scalerel}
\usepackage{graphicx}
\usepackage{amsmath}
\usepackage{amsfonts}
\usepackage{amssymb}
\usepackage{dsfont}
\usepackage{physics}
\usepackage{braket}
\usepackage{microtype}
\usepackage{bbold}
\usepackage{bm}
\usepackage{pifont}
\usepackage{mathtools}
\usepackage{cite}
\usepackage{cancel}
\usepackage{xcolor}
\usepackage{mdframed}
\usepackage{xpatch}
\usepackage{etoolbox}
\usepackage[normalem]{ulem}
\usepackage[colorlinks=true, linkcolor=blue, filecolor=blue, urlcolor=blue, citecolor=blue]{hyperref}
\usepackage[font=footnotesize,labelfont=bf,width=.9\textwidth]{caption}
\usepackage{mathtools}
\usepackage{slashed}    
\usepackage{enumitem}
\usepackage{bm}
\usepackage{orcidlink}
\usepackage{multirow}

\allowdisplaybreaks
\numberwithin{equation}{section}
\renewcommand{\title}[1]{\vbox{\center\LARGE{#1}}\vspace{5mm}}
\renewcommand{\author}[1]{\vbox{\center#1}\vspace{5mm}}
\newcommand{\address}[1]{\vbox{\center\footnotesize\em#1}}
\newcommand{\email}[1]{\vbox{\center\footnotesize\tt#1}\vspace{5mm}}

\DeclareMathOperator{\sgn}{sgn}

\def\le{\left}
\def\ri{\right}

\newcommand{\f}{\qty}
\newcommand{\m}{\bmqty}
\renewcommand{\r}{\frac}

\newcommand{\R}{\mathbb{R}}
\newcommand{\C}{\mathbb{C}}

\newcommand{\Z}{\mathbb{Z}}

\newcommand{\mc}{\mathcal}
\newcommand{\mr}{\mathrm}
\newcommand{\mf}{\mathfrak}
\newcommand{\ar}{\rightarrow}

\newcommand{\ol}{\overline}
\newcommand{\ti}{\tilde}

\newcommand{\bars}[1]{\mathbin{\bar{#1}}}

\begin{document}

\begin{titlepage}

\begin{center} 

\hfill \\
\hfill \\
\vskip 1cm

\title{Revisiting near-extremal and near-BPS black holes\\ in AdS$_3$ supergravity}


\author{Adam Bac\,\orcidlink{0000-0002-0650-9372}, Alejandra Castro\,\orcidlink{0000-0003-1485-8393}, Diksha Jain\,\orcidlink{0000-0002-7405-631X}}


\address{Department of Applied Mathematics and Theoretical Physics,  University of Cambridge, \\
Cambridge CB3 0WA, UK}

\email{ab3003@cam.ac.uk, ac2553@cam.ac.uk, dj428@cam.ac.uk}
\end{center}

\begin{abstract}

Despite the archetypal status of the BTZ background in quantifying quantum aspects of black holes, several features at low temperatures remain imprecise and incomplete. Here, we systematically investigate the behaviour of the Euclidean path integral at low temperatures in the context of AdS$_3$ supergravity, including an analysis of quantum fluctuations in both the near-horizon and asymptotic regions. We clarify and rectify aspects of the bosonic fluctuations, highlighting the role of boundary conditions in AdS$_3$, and show in particular that the gravitational path integral in the near-horizon region is inequivalent to that around BTZ at low temperature. We further account in detail for the contributions of Chern–Simons fields and spin-3/2 modes, thereby refining the disparities between the near-extremal and near-BPS limits at low temperature. Altogether, our analysis sharpens the distinction between near- and far-region dynamics and demonstrates a disagreement in the gravitational path integral at the quantum level.

\end{abstract}

\end{titlepage}
 
\tableofcontents
\newpage

\section{Introduction}

The BTZ black hole \cite{Banados:1992wn,Banados:1992gq} has long served as a canonical setting for exploring aspects of quantum gravity. Owing to its simplicity and its embedding in AdS$_3$ gravity, the BTZ geometry provides a tractable setup for studying the interplay between geometry, thermodynamics, and quantum effects. In particular, the Euclidean path integral formulated around BTZ saddles has played a central role in understanding thermodynamic properties, modular features, and connections to a dual conformal field theory. In this work, we revisit the analysis of this path integral with particular emphasis on its behaviour near extremality. 

Our motivation stems from renewed interest in gravitational path integrals (GPI) at low temperatures, where it has been shown that quantum effects can become dominant. Developments in a variety of contexts have highlighted that near-extremal geometries may exhibit enhanced contributions from a certain class of modes, which can qualitatively modify the semi-classical expansion \cite{Charles:2019tiu,Iliesiu:2020qvm,HeyIli20,Iliesiu:2022onk}. These effects often manifest as infrared sensitivities in the Euclidean path integral, requiring a careful treatment of fluctuations and boundary conditions. In certain cases, such corrections can even compete with or dominate over the classical saddle, thereby reshaping the thermodynamic interpretation of the system. See \cite{Turiaci:2023wrh} for a general discussion of these developments. 

Despite the central role of the BTZ background in such analyses, several aspects of its low-temperature behaviour remain incompletely understood. In particular, it is often implicitly assumed that the near-horizon region captures the relevant infrared physics, and that the corresponding path integral is equivalent to that defined in the full asymptotically AdS$_3$ geometry. However, the validity of this assumption has not been systematically examined, especially in settings where additional fields and boundary contributions may play an important role. Some of the recent works that have studied the low-temperature behaviour of the GPI evaluated on the BTZ saddle include \cite{HeyIli20,KapLaw24,KolMar24,Acito:2025hka,GhoMax19}. 

Our analysis centres around the GPI in AdS$_3$ supergravity at low temperatures, with the aim of testing these assumptions. In addition to the graviton, we incorporate the contributions of Chern-Simons fields and fermionic degrees of freedom, including spin-$3/2$ fields, which are intrinsic to the supergravity theory. This allows us to refine the comparison between near-extremal and near-BPS limits. We analyse quantum fluctuations in both the near-horizon and asymptotic regions, paying particular attention to the structure of the modes and the role of boundary conditions.

We show that, at low temperature, the gravitational path integral defined in the near-horizon region is \textit{typically} inequivalent to that obtained from fluctuations around the BTZ background; agreement occurs only within specific sectors of the GPI and relies on boundary conditions. More precisely, the following mismatches and agreements occur. The GPI near the horizon and near-extremality takes the form
\begin{equation}\label{eq:Znhg-intro}
    Z_{\rm ENHG}[T] \underset{T\to0}{\approx} Z_0[T] Z_{\rm z.m.}[T] Z_{\rm n.z.m.}[T] + \cdots ~.
\end{equation}
Here, by ENHG, we refer to the Euclidean near-horizon geometry of the black hole, including the first correction in temperature $T$.  $Z_0[T]$ is the classical contribution arising from evaluating the renormalised action on-shell on this background. $Z_{\rm n.z.m.}$ and $Z_{\rm z.m.}$ are one-loop corrections including the leading temperature effect, where ``n.z.m.'' stands for non-zero modes, which are modes with non-zero eigenvalue at $T=0$, and ``z.m.'' stands for zero modes, which are modes with zero eigenvalue at $T=0$. The dominant contribution as $T\to0$ arises from $Z_{\rm z.m}$ and it is the focus of our analysis. On the other hand, we can also consider the path integral evaluated on BTZ, without taking a near-horizon limit, where we would have
\begin{equation}\label{eq:Zbtz-intro}
    Z_{\rm BTZ}[T] \underset{T\to0}{\approx} Z_0[T] Z_{\rm low}[T] + \cdots ~.
\end{equation}
Here, $Z_0[T]$ is again the classical contribution, and it is precisely the same quantity as in \eqref{eq:Znhg-intro}. The contribution ${Z}_{\rm low}[T]$ comes from modes in the one-loop determinant whose eigenvalues, as a function of $T$, approach zero at extremality.  $Z_{\rm low}[T]$ would also encapsulate the dominant effect at low temperatures. The lore is that  $Z_{\rm low}[T]$ should capture the same physics as $ Z_{\rm z.m.}[T]$; however, here we show that 
\begin{equation}
    Z_{\rm low}[T]\neq Z_{\rm z.m.}[T]~,
\end{equation}
and only under specific circumstances, we have an agreement. The technical details of our findings are as follows: 
\begin{enumerate}
    \item For certain classes of modes, we find that their contribution to $Z_{\rm low}[T]$ and  $Z_{\rm z.m.}[T]$ is the same, and their profiles match nicely at low temperature. In the gravitational sector, these are the so-called tensor or Schwarzian modes. This agrees with the findings in \cite{KapLaw24}, which carefully examined this sector in both near and far regions. The fermionic contributions, either spin-1/2 or spin-3/2 fields, also agree in the far and near regions.
    \item In the gravitational sector of the GPI, there are additional modes to be considered. In the near region, these modes are usually called rotational or vector modes, and from the perspective of the ENHG, they comply with the same conditions as tensor modes at $T=0$: gauge fixing conditions, normalisable, and do not change the charges of the black hole.
    However, their response after turning on the temperature is starkly different relative to tensor modes in the near-horizon region. If we assume that turning on temperature is an adiabatic process, meaning that we can treat $T$ as a small coupling and use standard rules of perturbation theory, we find that rotational modes do not contribute to $Z_{\rm z.m.}[T]$. One would suspect, therefore, that these modes should be discarded. Rather surprisingly, we will show that there is a new set of normalizable modes in the whole geometry that contribute to $Z_{\rm low}[T]$, under appropriate boundary conditions.  The reason why $Z_{\rm z.m.}[T]$ does not capture this contribution is due to a breakdown of perturbation theory in $T$: although in the decoupling limit, as $T\to 0$, the modes supported on BTZ match the rotational modes perfectly, the first correction in $T$ to the eigenmodes is non-normalizable in the NHG.    
This illustrates that the GPI in the NHG is blind to certain physical modes supported in the whole geometry.
    
    \item Quantum fluctuations arising from the Chern-Simons gauge field are similar to the rotational modes, and the tensions between $Z_{\rm low}[T]$ and $Z_{\rm z.m.}[T]$ persist. In the ENHG analysis, the gauge zero modes are not affected by the temperature deformation, therefore, not making an imprint on the leading temperature effects in $Z_{\rm z.m.}[T]$. In the far region analysis, there is a non-trivial contribution to $Z_{\rm low}[T]$ coming from this sector, and this is expected from the dual CFT$_2$ \cite{HeyIli20,Murthy:2025moj,Ferko:2024uxi}. We can track this disagreement by noting that, although quantum fluctuations are normalizable around BTZ, they become non-normalizable in the ENHG when the temperature is turned on. 
\end{enumerate} 
Therefore, in the generic scenario, there is a disagreement between $Z_{\rm BTZ}[T] $ and $ Z_{\rm ENHG}[T] $ at low temperatures. 

It is useful to revisit why one would expect that the low temperature limit of $Z_{\rm BTZ}[T] $ to be equivalent to $ Z_{\rm ENHG}[T] $. The ENHG of BTZ contains an AdS$_2$ factor, which can be attributed to an incarnation of the attractor mechanism \cite{Ferrara:1996dd,Ferrara:1996um,Gupta:2008ki}. This mechanism tells us that the shapes and sizes of the ENHG are insensitive to the behaviours of fields in the asymptotically far region, and as a consequence, the Bekenstein-Hawking area law (or more generally, the Wald entropy) can be evaluated in the ENHG. It has also been clear that any dynamical process in AdS$_2$ breaks this fragile throat \cite{Maldacena:1998uz}; quantifying finite temperature effects necessitates moving away from the pristine AdS$_2$ background, leading to the concept of near-AdS$_2$ \cite{AlmPol14,MalSta16}. In \eqref{eq:Znhg-intro} we have taken into account this breakdown of the decoupling limit according to the rules of near-AdS$_2$. Taking all of this into account, one can see that  $Z_0[T]$  in \eqref{eq:Znhg-intro} and \eqref{eq:Zbtz-intro} is the same quantity for the class of theories under consideration. Our failure is more severe than this:  $Z_{\rm BTZ}[T] $ is typically inequivalent to $ Z_{\rm ENHG}[T] $ at the quantum level even after taking into account the dynamics of near-AdS$_2$. 

Another question we address concerns the holographic proposal of Kerr/CFT \cite{Guica:2008mu}.  This proposal has been applied to extremal BTZ in, for example, \cite{Balasubramanian:2009bg,Azeyanagi:2011zj}, where the idea is to characterise the set of large diffeomorphisms acting on the boundary of the NHG that could provide clues to a dual quantum description of the black hole.  Here, we show that the diffeomorphisms inspired by the Kerr/CFT correspondence do not contribute to either $Z_{\rm ENHG}$ or $Z_{\rm BTZ}$. This indicates that not every set of large diffeomorphisms that leads to an asymptotic symmetry group, even if it defines a reasonable classical phase space, satisfies the basic requirements to be included in the GPI. 

This paper is organised as follows. In Sec.\,\ref{sec:sugra-bhs} we collect classical properties of the supergravity theories we will consider, establishing basic conventions. We also review the properties of black holes embedded in this theory, including a discussion of extremal versus BPS black holes and their response to a temperature deformation. In Sec.\,\ref{sec:one-loop} we start our discussion of the gravitational path integral, where we review the one-loop contributions to the graviton, and discuss the one-loop determinants of Chern-Simons and spin-3/2 fields. In Sec.\,\ref{sec:near} we focus on evaluating the GPI at zero temperature in the near-horizon region, and then quantify how the GPI responds when we slightly turn on temperature. Our analysis is exhaustive in the bosonic and fermionic sectors of the GPI. With this, we show in Sec.\,\ref{sec:kerr-cft-no-go} that the diffeomorphisms proposed by Kerr/CFT do not play a role in the GPI of BTZ at the quantum level. 
In Sec.\,\ref{sec:far} we analyse the zero temperature limit of the GPI from the perspective of the whole geometry, which includes the asymptotically AdS$_3$ region. We discuss how our answers in the near and far regions disagree for rotational modes and gauge modes. We also discuss the role of boundary conditions at AdS$_3$ in the analysis of these zero modes. In Sec.\ref{sec:discussion}, we conclude our analysis with a discussion and future directions. App.\,\ref{app:conventions} contains conventions used throughout the text. App.\,\ref{sec:app-full-btz} covers the technical details of the analysis of quantum eigenmodes contributing to the one-loop determinants.  

\section{Black holes in \texorpdfstring{AdS$_3$}{AdS3} supergravity}\label{sec:sugra-bhs}

In this section, we review some basic aspects of AdS$_3$ supergravity, with different amounts of supersymmetry. We will overview the different versions of the BTZ black hole as solutions to the supergravity equations of motion, for both neutral and charged cases, and quantify the conditions under which the solution supports a Killing spinor. We will also highlight the differences and similarities of taking a near-extremal limit of supersymmetric against non-supersymmetric configurations. 

The supergravity theories we will discuss follow mainly from those in \cite{Achucarro:1987vz}; other useful references that discuss various aspects of AdS$_3$ supergravity are \cite{Henneaux:1999ib,Nishimura:1999gg,David:1999nr}.
The action of AdS$_3$ supergravities we will consider naturally splits into three contributions  
\begin{equation}\label{eq:action}
    S= \r{1}{16\pi G}\int_M \left({\cal L}_{\rm EH} + {\cal L}_{\rm CS} +{\cal L}_{ 3/2}\right)~,
\end{equation}
for the metric, two classes of non-Abelian Chern-Simons gauge fields, and two gravitini fields. Each Lagrangian, in Lorentzian signature, reads
\begin{equation}
    \begin{aligned}
\label{lor-sugra}
    {\cal L}_{\rm EH} &=  \star\f(R+\r{2}{\ell^2})~, \\
    {\cal L}_{\rm CS} &= - \ell \Tr\f(A\wedge\dd{A} + \r{2}{3}A\wedge A\wedge A) + \ell \Tr\f(A'\wedge\dd{A}' + \r{2}{3}A'\wedge A'\wedge A')~, \\
        {\cal L}_{3/2}&= \bar{\psi} \wedge \mc{D}\psi
    +  \bar{\psi'} \wedge \mc{D}'\psi'   ~.     
    \end{aligned}
\end{equation}
Here $\ell$ is the AdS$_3$ radius, and $G$ is the three-dimensional Newton's constant. The gauge fields $A$ and $A'$ are one-forms valued in the Lie algebras $\mf{g}$ and $\mf{g}'$, while the gravitini $\psi$ and $\psi'$ are spin-3/2 Majorana fields transforming in the fundamental representation of the corresponding groups $G$ and $G'$.
The covariant derivatives appearing in ${\cal L}_{3/2}$ are defined as
\begin{equation}\label{eq:def-covariant}
    \mc{D}_\mu = D_\mu - \r{1}{2\ell}e^{a}_\mu\gamma_a~,\quad
    \mc{D'}_\mu = D'_\mu + \r{1}{2\ell}e^{a}_\mu\gamma_a~,
\end{equation}
where $e^a_\mu$ is the spin coframe,
\begin{equation}\label{eq:def-covariant-1}
     D_\mu = \nabla_\mu + \r{1}{4}\omega^{ab}_\mu\gamma_{ab} + A_\mu ~,\quad D'_\mu = \nabla_\mu + \r{1}{4}\omega^{ab}_\mu\gamma_{ab} + A'_\mu ~,
\end{equation}
and $\nabla_\mu$ is the usual covariant derivative, containing the Christoffel symbol, acting on spacetime indices. The $\gamma$-matrices form the Clifford algebra with signature $(-++)$, where we choose to represent them as 
\begin{equation}
\gamma^0 = i\sigma_2~,\quad \gamma^1 = \sigma_1~,\quad \gamma^2 = \sigma_3~,    
\end{equation}
with $\sigma_i$ Pauli matrices, and $\gamma^{ab} \coloneqq \r{1}{2}[\gamma^a, \gamma^b]$. The two types of covariant derivatives in \eqref{eq:def-covariant} arise because in three spacetime dimensions there are two inequivalent two-dimensional irreducible representations of the Clifford algebra: $\gamma^a$ and $-\gamma^a$. This naturally defines two sectors in three-dimensional supergravity. More conventions, definitions and identities are presented in App.\,\ref{app:conventions}.

To avoid cluttering, it is convenient to introduce further details of the supergravity theory for specific groups. The cases we will cover are: 
\begin{itemize}[label=$\circ$]
\setlength\itemsep{-0.1em}
    \item the simplest case of ${\cal N}=(1,1)$, where the Chern-Simons gauge fields are absent,
    \item the $(p,q)$-supergravities introduced in \cite{Achucarro:1987vz}, where  $\mf{g}=\mf{so}(p)$ and $\mf{g}'=\mf{so}(q)$ and the corresponding groups are $SO(p)\times SO(q)$,
    \item ${\cal N}=(2,2)$ supergravity where the gauge group is $U(1)\times U(1)$,
    \item and  ${\cal N}=(4,4)$ supergravity, where the standard choice of group is  $SU(2)\times SU(2)$.

\end{itemize}
 Further variants and possible choices of gauge groups are listed in \cite{Henneaux:1999ib}. 

\paragraph{${\cal N}=(1,1)$ Supergravity.} In the absence of gauge fields, the theory at hand is described by ${\cal L}_{EH} + {\cal L}_{3/2}$ in \eqref{lor-sugra}, and the covariant derivatives in \eqref{eq:def-covariant} reduce to
\begin{equation}\label{eq:def-covariant-2}
    \mc{D}_\mu = \nabla_\mu + \r{1}{4}\omega^{ab}_\mu\gamma_{ab} - \r{1}{2\ell}e^{a}_\mu\gamma_a~,\quad
    \mc{D'}_\mu =  \nabla_\mu + \r{1}{4}\omega^{ab}_\mu\gamma_{ab} + \r{1}{2\ell}e^{a}_\mu\gamma_a~.
\end{equation}
The gravitini, $\psi=\psi_\mu \dd x^\mu$ and $\psi'=\psi'_\mu \dd x^\mu$, are one-form valued Majorana spinors. We define the Majorana conjugation on a spinor $\chi$ as
\begin{equation}
    \bar{\chi} \coloneqq \chi^T \gamma^0~,
\end{equation}
and the Majorana condition imposes the reality condition $\chi = \chi^*$. This theory is supersymmetric, with the corresponding transformations of the fields being
\begin{equation}\label{susy-trans-1}
\begin{aligned}
    \delta \psi_\mu &= \mc{D}_\mu \epsilon~,\quad
    \delta {\psi'}_\mu= \mc{D}'_\mu \epsilon'~,\\
    \delta e^a_\mu &= -\r{1}{2}\f(\bar{\epsilon} \gamma^a \psi_\mu +\bar{\epsilon}'\gamma^a \psi'_\mu)~,
\end{aligned}
\end{equation}
with $\epsilon$ and $\epsilon'$ Majorana spinors labelling the transformation. The notation ${\cal N}=(1,1)$ denotes that there is a single Majorana degree of freedom in each sector, and in the corresponding supersymmetric algebra, there is one super-charge for each sector \cite{Banados:1998pi}. 

\paragraph{$(p,q)$ Supergravity \cite{Achucarro:1987vz}.} A historically important example that includes the Chern-Simons gauge fields corresponds to the case where the Lie algebras are $\mf{g}=\mf{so}(p)$ and $\mf{g}'=\mf{so}(q)$, and the corresponding gauge groups are $G= SO(p)$ and $G'=SO(q)$. Let $T_I$, $T'_{I'}$ be generators of a particular choice of representations of the Lie algebras $\mf{so}(p)$ and $\mf{so}(q)$, respectively, so that
\begin{equation}
\begin{aligned}
  A &= A^I_\mu \,T_I \,\dd x^\mu ~,\qquad I= 1,\ldots,\r{p(p-1)}{2}~,\\
  A' &= A^{\prime I'}_\mu \,T'_{I'} \,\dd x^\mu ~,\qquad   I'= 1,\ldots,\r{q(q-1)}{2}~.
\end{aligned}
\end{equation}
To discuss supersymmetry of the action \eqref{eq:action}-\eqref{lor-sugra}, we need to specify a few more details of the algebra and representation used. We choose $T_I$ and $T'_{I'}$ to be such that
\begin{equation}\label{eq:lie-algebra}
  \begin{aligned}
    [T_I,T_J] &= \epsilon_{IJK}T_K~,\qquad \Tr(T_I T_J) = -C\delta_{IJ}~, \\
    [T'_{I'},T'_{J'}] &= \epsilon_{I'J'K'}T'_{K'}~,\quad \Tr(T'_{I'} T'_{J'}) = -C'\delta_{I'J'}~.
  \end{aligned}
\end{equation}
Here, $C$ and $ C'$ are constants that account for the normalisation of the Cartan-Killing metric, where $g_{IJ}=-C\delta_{IJ}$ and $g'_{I'J'}=-C'\delta_{I'J'}$. The fundamental representation of $\mf{so}(p)$ corresponds to real antisymmetric $(p\times p)$-matrices. With this, it should be understood that the gravitini are $p$-dimensional objects, and under a global $SO(p)$ transformation, we have
\begin{equation}\label{eq:gravitini-fund}
\psi^i_\mu \,\to \, \left(e^{\alpha^I T_I}\right)^i_{~j}\, \psi^j_\mu ~, \qquad i,j= 1,\ldots, p~,   
\end{equation}
and $\alpha^I$ are constants. In writing ${\cal L}_{3/2}$ this structure is implicit, where we are using 
\begin{equation}
 \bar \psi \wedge \psi = \sum_{i=1}^p \bar \psi^i \wedge \psi^i~,     
\end{equation}
and $\psi^i=\psi^i_\mu \dd x^\mu$, and $\psi^i_\mu$ satisfies the Majorana condition. Analogous expressions apply for the primed sector for which the gravitini $\psi'^{i'}$, with $i'=1,\ldots, q$, transforms under the fundamental representation of $SO(q)$.

The action \eqref{eq:action}-\eqref{lor-sugra} is endowed with supersymmetry. The infinitesimal supersymmetric transformations of the fields are
\begin{equation}\label{susy-trans}
\begin{aligned}
    \delta \psi_\mu^i &= \mc{D}_\mu \epsilon^i~,\quad
    \delta {\psi'}_\mu^{i'}= \mc{D}'_\mu \epsilon'^{i'}~,\\
    \delta A_\mu^I &= -\r{1}{\ell}\bar{\epsilon} \, T^I\, \psi_\mu~,\quad 
    \delta {A'}_\mu^{I'} = \r{1}{ \ell} \bar{\epsilon}'\,T'^{I'}\,{\psi'}_\mu~,\\
    \delta e^a_\mu &= -\r{1}{2}\f(\bar{\epsilon} \gamma^a \psi_\mu +\bar{\epsilon}'\gamma^a \psi'_\mu)~,
\end{aligned}
\end{equation}
 where $\epsilon^i$ and $\epsilon'^{i'}$ are Majorana spinors labelling the supersymmetric transformation, and for the generators we have $T^I = g^{IJ}T_J$, with analogous definition for $T'^{I'}$. To match the notation used in \cite{Achucarro:1987vz}, one uses the explicit form of the fundamental representation in terms of antisymmetric matrices and transcribes $A^I_\mu T_I \to A_\mu^{ij}$. This leads to the transformation of the gauge fields being $\delta A_\mu^{ij} = \r{1}{\ell}\bar{\epsilon}^{[i}\psi_\mu^{j]}$ as stated in \cite{Achucarro:1987vz}.

\paragraph{${\cal N}=(2,2)$ Supergravity.} Many aspects of what we discussed for $(p,q)$ supergravity follow here, where we emphasise the key differences. This is one of the simplest supergravities containing gauge fields, where the gauge group is $U(1)\times U(1)$, and will serve as a good representative of how our subsequent analysis is affected by the inclusion of electric charges. In this case, we have a single generator for $\mf{u}(1)$, and the appropriate fundamental representation is
\begin{equation}\label{eq:u1-rep-generator}
  T = T' = i~,
\end{equation}
so that $C = C' = 1$ in \eqref{eq:lie-algebra}. Since the algebra is one-dimensional, we will employ a standard abuse of notation and write $iA_\mu$ to denote the element of $u(1)$, rather than $A_\mu$ (as we do in the non-Abelian case), so that $A_\mu$ is real. The second line in \eqref{susy-trans} becomes
\begin{equation}
    \delta A_\mu = \r{i}{\ell}\bar{\epsilon}\, \psi_\mu~,\quad 
    \delta A'_\mu = -\r{i}{\ell}\bar{\epsilon}'\, \psi'_\mu~.
\end{equation}
The gravitini, $\psi_\mu$ and $\psi'_\mu$, do not carry an index in this case, but transform by a phase under the action of global $U(1)$. Effectively, in the ${\cal N}=(2,2)$ theory, the gravitini are treated as Dirac spinors. For comparison, we refer to \cite{Banados:2015tft}, where the theory is written in Chern-Simons language, and \cite{Izquierdo:1994jz}, which studies the ${\cal N}=(2,0)$ example. In relation to the $(p,q)$ supergravities, we could have also thought of this instance as a theory with gauge group $SO(2)\times SO(2)$, and the gravitini being again Majorana.

\paragraph{${\cal N}=(4,4)$ Supergravity.} In this case the choice of gauge group is $SU(2)\times SU(2)$, and we choose $T_I$ and $T'_I$ to generate the fundamental representation of $\mf{su}(2)$,\footnote{Note that while $\mf{so}(3)$ and $\mf{su}(2)$ are isomorphic, the exponentiation is important when introducing fermions.  For $\mf{so}(3)$, the fundamental representation acts on a 3-dimensional space by multiplication by real antisymmetric matrices, whereas for $\mf{su}(2)$ it acts on a 2-dimensional space by multiplication by complex traceless antihermitian matrices.
} where
\begin{equation}\label{eq:fund-su2}
  T_I = T'_I =  -\r{i}{2}\sigma_I~,\qquad
  I = 1,2,3~,
\end{equation}
which means that now $C=C'=\r{1}{2}$ in \eqref{eq:lie-algebra}, so $T^I = T^{I'} = i \sigma_I$. With this, the gravitini $\psi^i_\mu$ and $\psi'^{i}_\mu$ comply with an appropriate modification of \eqref{eq:gravitini-fund} where now $i=1,2$.   The supersymmetric transformations have the same structure as \eqref{susy-trans}, and in particular, the transformation of the gauge fields reads
\begin{equation}
  \delta A_\mu^I = -\r{i}{\ell}\bar{\epsilon}\sigma_I \psi_\mu~,\quad
  \delta A_\mu^{\prime I} = \r{i}{\ell}\bar{\epsilon}'\sigma_I \psi'_\mu~,
\end{equation}
in agreement with the expressions in, for example, \cite{Nishimura:1999gg,David:1999nr,Henneaux:1999ib}. Since the representations here are complex, we do not impose a Majorana condition on the spinors, resulting in four supercharges in each sector. Relative to \cite{Nishimura:1999gg}, the spinors used there have an additional global $SU(2)$ index, and a symplectic Majorana condition reduces the number of supercharges from eight to four. This theory is also commonly referred to as $SU(1,1|2)\times SU(1,1|2)$ supergravity, where the bosonic subgroup is $SO(2,2)\times SO(4)$.

In the following sections, we will primarily work with the Euclidean path integral. The Euclidean version of this supersymmetric action arises from (\ref{lor-sugra}) by Wick rotation, where $t=-it_{\mr{E}}$. The resulting Euclidean action is therefore
\begin{multline}\label{eq:Euclidean-3D-action}
    S_E = -\r{1}{16\pi G}\int_M \left[\star\f(R+\r{2}{\ell^2}) \right.
    + i\bar{\psi} \wedge \mc{D}\psi - i\ell \Tr\f(A\wedge\dd{A} + \r{2}{3}A\wedge A\wedge A) \\
    + \left. i\bar{\psi'} \wedge \mc{D}\psi' + i\ell \Tr\f(A'\wedge\dd{A}' + \r{2}{3}A'\wedge A'\wedge A') \right].
\end{multline}
In this signature, the $\gamma$-matrices now  are $\gamma^0 = -\sigma_2$, $\gamma^1 = \sigma_1$, and  $\gamma^2 = \sigma_3$ as the representation of the Clifford algebra with  $(+++)$; the adjoint in Euclidean is now $\bar \chi= \chi^\dagger$. For simplicity and clarity, in Euclidean signature, we will not impose the Majorana condition in intermediate steps; spinors will be treated as Dirac fields, and at the end, we will halve the degrees of freedom, when appropriate.  
With these choices, the supersymmetry transformation again has the form \eqref{susy-trans}.

\subsection{Classical black hole backgrounds}\label{sec:bhs}

Our main interest is quantifying the gravitational path integral for these supergravity theories at low temperatures. This requires carefully characterising the extremal and near-extremal background solutions and the role of supersymmetry in the analysis. In the following, we will collect the key features of the bosonic backgrounds that will be used in later sections. 

\paragraph{Non-extremal neutral and charged black hole.}
We start by establishing some basic notation and conventions by reviewing the properties of the BTZ black hole \cite{Banados:1992gq}. This can be viewed as a solution of AdS$_3$ supergravities described above, where the Chern-Simons gauge fields are trivial, i.e., a neutral solution. The line element is given by    
\begin{equation}\label{btzm1}
    ds^2_{\mathrm{BTZ}} = -\r{(r^2-r_+^2)(r^2-r_-^2)}{\ell^2 r^2}\dd{t}^2 +
    \r{\ell^2 r^2}{(r^2-r_+^2)(r^2-r_-^2)}\dd{r}^2 + r^2\f(\dd{\phi} - \r{r_+ r_-}{\ell r^2}\dd{t})^2~,
\end{equation}
where $r_-$ and $r_+$ are the inner and outer horizons of the black hole. The mass and angular momentum carried by the black hole are given by
\begin{equation}\label{btz-parameters}
    M_{\rm BTZ} = \frac{r_+^2+ r_-^2}{8G\ell^2}~, \qquad J_{\rm BTZ} =  \frac{r_+ r_-}{4G\ell}~.
\end{equation}
Conjugate to these variables, the temperature and angular velocity read 
\begin{equation}\label{eq:potentials}
    T = \frac{r_+^2 -r_-^2}{2 \pi r_+ \ell^2 }~,\qquad \Omega = \r{r_-}{\ell r_+}~.
\end{equation}
Without loss of generality, we assume $r_- > 0$, which corresponds to choosing the direction of rotation, i.e., the sign of angular momentum.

In Euclidean signature, we will take $t=-it_{\mr{E}}$ and introduce the complex coordinates,
\begin{equation}\label{eq:wwbar-euc}
    w = i \frac{t_{\mr{E}}}{\ell} +\phi~, \qquad \bar w = -i \frac{t_{\mr{E}}}{\ell} +\phi~,
\end{equation}
where the periodicity of $w$ is $w\sim w+ 2\pi$ around the spatial cycle, wrapping the horizon of the black hole, and around the thermal cycle
\begin{equation}\label{eq:thermal-cycle}
     w\sim w+ 2\pi \tau~, \qquad \bar w\sim \bar w+ 2\pi \bar\tau~, 
\end{equation}
with $\tau$ and $\bar\tau$ are functions of the potentials \eqref{eq:potentials},
\begin{equation}\label{eq:taus-new}
    \tau = \frac{i\beta}{2\pi\ell}(1-\ell\Omega)~,\qquad \bar\tau = -\frac{i\beta}{2\pi\ell}(1+\ell\Omega)~,
\end{equation}
and $\beta=1/T$ the inverse temperature. We will also use the notations for left/right temperatures 
\begin{equation}\label{eq:lr-tempratures}
    \r{1}{T_L} \coloneqq  2\pi i \ell \bar{\tau}~,\quad
    \r{1}{T_R} \coloneqq  -2\pi i \ell \tau~,
\end{equation}
and their inverses, $\beta_{L,R}=1/T_{L,R}$, satisfying $\beta_L + \beta_R = 2\beta$.

Turning on $A$ and $A'$ is rather simple in three dimensions, and we follow the discussion in \cite{Kraus:2006wn,Banados:2015tft}. The equations of motion for the Chern-Simons gauge fields do not involve the metric, hence the line element \eqref{btzm1} remains unchanged. The task is to construct flat connections $A$ and $A'$ that satisfy appropriate boundary conditions on the asymptotic boundary of AdS$_3$ and are regular at the horizon.  To start, we consider the simplest case where the gauge group is $U(1)\times U(1)$. A suitable choice of flat connections in Euclidean signature are 
\begin{equation} \label{eq:u(1)-gauge}
    A_{\mf{u}(1)}=  a_{\bar w} \,\dd \bar w + a_w\,\dd w~, \qquad A'_{\mf{u}(1)}= a'_{w} \,\dd w + a'_{\bar w}\,\dd \bar w~,
\end{equation}
where $a_{\bar w}$, $a_w$ and primed counterparts are constant. The electrostatic potentials $(\mu,\mu')$ are carried by $a_{w}$ and $a'_{\bar w}$, i.e.,
\begin{equation}\label{eq:aw-potential}
    a_{w} = \frac{\mu}{2}~, \qquad a'_{\bar w} = \frac{\mu'}{2}~,
\end{equation}
and the electric charges are supported along the spatial direction,
\begin{equation}\label{eq:aphi-charge}
    a_{\phi} = GQ ~,\qquad a'_{\phi} = GQ'~.
\end{equation}
 The potentials are fixed by charges by demanding that the holonomy of each gauge field is trivial along the thermal cycle \eqref{eq:thermal-cycle}, which gives 
 \begin{equation}\label{eq:u(1)-potential}
    GQ = \frac{\mu}{1+\ell\Omega} ~,\qquad GQ' = \frac{\mu'}{1-\ell \Omega}~.
\end{equation}

Although $A$ and $A'$ do not alter the line element in \eqref{btzm1}, the definition of mass and angular momentum in \eqref{btz-parameters} is affected by the presence of Chern-Simons terms. As explained in \cite{Kraus:2006nb,Kraus:2006wn}, this is because the Chern-Simons fields contribute to the boundary stress tensor via boundary terms. The outcome is that the mass and angular momentum for the charged BTZ black hole are now given by 
\begin{equation}
M=  M_{\rm BTZ} + \frac{G}{4} \left(Q^2+ Q'^2\right) ~,\qquad  J=  J_{\rm BTZ} - \frac{G\ell}{4} \left(Q^2- Q'^2\right)~,
\end{equation}
where $M_{\rm BTZ}$ and $J_{\rm BTZ}$ are given in \eqref{btz-parameters}. With this, the first law of thermodynamics is
\begin{equation}\label{eq:first-law}
    dM = T dS_{\rm BH} + \Omega dJ + \frac{\mu}{2} dQ + \frac{\mu'}{2} dQ'~,
\end{equation}
with $S_{\rm BH}= \pi r_+/2G$ the Bekenstein-Hawking area law.

Finally, we can generalise the discussion for $U(1)\times U(1)$  to a non-Abelian theory where the gauge group is $SU(2)\times SU(2)$. In this case, we will label the generators as \eqref{eq:fund-su2}, and select, without loss of generality, $A^I$ and $A^{\prime I}$ to only have a component along $T_3$.\footnote{In non-abelian cases, we identify the appropriate electric charges as the eigenvalues of the holonomy around the non-contractible (spatial) cycle. Therefore, the simplest gauge to write $A$ and $A'$ is one where they have support only on the Cartan elements of the Lie algebra. For $\mathfrak{su}(2)$ this is $T_3$.} That is,
\begin{equation} \label{eq:gauge-fields-bh}
    A_{\mf{su}(2)}^{I=3}=  a_{\bar w}\, \dd \bar w + a_w\,\dd w~, \qquad A'^{I=3}_{\mf{su}(2)}= a'_{ w} \,\dd w + a'_{\bar w}\,\dd \bar w~.
\end{equation}
The rest of the analysis follows in a straightforward way to the Abelian case. See, for example, \cite{Larsen:2021wnu} for further details on the thermodynamical properties of the charged BTZ black hole, which are applicable for both the non-Abelian and Abelian cases.  

\paragraph{Extremal and supersymmetric black holes.} The notion of extremality is simple to specify and implement: it is the confluence of the inner and outer horizon, $r_+=r_-$. In terms of the thermodynamical variables, it follows from \eqref{eq:potentials} and \eqref{eq:u(1)-potential} that 
\begin{equation}
    T_{\rm ext}= 0 ~,\qquad \ell\Omega_{\rm ext} =1~, \qquad \mu_{\rm ext} = 2 G Q~. 
\end{equation}
Moreover, from \eqref{btz-parameters}, we have $\ell M_{\rm BTZ,ext} = J_{\rm BTZ,ext}$, and $T_L=0$ in \eqref{eq:lr-tempratures} while $T_R$ is finite. However, setting $r_+=r_-$ in the charged black holes requires further input: to assure regularity of both $A$ and $A'$, while keeping the electric charges $Q$ and $Q'$ finite in \eqref{eq:u(1)-potential}, we will also demand that at extremality 
\begin{equation}
    \mu'_{\rm ext}=0~.
\end{equation}
With this, the extremal bound on the charges of the black hole reads
\begin{equation}\label{eq:ext-bound}
    \ell M_{\rm ext} - J_{\rm ext}= \frac{G\ell}{2} \, Q^2~,
\end{equation}
in agreement with \cite{Banados:2015tft,Larsen:2021wnu}.

The notion of supersymmetry of the black hole is defined by the existence of a spinor $\epsilon$ such that, under the transformations \eqref{susy-trans}, the transformation of the fermionic fields is trivial. For both neutral and charged BTZ black holes, the analysis of supersymmetry has been discussed in, for example, \cite{Coussaert:1993jp,balasubramanian_supersymmetric_2001,Banados:2015tft}. Here we will provide a self-contained derivation that adapts to our conventions.

To start, let us consider solving for the condition $\delta \psi =0$ in \eqref{susy-trans-1}, which reads 
\begin{equation}\label{eq:kilspin-eq}
    \mc{D} \epsilon = 0~,
\end{equation}
where the covariant derivatives are defined in \eqref{eq:def-covariant}-\eqref{eq:def-covariant-1}. A solution to this equation defines a Killing spinor, and the number of independent solutions determines the amount of supersymmetry preserved by the background. (We can also require that $\delta \psi' =0$, and hence solve for $ \mc{D}'\epsilon' = 0$, which we will discuss momentarily.)  
For the neutral black hole, effectively the ${\cal N}=(1,1)$ theory, the solutions to \eqref{eq:kilspin-eq} are as follows.  The radial component of the Killing spinor equation reads
\begin{equation}\label{eq:ungauged-kilspin-radial}
    \partial_r\epsilon = \frac{r^2 + r_+r_-}{2r\sqrt{(r^2-r_+^2)(r^2-r_-^2)}}\gamma_1 \epsilon ~,
\end{equation}
while the time and spatial components are
\begin{equation}\label{eq:ungauged-kilspin-angular1}
\begin{aligned}
    \partial_{+} \epsilon &= G^a(r)\gamma_a \epsilon ~,\\
    \partial_{-} \epsilon &= 0 ~,
\end{aligned}
\end{equation}
where we introduced the null coordinates $x^{\pm} \coloneqq \phi \pm t/\ell$ and 
\begin{equation}
    G_a(r)\gamma^a:= \frac{1}{2\ell r}\f[\sqrt{(r^2-r_+^2)(r^2-r_-^2)}\gamma_0 + \f(r^2 - r_+r_-)\gamma_2]~.
\end{equation}
The general solution to (\ref{eq:ungauged-kilspin-angular1}) is
\begin{equation} \label{eq:spinor-gen}
    \epsilon(x) = \f[\cosh\f(\r{r_+ - r_-}{2\ell}x^+) + \frac{2\ell}{r_+-r_-} \sinh(\r{r_+ - r_-}{2\ell}x^+)G^a(r)\gamma_a]\tilde{\epsilon}(r)~,
\end{equation}
where the radial profile $\tilde{\epsilon}(r)$ is further fixed by the radial equation (\ref{eq:ungauged-kilspin-radial}). Requiring the spinor to be (anti-)periodic over the spatial circle $\phi \sim \phi +2\pi$ forces $r_+ = r_-$, i.e., it requires the black hole to be extremal. Finally, solving \eqref{eq:ungauged-kilspin-radial}, with $r_+=r_-$, gives  
\begin{equation}\label{eq:spinor-N=1}
    \epsilon(x) = \sqrt{\frac{r^2-r_+^2}{\ell r}}\bmqty{1 \\ 1}~,
\end{equation}
which agrees with the Killing spinor first constructed in \cite{Coussaert:1993jp}.

One could also impose that the primed sector is supersymmetric, by demanding
\begin{equation}\label{eq:kilspin-eq-primed}
  \delta \psi'=  \mc{D}'\epsilon' = 0~.
\end{equation}
This looks almost identical to the above analysis, with some changes arising from the differences in the definitions of the covariant derivative in \eqref{eq:def-covariant}. The final result is that a (anti-)periodic solution to that equation exists if and only if $r_+ + r_- = 0$, which can only be satisfied for the case of an extremal black hole rotating in the opposite direction ($\ell M_{\rm BTZ, ext}=-J_{\rm BTZ, ext} $). Since we are assuming $J_{\rm BTZ}$ (or equivalently $r_-$) to be positive, the primed sector does not preserve supersymmetry for non-zero mass and angular momentum. One should also note that in the special case of the massless black hole ($M_{\rm BTZ}=J_{\rm BTZ} = 0$), both (\ref{eq:kilspin-eq}) and (\ref{eq:kilspin-eq-primed}) have a solution---this is the only uncharged ``extremal'' black hole with more than one Killing spinor.

We next consider charged black holes, in which the gauge fields $A$ and $A'$ are non-trivial. Following the conventions in \eqref{eq:gauge-fields-bh}, we choose a gauge where $A$ only has a non-zero component along one of the generators of the Cartan subalgebra of the gauge algebra, which we will denote by $T_c$, and use $A^IT_I= A_{\mf{u}(1)}T_c$ with $A_{\mf{u}(1)}$ given by \eqref{eq:u(1)-gauge}. For example, for $U(1)$, we have $T_c = i$, and for $SU(2)$,  $T_c = T_3 = -\r{i}{2}\sigma_3$.  The temporal and angular equations in \eqref{eq:ungauged-kilspin-angular1} then become
\begin{equation}\label{eq:u1-kilspin-angular1}
\begin{aligned}
    \partial_{+} \epsilon &= \left(G^a(r)\gamma_a - a_{+}\,T_c\right) \epsilon ~,\\
    \partial_{-} \epsilon &= - a_{-}\,T_c \,\epsilon ~,
\end{aligned}
\end{equation}
where, using \eqref{eq:aw-potential}-\eqref{eq:u(1)-potential}, we have
\begin{equation}
  a_+ = \frac{GQ}{2} (1-\ell \Omega)~,\quad a_- = \frac{GQ}{2} (1+\ell \Omega)~.
\end{equation}
The radial component of the spinor equation \eqref{eq:ungauged-kilspin-radial} remains unchanged.
Therefore, for non-Abelian groups, the solution reads
\begin{equation}
    \epsilon^i(x) = U^i_{~j}(x) \, \varepsilon^j  \,\epsilon(x)~,
\end{equation}
where $\epsilon(x)$ is given by \eqref{eq:spinor-N=1}, and $\varepsilon^j$ are real constants for $(p,q)$ supergravity, while it is complex for ${\cal N}=(4,4)$. The group element appearing here is the holonomy of $A$, given by the path-ordered exponential of the gauge field:  
\begin{equation}\label{eq:charged-kilspin-ang-sol}
   U(x)= {\rm P}\exp\left(\int_{\gamma(x)} A_{\mf{u}(1)}T_c \right)~.
\end{equation}
Here $\gamma(x)$ is a curve connecting $x^\mu$ to some reference point, which can be adjusted by a choice of $\varepsilon^j$. In the case of $U(1)$, we are simply complexifying our variables, and hence it is more convenient to write the solution as
\begin{equation}
    \epsilon_{\rm U(1)}(x)= \exp\left(i\int_{\gamma(x)} A_{\mf{u}(1)} \right)\sqrt{\frac{r^2-r_+^2}{\ell r}}\bmqty{1 \\ 1}~,
\end{equation}
up to an overall complex normalisation.

Note that  $\epsilon^j(x)$ and $\epsilon_{\rm U(1)}(x)$ are proportional to the spinor \eqref{eq:spinor-N=1}, and therefore, portions of them are periodic as we transport them along the spatial circle $\phi\sim\phi+2\pi$ since we have imposed $r_+=r_-$ at this stage.
However, extremality, while necessary, is not sufficient. Demanding (anti-)periodicity along the spatial direction imposes a condition on the holonomy of $A$: if $\gamma(x)$ loops the spatial direction, we demand
\begin{equation}\label{eq:quanta}
    U(\phi\sim\phi+2\pi)=
    \begin{cases}
      ~\mathbb{1} &\text{periodic (R)} \\
      -\mathbb{1} &\text{anti-periodic (NS)}
    \end{cases}~,
\end{equation}
where we have identified the anti-periodic condition as Neveu-Schwarz (NS) or the periodic one as Ramond (R). Given our choices for the generators \eqref{eq:u1-rep-generator} and \eqref{eq:fund-su2}, for the case of $U(1)$ we have
\begin{equation}\label{eq:susy-charge-U(1)}
    GQ \in
    \begin{cases}
       \mathbb{Z} &\text{periodic (R)} \\
      \mathbb{Z} + \frac{1}{2} &\text{anti-periodic (NS)}
    \end{cases}~,
\end{equation}
and when the gauge group is $SU(2)$
\begin{equation}\label{eq:susy-charge-SU(2)}
    GQ \in
    \begin{cases}
       2\mathbb{Z} &\text{periodic (R)} \\
      2\mathbb{Z} + 1 &\text{anti-periodic (NS)}
    \end{cases}~.
\end{equation}
Similar conclusions can be drawn for $SO(p)$. As discussed in detail in \cite{Banados:2015tft}, the quantisation condition \eqref{eq:quanta} combined with the extremal bound \eqref{eq:ext-bound} perfectly agrees with unitarity bounds derived from superconformal algebras in CFT$_2$. In particular, the relation to the notation used there is
\begin{equation}
    h-\frac{c}{24}=\frac{1}{2}(\ell M-J)~, \qquad q = \frac{\ell}{2} Q~.
\end{equation}
With this, the integer $\mathbb{Z}$ appearing in \eqref{eq:susy-charge-U(1)} and \eqref{eq:susy-charge-SU(2)} represents spectral flow sectors, which intersect once the bound \eqref{eq:ext-bound}.

\paragraph{Near-horizon region of near-extremal and near-BPS black holes.} The last property we will discuss is how the extremal and supersymmetric black holes respond as we turn on a small amount of temperature. This will define the near-extremal and near-BPS backgrounds.

Both extremal and supersymmetric black holes share the feature of being solutions at zero temperature, and their line element is identical. A convenient way to describe the decoupling limit that zooms into their near-horizon geometry is to start from the non-extremal BTZ and use temperature as the decoupling parameter; see, for example, \cite{KapLaw24,KolMar24}. Explicitly, for the Euclidean BTZ black hole, with $r_-$ real, we introduce the coordinate system 
\begin{equation} \label{eq:decouple-coords}
r= r_+ + \frac{\pi  \ell^2}{2} T \left( \cosh \eta -1\right)~,\qquad t_{\mr{E}} = \frac{\hat t_{\mr{E}}}{2 \pi  T}~, \qquad \phi = \hat\phi-i \hat t_{\mr{E}} \left(\frac{1}{2 \pi  \ell T}-\frac{\ell}{r_++r_-}\right)~.
\end{equation}
Note that the periodicities of the spatial and thermal cycles of Euclidean BTZ in this coordinate system read 
\begin{equation}\label{eq:nhg-trivial-periodicity}
    (\hat{t}_E,\hat{\phi})
    \sim(\hat{t}_E,\hat{\phi} + 2\pi)
    \sim(\hat{t}_E+2\pi,\hat{\phi})~.
\end{equation}
Thus, we are placing all the information about temperature and potentials in the line element, which facilitates taking a zero temperature limit.
In that regard, expanding the parameters of BTZ at low temperature and fixed $J_{\rm BTZ}$ we will use
\begin{equation}\label{eq:rpm-btz}
    r_\pm = \ell \sqrt{4 G M_{\rm BTZ}} \left(1\pm\sqrt{1-\frac{J_{\rm{BTZ}}^2}{\ell^2 M_{\rm BTZ}^2}}\right)^{\frac{1}{2}}~, 
\end{equation}
and
\begin{equation}\label{eq:low-t-mass}
    M_{\rm BTZ} = \frac{J_{\rm{BTZ}}}{\ell}+\frac{\pi ^2 \ell^2 T^2}{8G}+ \frac{\pi ^3  \ell^4 T^3}{16 G \sqrt{G J_{\rm{BTZ}} \ell}}+ \frac{3 \pi ^4 \ell^5 T^4}{2 (8G)^2 J_{\rm{BTZ}}} + O(T^5)~.
\end{equation}
Implementing the coordinate transformation \eqref{eq:decouple-coords} on \eqref{btzm1} and  taking the limit $T\to 0$, with $J_{\rm BTZ}$ fixed, gives to leading order in temperature
\begin{equation}\label{ebtz-expansion}
    ds^2_{\mr{EBTZ}} = ds^2_{\mr{ENHG}} + T\, ds^2_{1} + O(T^2)~.
\end{equation}
The first contribution corresponds to the Euclidean Near Horizon Geometry (ENHG),
\begin{equation}\label{Enhorizon}
    ds^2_{\mathrm{ENHG}} =  \frac{\ell^2}{4}\left(\sinh^2 \eta ~ \dd{\hat t_{\mr{E}}}^2 +
    \dd{\eta}^2\right) + r_0^2\left(~ \dd{\hat\phi} - i\frac{\ell}{2r_0}\left(\cosh\eta - 1\right) ~\dd \hat t_{\mr{E}}\right)^2~, 
\end{equation}
where $r_0^2= 4G \ell J_{\rm BTZ}$. 
This portion contains the AdS$_2$ factor that is characteristic of the NHG of extremal black holes, with radius $\ell_2 = \ell/2$.  Accompanying it is the $S^1$ fibration of length $r_0$, the extremal horizon; for the case of BTZ, this fibration has an imaginary part, making the ENHG intrinsically a complex background. This last fact will be important in our subsequent analysis of zero modes in Sec.\,\ref{sec:near}. 
The leading response in temperature in \eqref{ebtz-expansion} is given by
\begin{multline}\label{nhg-correction}
 ds^2_1 =  \frac{ \pi  \ell^4 }{8 r_0} (\cosh \eta +2) \tanh ^2\left(\frac{\eta }{2}\right) \dd\eta^2+\frac{\pi  \ell^4 }{2 r_0} \sinh ^4\left(\frac{\eta }{2}\right)
 \dd{\hat t_{\mr{E}}}^2 \\+\pi   \ell^2 r_0 \cosh \eta  \dd\hat \phi^2   
 -\frac{i}{2}  \pi  \ell^3 \sinh ^2\left(\frac{\eta }{2}\right) (\cosh \eta -3) \dd{\hat t_{\mr{E}}} \dd\hat\phi~.
\end{multline}

For charged black holes, we also have to inspect the behaviour of the background gauge fields. To start, consider the simplest case of $u(1)\times u(1)$; implementing the coordinate transformation \eqref{eq:decouple-coords} on the connections \eqref{eq:u(1)-gauge} gives 
\begin{equation}\label{eq:A-enhg}
    \begin{aligned}
        A_{{\mf{u}(1)}} = GQ\, \dd \hat \phi ~,\qquad A'_{{\mf{u}(1)}} = GQ' \,\dd \hat \phi ~,
    \end{aligned}
\end{equation}
where we used as well \eqref{eq:u(1)-potential}.  Generalisation to non-Abelian fields is straightforward.   
Note that in an ensemble of fixed $(Q,Q')$, there is no temperature dependence  \eqref{eq:A-enhg}. This implies that the gauge fields will play no explicit role in the classical backreaction in the canonical ensemble. Its point of entry in subsequent sections will arise from the quantisation condition \eqref{eq:quanta} imposed by supersymmetry, which will affect the periodicity conditions on the fluctuations considered.

\section{GPI at one loop}\label{sec:one-loop}

In this section, we collect properties of the Euclidean gravitational path integral (GPI), including one-loop effects. We will be working with the supergravity theories introduced in Sec.\,\ref{sec:sugra-bhs}, hence the GPI we will consider reads
\begin{align}\label{eq:path-integral}
    Z = \int \left[ Dg \right]\left[DA_{\rm CS}\right] \left[D\psi_{3/2}\right] e^{-S_{\rm 3D}}\,,\qquad  
S_{\rm 3D}= S_E+S_{\text{boundary}}+S_{\text{gauge}}\,.
\end{align}
Here, $S_E$ is the Euclidean action introduced in \eqref{eq:Euclidean-3D-action}, depending on metric $g_{\mu\nu}$, gauge fields $A_{\rm CS}= \{A, A'\}$, and gravitini $\psi_{3/2}=\{\psi,\psi'\}$. The term $S_{\text{boundary}}$ imposes Dirichlet or Neumann boundary conditions on the fields as appropriate; the term $S_{\text{gauge}}$ ensures gauge-fixing of the metric, the gauge fields, and the gravitini.

We will evaluate $Z$ by saddle point approximation, and focus our attention on the leading quantum effects appearing at one-loop. Concretely, we will consider a bosonic solution to the equation of motion, given by $(\bar g\,,\bar A)$ and decompose the bosonic fields as
\begin{equation}
    g = \bar g + h ~, \qquad A_{\rm CS} = \bar A + a~, 
\end{equation}
with $h$ and $a$ the quantum fluctuations. 
To quadratic order in the fluctuations, the GPI around this saddle point takes the form 
\begin{align}
    Z \approx Z_0[\bar g, \bar A ] Z_{\rm grav} Z_{\rm CS} Z_{3/2} + \cdots~.
\end{align} 
The classical contribution is encoded in $Z_0$, 
\begin{equation} \label{eq:tree-level}
    Z_0=\exp(S_{\mr{3D}}[\bar g, \bar A])~,
\end{equation}
which evaluates the on-shell action with appropriate boundary conditions, while the quadratic fluctuations are split into three integrals
\begin{equation}
\begin{aligned}\label{eq:one-loop}
    Z_{\rm grav} [\bar g]&=  \int \left[ Dh \right] e^{-S_{\rm grav}^{(2)}[h;\bar g]}~,\\
    Z_{\rm CS}[ \bar A, \bar g] &=  \int \left[ Da \right] e^{-S_{\rm CS}^{(2)}[a; \bar g, \bar A]}~,\\
    Z_{3/2}[ \bar A, \bar g] &=  \int \left[ D\psi_{3/2} \right] e^{-S_{\rm 3/2}^{(2)}[\psi_{3/2};\bar g, \bar A]}~.
\end{aligned}
\end{equation}
Notice that there is no mixing between the fluctuations of the bosonic and fermionic fields. This is due to the topological nature of the three-dimensional action we are considering in \eqref{lor-sugra}. In the following, we discuss the relevant formulas and definitions that enter into each of the terms in \eqref{eq:one-loop}. In particular, we will review the ingredients behind each of the quadratic actions appearing and their accompanying ghost fields, which, for brevity, have been omitted in \eqref{eq:one-loop}.

\subsection{Graviton}\label{sec:graviton-oneloop}
We start by quantifying the gravitational fluctuations in $Z_{\rm grav}$. This follows the standard references such as \cite{gibbons_quantizing_1978,christensen_quantizing_1980}; see also \cite{Giombi:2008vd,David:2009xg} for a discussion focused on three dimensions. The quadratic fluctuations appearing in \eqref{eq:one-loop} arise from the Einstein-Hilbert action
\begin{equation}
S_{\mr{EH}}[g] \coloneqq -\r{1}{16\pi G}\int \star \left(R+\r{2}{\ell^2}\right)~.
\end{equation}

Expanding up around a general background solution $\bar{g}_{\mu\nu}$ gives
\begin{equation}
    S^{(2)}_{\rm EH}[h;\bar{g}] = \r{1}{16\pi G} \int\bar{\star}
    \left[ 
    \ti{h}^{\mu\nu}(\bar{\Delta}_{\mr{grav}} h)_{\mu\nu}
    - \r{1}{2}(\bar{\nabla}^\nu \ti{h}_{\nu\mu})(\bar{\nabla}^\lambda \ti{h}_{\lambda}{}^\mu)
    \right]
~,
\end{equation}
where
\begin{equation}
\ti{h}_{\mu\nu} \coloneqq h_{\mu\nu} - \r{1}{2}\bar{g}_{\mu\nu} h^\lambda{}_\lambda~,
\end{equation}
and
\begin{equation}\label{eq:lichnerowicz-def}
    (\bar{\Delta}_{\mr{grav}} h)_{\mu\nu} \coloneqq 
    \r{1}{4}
    \f(-\bar{\Box} h_{\mu\nu} + 2\bar{R}_{\lambda(\mu} h^\lambda{}_{\nu)} - 2\bar{R}_{\mu\lambda\nu\sigma}h^{\lambda\sigma})
    - \f(\bar{R}_{\lambda(\mu} - \r{1}{4}\bar{g}_{\lambda(\mu}\bar{R})h^\lambda{}_{\nu)} + \r{ h_{\mu\nu}}{2\ell^2}~.
\end{equation}
Here all indices are raised and lowered with $\bar{g}_{\mu\nu}$, and the symbol with bar on top, such as $\bar \nabla_\mu$, $\bar\star$, and $\bar\Box$, represent operators with respect to $\bar{g}_{\mu\nu}$.\footnote{Note that there is a slight abuse of notation. For spinors, the bar means adjoint, while on bosonic operators and fields, it means background solution. We hope that the context makes clear the meaning.}  To remove gauge redundancies, we will be using the harmonic gauge
\begin{equation}\label{eq:eh-gauge}
    \bar{\nabla}^\mu \ti{h}_{\mu\nu} = 0~.
\end{equation}
This adds the following gauge-fixing and ghost terms to the action
\begin{equation}
  S_{\mr{gauge,grav}}[h,\eta,\eta^*;\bar{g}] \coloneqq \r{1}{32 \pi G}\int\bar{\star}
    \left[
        (\bar{\nabla}^\nu \ti{h}_{\nu\mu})(\bar{\nabla}^\lambda \ti{h}_{\lambda}{}^\mu)
        + {\eta^*}^\mu(\bar{\Delta}_{\mr{gh}} \eta)_\mu
    \right]~,
\end{equation}
with $\eta_\mu$ and $\eta^*_\mu$ the fermionic ghost fields, where we have 
\begin{equation}
    (\bar{\Delta}_{\mr{gh}} \eta)_\mu \coloneqq 
    -\f(\bar{\Box} \eta_\mu + \bar{R}_{\mu\nu}\eta^\nu)~.
\end{equation}
At this stage, it is convenient to rescale our fields by $h\to \kappa h$ and $\eta\to \kappa \eta$ with $\kappa^2=32\pi G$. Separating the metric in a traceless and trace part, i.e., $h_{\mu\nu} = \hat h_{\mu\nu} + g_{\mu\nu} \phi$ and $ \hat h^\mu_{~\mu}=0$, the graviton contribution to the one-loop partition function is
\begin{equation}\label{eq:eh-partition-function}
\begin{split}
    Z_{\mr{grav}} &= \int [D \hat h] [D \phi] [D \eta][D{\eta^*}] e^{-S^{(2)}_{\mr{EH}}[h;\bar{g}]-S_{\mr{gauge,EH}}[h,\eta,\eta^*;\bar{g}]} \\
    &= \r{\det(\bar{\Box}_1 - \r{2}{\ell^2})_{\rm gh}}{\sqrt{\det(\bar{\Box}_2 + \r{2}{\ell^2})_{\hat h} \det(\bar{\Box}_0 - \r{4}{\ell^2})_\phi}} ~,
\end{split}
\end{equation}
where $\bar\Box_{j}=\bar\nabla_\mu \bar\nabla^\mu$ is the Laplacian acting on a rank-$j$ tensor. 
In simplifying these expressions we have used that $\bar{g}_{\mu\nu}$ is a solution to Einstein equations, i.e., $\bar{R}_{\mu\nu} = -\r{2}{\ell^2}\bar{g}_{\mu\nu}$. 
The determinants in \eqref{eq:eh-partition-function} are taken over the space of all symmetric-traceless modes $\hat h_{\mu\nu}$, scalars $\phi$, and vectors $\eta_\mu,{\eta}^*_\mu$, that are normalizable and satisfy appropriate boundary conditions. That said, if one decomposes $\hat h_{\mu\nu}$ into the transverse-traceless (TT) and longitudinal parts, and similarly decomposes $\eta_\mu$,${\eta}^*_\mu$ into a divergence-free and a gradient part, the determinants then factorise into products of     determinants on these subspaces. This cancels  some common contributions in each determinant and leads to the standard result 
\begin{equation}\label{eq:eh-partition-function-reduced}
    Z_{\mr{grav}}[\bar{g}] = \sqrt{\r{\det(\eval{(\bar{\Box}_1 - \r{2}{\ell^2})}_{\mr{T}})}{\det(\eval{(\bar{\Box}_2 + \r{2}{\ell^2})}_{\mr{TT}})}}~,
\end{equation}
where $\eval{\cdot}_{\mr{TT}}$ denotes the restriction of the operator to the transverse traceless perturbations, where $\nabla^\mu h_{\mu\nu} = h^\mu{}_\mu = 0$, and $\eval{\cdot}_{\mr{T}}$ to the divergence-free perturbations, that is $\nabla^\mu \eta_\mu = 0$. One important aspect that is implicit in \eqref{eq:eh-partition-function} and \eqref{eq:eh-partition-function-reduced} is the potential appearance of ``zero modes,'' that is, the possibility that the spectrum of the operators appearing in the determinants might have normalizable eigenmodes with zero eigenvalue. This is a key part of our subsequent analysis and will be discussed in detail in Sec.\,\ref{sec:near} and Sec.\,\ref{sec:far}.

\subsection{Chern-Simons gauge fields}\label{sec:gauge-oneloop}
The second bosonic contribution appearing in \eqref{eq:one-loop}  is due to the two copies of the Chern-Simons terms, for the fields $A$ and $A'$ in \eqref{lor-sugra}. For simplicity, we will focus on just the contribution of a single gauge field, and our discussion here follows \cite{Witten:1988hf,bar-natan_perturbative_1991,nash_differential_1991,grabovsky_chern_2022,Porrati:2019knx}. The action is given by 
\begin{equation}\label{eq:cs-euc-action}
   S_{\mr{CS}}[A] \coloneqq \r{i\ell}{16\pi G}\int_M \Tr(A\wedge \dd{A} + \r{2}{3}A\wedge A\wedge A)~,
\end{equation}
with $A$ a one-form valued in the Lie-algebra $\mathfrak{g}$.
The saddle points (classical solutions) of $S_{\mr{CS}}[A]$ are flat connections, which we denote as $\bar{A}$. The expansion of \eqref{eq:cs-euc-action} around such a solution is
\begin{equation}
    S_{\mr{CS}}[\bar{A} + a] = S_{\mr{CS}}[\bar{A}] + S^{(2)}_{\mr{CS}}[a;\bar{A}] + \mc{O}(a^3)~,
\end{equation}
where $S_{\mr{CS}}[\bar{A}]$ is the on-shell action, which would be included in the classical contribution \eqref{eq:tree-level}, and the quadratic fluctuations contribute as
\begin{equation} \label{eq:two-CS}
    S^{(2)}_{\mr{CS}}[a;\bar{A}] \coloneqq \r{i\ell}{16\pi G}\int_M \Tr(a\wedge \bar{D} a)~.
\end{equation}
Here $\bar{D}$ is the covariant exterior derivative acting on forms valued in the algebra,
\begin{equation}\label{eq:ext-cov-der}
    \bar{D} \omega \coloneqq \r{1}{p!}\bar{D}_\mu \omega_{\nu_1\cdots\nu_p} \dd{x^\mu}\wedge\dd{x^{\nu_1}}\wedge\ldots\wedge\dd{x^{\nu_p}}~,
\end{equation}
for a $p$-form $\omega$, where
\begin{equation}\label{eq:def-D-A}
    \bar{D}_{\mu} \coloneqq \bar\nabla_{\mu} + [\bar{A}_\mu, \cdot]~,
\end{equation}
acting on general $\mathfrak{g}$-valued tensor. 
Similarly to the case of the Einstein-Hilbert action, to calculate the path integral, one has to employ a gauge-fixing procedure. Following \cite{Witten:1988hf,bar-natan_perturbative_1991}, we impose the Lorentz gauge
\begin{equation}\label{eq:cs-gauge}
    \bar{D}_\mu a^\mu = 0~,
\end{equation}
by introducing a Lagrange multiplier in the form of a 3-form $\phi$, as well as ghost fields $c,c^*$.  This results in the action
\begin{equation}\label{eq:gauge-CS}
     S_{\mr{gauge,CS}}[a,\phi,c,c^*;\bar{A},\bar{g}] \coloneqq \int_M \Tr(\r{2i \ell}{16\pi G}\phi\bars{\star} \bar{D}\bars{\star} a + c^*\,\bar{D}\bars{\star} \bar{D} c)~.
\end{equation}
Combining \eqref{eq:two-CS} and \eqref{eq:gauge-CS}, the  one-loop path integral for the Chern-Simons fields reads
\begin{equation}\label{eq:CS-oneloop1}
\begin{split}
    Z_{\mr{CS}}[\bar{A},\bar{g}] &= \int[D a][D\phi][D c][D c^*] e^{-S_{\mr{CS}}^{(2)}[a;\bar{A}]-S_{\mr{gauge,CS}}[a,\phi,c,c^*;\bar{A},\bar{g}]} \\
    &= \int[D H][D c][D c^*] \exp[-\int_M\Tr(i H\wedge\bars{\star}\bar{L}_- H - c^*\,\bar{D}^\mu\bar{D}_\mu c)] \\
    &= e^{i\theta[\bar{g},\bar{A}]}\r{\det(-(\bar{D}^\mu\bar{D}_\mu)_0)}{\sqrt{\abs{\det(\bar{L}_-)}}}~,
\end{split}
\end{equation}
where the subscript in $-(\bar{D}^\mu\bar{D}_\mu)_0$ is meant to denote that the operator is acting on $\mathfrak{g}$-valued scalars, and the field $H =(a,\phi)$ belongs to the direct product of the spaces of $\mathfrak{g}$-valued one-forms and three-forms. Note that we have rescaled $a$ and $\phi$ by an appropriate factor of $G/\ell$, and in the second line of \eqref{eq:CS-oneloop1} we have included the framing anomaly contained in $\theta[\bar{g},\bar{A}]$.  The operator $\bar{L}_-$ acts on $H$ and it is defined as
\begin{equation}
    \bar{L}_- \coloneqq (\bars{\star}\bar{D} + \bar{D}\bars{\star})\circ J~,
\end{equation}
where $J(a,\phi) \coloneqq (a,-\phi)~$, or less abstractly, we have
\begin{equation}\label{eq:lminus-explicit}
    \bar{L}_-(a,\phi) \coloneqq (\bars{\star}\bar{D} a - \bar{D}\bars{\star}\phi,~\bar{D}\bars{\star}a)~.
\end{equation}
To further simplify \eqref{eq:CS-oneloop1}, we want to decompose $\det(\bar{L}_-)$ by analysing its invariant subspaces. We can use the fact that for the exterior covariant derivative \eqref{eq:ext-cov-der} with a flat connection $\bar{A}$ of a compact Lie group acting on forms valued in the adjoint representation, Hodge theory holds \cite{atiyah_yang_1983,nash_differential_1991}. In particular any $\mathfrak{g}$-valued form can be decomposed into exact (with respect to $\bar{D}$), coexact (with respect to $\bars{\delta} = (-1)^p\bars{\star}\bar{D}\bars{\star}$ acting on $p$-forms) and harmonic (with respect to $\bars{\delta}\bars{D} + \bars{D}\bars{\delta}$) components.
Using this the space $V$ of all fields $H = (a,\phi)$ can be decomposed into subspaces invariant under $\bar{L}_-$, that is, $V = V_{\mr{T}} \oplus V_{\mr{g}}$, where $V_{\mr{T}}$ is the gauge-fixed part, and $V_{\mr{g}}$ contains gauge transformations:
\begin{equation}
  V_{\mr{T}} = \set{(a,0) | \bar{D}^\mu a_\mu = 0}~,\quad
  V_{\mr{g}} = \set{(\bar{D} f,\phi ) | f \in \Omega^0(M,\mathfrak{g}), \phi \in \Omega^3(M,\mathfrak{g})}~.
\end{equation}
The determinants of $\bar{L}_-$ on these spaces satisfy
\begin{equation}\label{eq:det-lminus-restricted}
  \abs{\det(\bar{L}_-\big|_{V_\mr{T}})} = \abs{\det(\bars{\star}\bar{D}\big|_{\mr{T}})}~,\quad
  \abs{\det(\bar{L}_-\big|_{V_\mr{g}})} = \det(-(\bar{D}^\mu\bar{D}_\mu)_0)~,
\end{equation}
where $\eval{\cdot}_{\mr{T}}$ denotes the restriction to the space of divergence-free perturbations, i.e., those satisfying the gauge \eqref{eq:cs-gauge}. The first of the above equalities follows immediatly from inspecting \eqref{eq:lminus-explicit} restricted to $\phi = 0$ and $\bar{D}\bars{\star}a = \bars{\star}\bar{D}^\mu a_\mu = 0$. The second can be shown by examining the eigenvalue equation
\begin{equation}\label{eq:lminus-vg-eigeneq}
  \lambda(\bar{D}f,\phi) = \bar{L}_-(\bar{D}f,\phi) = (-\bar{D}\bars{\star}\phi, \bars{\star} \bar{D}^\mu\bar{D}_\mu f)~,
\end{equation}
which implies that (1) $\lambda f + \bars{\star}\phi$ is closed with respect to $\bar{D}$, and (2) $\lambda \phi = \bars{\star}\bar{D}^\mu\bar{D}_\mu f$. The space of solutions to \eqref{eq:lminus-vg-eigeneq} is spanned by the modes with $\lambda \ne 0$
\begin{equation}\label{eq:lminus-modes1}
  (\bar{D} f, -\lambda \bars{\star} f) \in V_{\mr{g}}~,\quad\text{where}\quad \bar{D}^\mu\bar{D}_\mu f = -\lambda^2 f~,
\end{equation}
which represent local gauge transformations $a \rightarrow a + \bar{D}f$ and have eigenvalue $\lambda$ with respect to $\bar{L}_-$, as well as zero modes of $\bar{L}_-$
\begin{equation}\label{eq:lminus-modes2}
  (0,\bars{\star} \alpha) \in V_{\mr{g}}~,\quad \bar{D}^\mu\bar{D}_\mu \alpha = 0~,
\end{equation}
since for 0-forms $\ker\bar{D} = \ker(-\bar{D}^\mu\bar{D}_\mu)_0$. The second equation in \eqref{eq:det-lminus-restricted} then follows from noting that every mode non-zero mode of $(-\bar{D}^\mu\bar{D}_\mu)_0$ with eigenvalue $\lambda^2$ gives rise to exactly two modes of $\eval{\bar{L}_-}_{V_{\mr{g}}}$ \eqref{eq:lminus-modes1} with eigenvalues $\pm \lambda$, while every zero-mode of $(-\bar{D}^\mu\bar{D}_\mu)_0$ (if there are any) gives rise to a unique zero mode of $\eval{\bar{L}_-}_{V_{\mr{g}}}$ of the form \eqref{eq:lminus-modes2}. The identities \eqref{eq:det-lminus-restricted} allow us to further simplify the ratios of determinants in \eqref{eq:CS-oneloop1} to the following expression
\begin{equation}\label{eq:cs-partition-function}
    Z_{\mr{CS}}[\bar A, \bar g] = e^{i\theta[\bar{g},\bar{A}]}\sqrt{\r{\det(-(\bar{D}^\mu\bar{D}_\mu)_0)}{\abs{\det(\bar{\star}\bar{D}\big|_{\mr{T}})}}}~.
\end{equation}
A similar expression holds for the gauge field $A'$.

\subsection{Gravitini}\label{sec:gravitini-oneloop}

For this portion of the analysis, we will be following the discussion in \cite{David:2009xg,Creutzig:2011fe}. The fermionic parts of the action entering in \eqref{eq:one-loop} are
\begin{equation}\label{eq:fermionic-action}
\begin{aligned}
    S_{\mr{3/2}}[\psi,A,g] &\coloneqq -\r{i}{16\pi G}\int_M \bar{\psi}\wedge{\mc{D}}\psi~,\\
    S'_{\mr{3/2}}[\psi',A',g] &\coloneqq -\r{i}{16\pi G}\int_M \bar{\psi}'\wedge{\mc{D}}'\psi'~,
\end{aligned}
\end{equation}
where the covariant derivatives are defined in \eqref{eq:def-covariant}-\eqref{eq:def-covariant-1}.
These actions are already quadratic in $\psi$ and $\psi'$, so they can be directly used to compute the one-loop expansion of the path integral around a fixed bosonic background $(\bar A, \bar g)$. However, one needs to take into account that when the background fields $\bar{g}$, $\bar{A}$ satisfy the classical equations of motion, the supersymmetry transformation \eqref{susy-trans} only affects the fermions, and thus around such saddles the variations
\begin{equation}
    \delta \psi_\mu = \bar{\mc{D}}_\mu \epsilon~,\quad
    \delta \psi'_\mu = \bar{\mc{D}}'_\mu \epsilon'~,
\end{equation}
do not change the value of the action. This redundancy has to be removed from the path integral using the Fadeev-Popov procedure, analogous to the diffeomorphisms and Chern-Simons gauge transformations in the previous subsections. The process involves decomposing the Rarita-Schwinger field into a ``traceless'' and ``pure trace'' parts,
\begin{equation}
    \psi_a = \hat{\psi}_a + \r{\gamma_a}{3}\gamma^b\psi_b~,\quad
    \gamma^a \hat{\psi}_a = 0~,
\end{equation}
and introducing bosonic Fadeev-Popov ghosts to fix $\gamma^a \psi_a = 0$. The traceless part can further be separated into a transverse and a longitudinal part as follows
\begin{equation}
    \hat\psi_a = \hat\psi^{\perp}_a + \left(\bar{\mc{D}}_a - \frac{1}{3}\gamma_a \bar{\slashed{\mc{D}}}\right)\xi, ~~~~ \text{where} \qquad D^a\hat\psi_a^{\perp} = \gamma^a\hat\psi_a^{\perp}=0~.
\end{equation} 
After including the one-loop contributions from ghosts and appropriately rescaling the fields, we finally obtain \cite{Creutzig:2011fe}
\begin{equation}
    Z_{3/2} [\bar g,\bar A]= \sqrt{\r{\det({\slashed{\bar D}}-\r{1}{2\ell})_{\rm TT}}{\det({\slashed{\bar D}} - \r{3}{2\ell})_{\mr{gh}}}}  ~,
\end{equation} 
where $.|_{\rm TT}$ refers to transverse $\gamma$-traceless modes. Here we focused on $\psi$; the analogous expression for $\psi'$ is straightforward and including that leads to
\begin{equation}\label{psipsibarf}
\begin{aligned}
    Z[\bar g, \bar A, \bar A'] &= Z_{3/2} [\bar g, \bar A]\, Z'_{3/2} [\bar g, \bar A']\\ &=\sqrt{\r{\det({\slashed{\bar D}}-\r{1}{2\ell})_{\rm TT}}{\det({\slashed{\bar D}} - \r{3}{2\ell})_{\rm gh}}
    \r{\det({\slashed{\bar D'}}+\r{1}{2\ell})_{\rm TT}}{\det({\slashed{\bar D'}} + \r{3}{2\ell})_{\rm gh}}}~.
\end{aligned}
\end{equation}

While \eqref{psipsibarf} is the expression we will use to identify the contribution from the zero modes, it is worth commenting on how it relates to the result in \cite{David:2009xg} where the background gauge fields are zero, i.e., just ${\cal N}=(1,1)$ supergravity. In that case, \eqref{psipsibarf}  simplifies to 
\begin{equation}\label{eq:ferm-one-loop-N=1}
    Z_{\mathcal{N}=(1,1)} [\bar g]= \sqrt{\r{\det(\bar{\Box}_{3/2}+\r{9}{4\ell^2})_{\rm TT}}{\det(\bar{\Box}_{1/2} - \r{3}{4\ell^2})_{\mr{gh}}}}~,
\end{equation}
which is the form used in \cite{David:2009xg} to derive the familiar expression for the partition function in terms of the super-Virasoro characters.

\section{GPI at low-temperature: near-horizon analysis}\label{sec:near}

In this section, we will study the behaviour of the GPI at zero temperature in the near-horizon region of the black hole solutions considered in Sec.\,\ref{sec:bhs}, and then quantify how the GPI responds when we slightly turn on temperature in the canonical ensemble, where electric charges and angular momentum are held fixed.   Our approach in this section closely follows the methods and ideas presented in \cite{Sen2012b,Iliesiu:2022onk,Banerjee:2023quv}, which we briefly summarise. 

As we reviewed in Sec.\,\ref{sec:one-loop}, the one-loop contribution to the GPI schematically reads
\begin{equation}
    Z_{\rm 1-loop} \sim \frac{1}{\sqrt{\det(\bar\Delta)}}~,
\end{equation}
where  $\bar \Delta$ is a quadratic operator, controlled by the background solution, acting on a bosonic field---for fermionic fields, the determinants are placed on the numerator, and the operator is linear in derivatives---and appropriate ghost factors are included to remove local gauge redundancies. There are multiple ways to evaluate $Z_{\rm 1-loop}$. Here, we want to think of it as a spectral problem, where the task at hand is to solve for eigenfunctions $\varphi_n$ and eigenvalues $\lambda_n$ of the operator, i.e., 
\begin{equation}\label{eq:eigenval-eq}
    \bar\Delta \, \varphi_n = \lambda_n\, \varphi_n ~.
\end{equation}
The eigenfunctions are {\it normalizable} and satisfy the appropriate {\it periodicity} and {\it regularity} conditions dictated by the background. We should also impose {\it boundary conditions} on the eigenfunctions that are compatible with those imposed on the GPI to start with; for the time being, this means that we will be working mainly in the canonical ensemble. \footnote{We will discuss the role of boundary conditions in more detail in Sec.\,\ref{sec:far}, as the interpretation of them is more evident when the modes are inspected from far away.} Therefore, the one-loop determinant is formally 
\begin{equation}
    Z_{\rm 1-loop} \sim \prod_n \lambda_n^{-1/2}~,
\end{equation}
where the infinite product needs to be appropriately regulated, and we have assumed a suitable definition of inner product, with
\begin{equation}
    \langle \varphi_n | \varphi_m \rangle < \infty~.
\end{equation}

The background solution we will work with in this section is the near-horizon metric \eqref{ebtz-expansion}-\eqref{nhg-correction}, together with background gauge fields \eqref{eq:A-enhg}.  The idea is to quantify the leading order response of $\lambda_n$ starting from $T=0$, hence we will write 
\begin{equation}\label{eq:expand-operator}
\begin{aligned}
 \bar\Delta = \bar\Delta^{(0)} + T\, \bar\Delta^{(1)} + O(T^2)~.
\end{aligned}
\end{equation}
At zeroth order, the operator $\bar\Delta^{(0)}$ is evaluated on the ENHG \eqref{Enhorizon} and $\delta \bar \Delta = T \bar\Delta^{(1)}$ is the contribution arising from correcting the background by \eqref{nhg-correction}. Similarly, expanding the eigenmodes and eigenvalues,
\begin{equation}
\begin{aligned}
\varphi_n = \varphi_{n}^{(0)} + T\, \varphi_n^{(1)}+ O(T^2)~,\\
\lambda_n = \lambda_{n}^{(0)} + T\, \lambda_n^{(1)}+ O(T^2)~,  
\end{aligned}
\end{equation}
where $\bar\Delta^{(0)}\varphi_{n}^{(0)}=  \lambda_{n}^{(0)} \varphi_{n}^{(0)}$, and the modes $\varphi_{n}^{(0)}$ are normalizable, the leading order in temperature effect is controlled by
\begin{equation}\label{eq:eigenval-expansion}
    \bar\Delta^{(0)} \varphi_n^{(1)} + \bar\Delta^{(1)} \varphi_{n}^{(0)} = \lambda_{n}^{(1)}\varphi_n^{(0)} +\lambda_{n}^{(0)}\varphi_n^{(1)} ~.
\end{equation}
Hence, the first-order correction to the eigenvalues is
\begin{equation}\label{eq:lambda2}
\lambda_{n}^{(1)}=
\frac{\langle \varphi_{n'}^{(0)} \,|\, \bar\Delta^{(1)} \varphi_n^{(0)}\rangle}
{\langle \varphi^{(0)}_{n'}|\varphi^{(0)}_n\rangle}
+
\frac{\langle \varphi_{n'}^{(0)} \,|\, \bar\Delta^{(0)} \varphi_n^{(1)}\rangle}
{\langle \varphi^{(0)}_{n'}|\varphi^{(0)}_n\rangle} - \frac{\lambda_{n}^{(0)}\langle\varphi^{(0)}_{n'}|\varphi_n^{(1)}\rangle}{\langle \varphi^{(0)}_{n'}|\varphi^{(0)}_n\rangle}~,    
\end{equation}
where $\varphi^{(0)}_{n'}$ is the only mode with a non-trivial overlap with $\varphi^{(0)}_n$.
Assuming that $\varphi_n^{(1)}$ is normalisable, hence a linear superposition of $\{ \varphi^{(0)}_n\}$,  the second and third term cancel against each other. This therefore gives
\begin{equation}\label{eq:lambda3}
\lambda_{n}^{(1)}=
\frac{\langle \varphi_{n'}^{(0)} \,|\, \bar\Delta^{(1)} \varphi_n^{(0)}\rangle}
{\langle \varphi^{(0)}_{n'}|\varphi^{(0)}_n\rangle}~.
\end{equation}
 In this section, we will first quantify the sets of modes $\varphi_n^{(0)}$ for which $\lambda_{n}^{(0)}=0$ on the near-horizon geometry \eqref{Enhorizon}. These are the so-called {\it zero modes}. For the zero modes, we will solve for $\lambda_{n}^{(1)}$ via \eqref{eq:lambda3}. This analysis will be performed for all three contributions to the GPI in \eqref{eq:one-loop}: graviton, Chern-Simons fields, and gravitini one-loop determinants. We will contrast cases where the background is near-extremal versus near-BPS at intermediate steps, and in Sec.\,\ref{sec:final-GPInear} we will summarise the imprint of these corrections in the GPI defined near the horizon.  

We are being overscrupulous in writing these expressions and defining the method carefully. The reason is the tension we will encounter in the next section, where we will see that an analogous analysis in the asymptotically AdS$_3$ region breaks some of the assumptions laid out here.

\subsection{Gravitational zero modes}\label{sec:nhg-grav}

There are at least three sources of bosonic zero modes in the near-horizon region: {\it tensor modes}, which are large diffeomorphisms acting on the AdS$_2$ portion of the background \eqref{Enhorizon}, {\it rotational modes}, which are large diffeomorphism acting on the fibre of \eqref{Enhorizon}, and {\it gauge modes} due to the large gauge transformation of the gauge fields $A$ and $A'$. In this portion, we will focus on tensor and rotational modes, which are those contributing to $Z_{\rm grav}$ in \eqref{eq:eh-partition-function-reduced}. We will also investigate the possibility of additional zero modes in $Z_{\rm grav}$, which relate to the analysis of large diffeomorphisms in Kerr/CFT.   

Tensor modes have been extensively studied in the literature, and for BTZ black holes in particular \cite{HeyIli20,KapLaw24,KolMar24,Acito:2025hka}. They are also known as the Schwarzian modes, which are prominent in the analysis of quantum features of near-extremal black holes. These modes can be tied directly to the low-temperature behaviour of the dual CFT$_2$ partition function. We will review them here for completeness and contrast with the other modes present in the analysis. 

Rotational modes suffer from a less prestigious reputation relative to tensor modes. There are arguments to disregard them in the path integral based on the dual CFT$_2$ partition function \cite{GhoMax19,Pal:2023cgk}, the reconstruction of these modes from afar \cite{KapLaw24,KolMar24,Acito:2025hka}, and thermodynamic properties \cite{KolMar24,Betzios:2025sct}. Here, we will revisit this debate and show explicitly what the near-horizon geometry offers in this regard. From the standpoint of the near-horizon region analysis, they seemingly have a similar standing as the tensor modes. 

The starting point of the construction of zero modes is to inspect \eqref{eq:eh-partition-function-reduced},
\begin{equation}\label{eq:grav-one-loop}
    Z_{\mr{grav}}[\bar{g}^{(0)}] = \sqrt{\r{\det(\eval{\bar{\Delta}_{\mr{gh}}}_{\mr{T}})}{\det(\eval{\bar{\Delta}_{\mr{grav}}}_{\mr{TT}})}}~,
\end{equation}
where $\bar{g}^{(0)}=g_{\rm ENHG}$ is the line element \eqref{Enhorizon} for the near horizon, and define the geometry at zero temperature according to the expansion in \eqref{eq:expand-operator}---in the subsequent equations we will drop the index ${(0)}$ to avoid cluttering and it will be restored once we turn on temperature.
The zero-modes appear as solutions to $\eval{\bar{\Delta}_{\mr{grav}}}_{\mr{TT}}$ equal to zero, which remain after imposing the maximum possible set of gauge conditions. Locally, these will be perturbations appearing as a diffeomorphism $x^\mu\to x^\mu +\xi^\mu$,
\begin{equation}\label{eq:h-diffeo}
    h_{\mu\nu} = \mc{L}_\xi \bar{g}_{\mu\nu} = 2\bar{\nabla}_{(\mu}\xi_{\nu)}~,
\end{equation}
and satisfying the gauge conditions
\begin{equation}\label{eq:h-gauge-cond}
    \bar{\nabla}^\mu h_{\mu\nu} = 0~,\qquad h^\mu{}_\mu = 0~,
\end{equation}
equivalent to
\begin{equation}\label{eq:nhg-xi-eigenproblem}
  (\bar{\Box} - \r{2}{\ell^2})\xi_\mu = 0~,\quad \bar{\nabla}^\mu \xi_\mu = 0~.
\end{equation}
In addition, we require $h_{\mu\nu}$ to be periodic under \eqref{eq:nhg-trivial-periodicity} and normalizable, where the inner product is defined as
\begin{equation}\label{eq:norm-h}
     \langle h_{1} | h_2 \rangle = \int \dd^3x \sqrt{\bar g}\, (h_1)^{\mu\nu} (h_2)_{\mu\nu}~.
\end{equation}

There are multiple ways to find the set of diffeomorphisms satisfying  \eqref{eq:h-gauge-cond}. In the following, we will take an approach that can be extended to higher dimensions; in Sec.\,\ref{sec:kerr-cft-no-go}, we will present an approach catered to AdS$_3$.
First, we note that from \eqref{eq:h-diffeo} that the trace condition gives
\begin{equation}
  \dd{\bars{\star} \xi} = 0~.
\end{equation}
Since the topology of the ENHG \eqref{Enhorizon} does not admit closed, non-exact 2-forms, this implies that
\begin{equation}\label{eq:nhg-xi-from-omega}
    \xi^\mu = \bar{\epsilon}^{\mu\nu\lambda}\bar{\nabla}_\nu w_\lambda~,
\end{equation}
for a one-form $w$. 
With this, the transverse condition in \eqref{eq:h-gauge-cond} then becomes
\begin{equation}\label{eq:nhg-omega-eq}
    \dd{[(\bar{\Box} - \r{2}{\ell^2})w]} = 0~.
\end{equation}
Solving this third-order differential equation in full generality is cumbersome. We start first by assuming that $w$ is independent of the angular variable $\hat \phi$ in \eqref{Enhorizon}, and consider the ansatz
\begin{equation}\label{eq:wn-def}
    w = \sum_{n\in \mathbb{Z}} e^{in \hat{t}_{\mr{E}}}\ti{w}^{(n)}_a\, \bar{e}^a~,
\end{equation}
where $\bar{e}^a$ are the orthonormal frames given in \eqref{nhframe} and $\ti{w}^{(n)}_a$ depend only on the radial coordinate $\eta$. Moreover, noting that shifting $w$ by a closed form does not change the resulting diffeomorphism generator $\xi$ in \eqref{eq:nhg-xi-from-omega}, we can choose $\ti{\omega}^{(n)}_{0}$ to be zero. After some manipulation, equation \eqref{eq:nhg-omega-eq} then leads to an algebraic expression for $\ti{w}^{(n)}_{1}$ in terms of $\ti{w}^{(n)}_{2}$ and its derivatives, as well as a fourth-order ODE for $\ti{w}^{(n)}_{2}$, resulting in four linearly independent generators of diffeomorphisms. The general solution reads
\begin{equation}\label{eq:wn-final}
   \ti{w}^{(n)}_{2} = \tanh\left(\frac{\eta}{2}\right)^{|n|}\left(c_1 + c_2 \cosh\eta  \right) + \tanh\left(\frac{\eta}{2}\right)^{-|n|}  \left(b_1 + b_2 \cosh\eta  \right)~,
\end{equation}
where $b_i$ and $c_i$ are constants. Demanding that the norm \eqref{eq:norm-h} is finite sets $b_1=b_2=0$. The remaining integration constants, $c_i$, reflect that we will have two classes of zero modes in this sector: {\it tensor} and {\it rotational modes}.

Tensor modes correspond to solutions with $c_1= n c_2$ in \eqref{eq:wn-final}. Using \eqref{eq:nhg-xi-from-omega}, the resulting diffeomorphism is
\begin{multline}\label{eq:xi-tensor}
  \xi_n^{\mr{tensor}} = \r{e^{in\hat{t}_{\mr{E}}}}{2\pi\sqrt{r_0\abs{n}(n^2-1)}}\tanh^{\abs{n}}\f(\r{\eta}{2})\left[(\cosh\eta + \abs{n})\f(\r{\abs{n}}{\sinh\eta}\partial_\eta + \r{in}{\sinh^2\eta}\partial_{\hat{t}_{\mr{E}}})\right.\\
  \left. + \sgn(n)\f(i\partial_{\hat{t}_{\mr{E}}} + \r{\cosh\eta - n^2 + \abs{n}+1}{\cosh\eta + 1}\r{\ell}{2r_0}\partial_{\hat{\phi}}) \right]~,
\end{multline}
and the metric perturbation reads
\begin{equation}\label{eq:nhg-sch} 
    h_n^{\mr{tensor}} = \r{\ell^2}{4\pi}\sqrt{\r{\abs{n}(n^2-1)}{r_0}}e^{in\hat{t}_{\mr{E}}}\tanh^{\abs{n}}\f(\r{\eta}{2})\f(\r{\dd{\eta}^2}{\sinh^2\eta} + \r{2i\sgn(n)\dd{\eta}\dd{\hat{t}_{\mr{E}}}}{\sinh\eta} - \dd{\hat{t}_{\mr{E}}}^2)~.
\end{equation}
Note that for $\abs{n} < 2$ the generators $\xi_n^{\rm tensor}$, after rescaling, are just Killing vectors of the ENHG associated to the isometries of AdS$_2$ and $h_n^{\mr{tensor}}$ vanishes. Therefore, tensor modes  $h_n^{\mr{tensor}}$ are defined for $\abs{n} \ge 2$. As has been discussed extensively in the literature, the diffeomorphisms $\xi_n^{\mr{tensor}}$ have a natural interpretation: they act as reparametrizations of time at the boundary of the AdS$_2$ factor in the ENHG.  In the context of 2D gravity, the dynamics of these diffeomorphisms is captured by a Schwarzian effective action. For this reason, tensor modes are also known as Schwarzian modes. For more comments on this regard, we refer to \cite{Iliesiu:2022onk}. Finally, the vector $\xi_n^{\rm tensor}$ is usually described as a large diffeomorphism, since the norm $ \langle \xi_n^{\rm tensor} | \xi_m^{\rm tensor} \rangle$ is infinite, although $h_n^{\mr{tensor}} $ is perfectly regular and normalizable. In particular, we have
\begin{equation}
    \langle h_{-n}^{\rm tensor} | h_m^{\rm tensor} \rangle    = \ell^2\delta_{nm}~.
\end{equation}

Rotational modes correspond to solutions with $c_2=0$ in \eqref{eq:wn-final}. The associated diffeomorphism in this case reads
\begin{equation}\label{eq:xi-rot}
    \xi_n^{\mr{rot}} = \r{e^{in\hat{t}_{\mr{E}}}}{2\pi n}\sqrt{\r{\abs{n}}{r_0}}\tanh^{\abs{n}}\f(\r{\eta}{2})\f(-\r{\abs{n}}{\sinh\eta}\partial_\eta - \r{in}{\sinh^2\eta}\partial_{\hat{t}_{\mr{E}}} + \sgn(n)\r{\cosh\eta+1 + \abs{n}}{\cosh\eta+1}\r{\ell}{2r_0}\partial_{\hat{\phi}})~,
\end{equation}
and the metric perturbation is
\begin{multline}\label{eq:h-rot}
    h^{\mr{rot}}_n = \r{\ell^2 ne^{in\hat{t}_{\mr{E}}}}{4\pi\sqrt{r_0\abs{n}}}\tanh^{\abs{n}}\f(\r{\eta}{2})\left[\r{(\cosh\eta-\abs{n})\dd{\eta}^2}{\sinh^2\eta} + \r{2i\sgn(n)(1-\abs{n})\dd{\eta}\dd{\hat{t}_{\mr{E}}}}{\sinh\eta} \right. \\
    \left. + (\cosh\eta + \abs{n} - 2)\dd{\hat{t}^2_{\mr{E}}} + \r{4ir_0\dd{\hat{\phi}}}{\ell}\f(\dd{\hat{t}_{\mr{E}}} - \r{i\sgn(n)\dd{\eta}}{\sinh\eta}) \right]~,
\end{multline}
which are non-vanishing for $\abs{n} \ge 1$. After an appropriate rescalling, $n=0$ in \eqref{eq:xi-rot} correspond to the Killing vector $\partial_{\hat{\phi}}$ associated to the axisymmetry of the ENHG. The modes are normalised so that
\begin{equation}
 \langle h_{-n}^{\rm rot} | h_m^{\rm rot} \rangle = \ell^2\delta_{nm}~.
\end{equation}
The nature of the rotational modes can be better understood by looking at the generator of the large diffeomorphisms \eqref{eq:xi-rot}: at the boundary of the ENHG, the vector becomes 
\begin{equation}
    \lim_{\eta\ar \infty}\xi^{\mr{rot}}_n \propto e^{in\hat{t}_{\mr{E}}}\partial_{\hat{\phi}}~.
\end{equation}
This corresponds to a shift of the $\hat{\phi}\to \hat{\phi} + f(t)$, with $f(t)$ a periodic function. Interpreted in 2D gravity, this would correspond to gauge transformations of a Maxwell field obtained by Kaluza-Klein reduction from 3D to 2D. 

In a similar fashion, one can also verify that there are no zero modes in the determinant associated with ghosts in \eqref{eq:grav-one-loop}. The only zero modes in the ENHG are due to the graviton determinant in the denominator of \eqref{eq:grav-one-loop}, and are described by tensor and rotational modes.

The next step in our analysis is to quantify the contribution of $h_m^{\rm tensor}$ and $h^{\mr{rot}}_n$ to the GPI.  The modes comply with all of the regularity conditions arising from $\bar{g}^{(0)}=g_{\rm ENHG}$, with the feature that their associated eigenvalue is zero. Taking the approach of \eqref{eq:expand-operator}, we can easily quantify how they respond under a deformation of the ENHG, where the deformation parameter is the temperature. We will write 
\begin{equation}\label{eq:g0+g1}
    \bar {g} = \bar{g}^{(0)} + T \,\bar{g}^{(1)}~,
\end{equation}
with $ \bar{g}^{(1)}$ given by \eqref{nhg-correction}, which leads to a correction of the Laplacian 
\begin{equation}
    \bar{\Delta}_{\mr{grav}} = \bar{\Delta}_{\mr{grav}, (0)}+T\,\bar{\Delta}_{\mr{grav}, (1)} + \cdots ~.
\end{equation}
With this, and the analysis leading to \eqref{eq:lambda3}, we find that the correction to the  eigenvalue is 
\begin{equation}\label{eq:nhg-sch-eigenval}
  \lambda^{(1)\rm tensor}_n = \frac{1}{\langle h_{-n}^{\rm tensor} | h_n^{\rm tensor} \rangle }\int\dd[3]{x}\sqrt{\bar{g}^{(0)}}\, (h_{-n}^{\mr{tensor}})^{\mu\nu}(\bar{\Delta}_{\mr{grav}, (1)} h_{n}^{\mr{tensor}})_{\mu\nu} = \r{\abs{n}\pi }{2r_0} ~,
\end{equation}
for tensor modes, and for rotational modes, we have 
\begin{equation}\label{eq:nhg-rot-eigenval}
   \lambda^{(1)\rm rot}_n = \frac{1}{\langle h_{-n}^{\rm rot} | h_n^{\rm rot} \rangle } \int\dd[3]{x}\sqrt{\bar{g}^{(0)}}\, (h_{-n}^{\mr{rot}})^{\mu\nu}(\bar{\Delta}_{\mr{grav}, (1)} h_{n}^{\mr{rot}})_{\mu\nu} = 0~,
\end{equation}
and the corresponding contributions to the change in action are 
\begin{equation}\label{actionchange:graviton}
    \delta \Lambda_n^{\rm tensor/ rot} = \ell^2 T \lambda^{(1)\rm tensor/rot}_n~.
\end{equation}
The most striking difference between \eqref{eq:nhg-sch-eigenval} and \eqref{eq:nhg-rot-eigenval} is that one correction is positive while the other is zero. This raises some concerns about the standing of rotational modes relative to tensor modes, which requires a closer inspection of the perturbative problem we are studying.

One aspect that \eqref{eq:lambda3} overlooks, used in \eqref{eq:nhg-sch-eigenval} and \eqref{eq:nhg-rot-eigenval}, is that the corrected eigenmodes for the graviton, $h^{(1)}$, are transverse and traceless. To assess whether the problem we are solving is free of pathologies, we now inspect these conditions at linear order in temperature. The gauge conditions are given by
    \begin{equation}
         \nabla^\mu h_{\mu\nu}=0  , \qquad \qquad h^{\mu}{}_{\mu} =0~.
    \end{equation}
    At linear order in temperature, we obtain
    \begin{equation}\label{gc-linearT-0}
    \begin{split}
        \bar \nabla^{(1)\mu}h^{(0)}_{\mu\nu} + \bar \nabla^{(0)\mu} h^{(1)}_{\mu\nu}&=0~,\\
        \bar g^{(0)\mu\nu}h^{(1)}_{\mu\nu} +  \bar g^{(1)\mu\nu}h^{(0)}_{\mu\nu} &=0~,   
    \end{split}
    \end{equation}
    where we used the fact that the $h^{(0)}_{\mu\nu}$ are transverse and traceless with respect to the near-horizon geometry. The eigenvector correction $h^{(1)}_{\mu\nu}$ can be further separated into two pieces: a propagating part and a pure gauge part, as follows
    \begin{equation}\label{eq:split-h}
       h^{(1)}_{\mu\nu}= \tilde{h}^{(1)}_{\mu\nu} + h^{(1)}_{\mu\nu,\rm{pg}}= \sum_m c_m h^{(0)}_{\mu\nu,m}  + \bar\nabla^{(0)}_{(\mu}\zeta^{(1)}_{\nu)}~.
    \end{equation}
    Here, $\tilde{h}^{(1)}_{\mu\nu} $ is the propagating part of $h^{(1)}_{\mu\nu}$; by the standard assumptions of perturbation theory, it is normalisable and hence can be expressed as a linear superposition of the eigenmodes of near-horizon geometry, as reflected in the second equality. $h^{(1)}_{\mu\nu,\rm{pg}}$ is a pure gauge component, with $\zeta^{(1)}_\mu$ the corresponding diffeomorphism linear in $T$; this term ensures that the gauge fixing conditions are satisfied. Using this expansion, the gauge conditions \eqref{gc-linearT-0} simplify to 
\begin{equation}\label{gc-linearT}
    \begin{split}
        \bar\nabla^{(1)\mu}h^{(0)}_{\mu\nu} + \bar \nabla^{(0)\mu} (\bar\nabla^{(0)}_{(\mu}\zeta^{(1)}_{\nu)})&=0~,\\
        \bar g^{(1)\mu\nu}h^{(0)}_{\mu\nu} +  \bar g^{(0)\mu\nu}(\bar\nabla^{(0)}_{(\mu}\zeta^{(1)}_{\nu)}) &=0~.
    \end{split}
    \end{equation}
    From here, it is straightforward to solve for $\zeta^{(1)}_\mu$ for tensor and rotational modes. The telling feature of this solution is how they behave at large $\eta$. More precisely, we find
    \begin{equation}\label{eq:grow-tensor}
        \begin{split}
         &h^{(1), \rm{tensor}}_{\hat{t}_E\hat{t}_E,\rm{pg}} = \mathcal{O}(\eta e^{2 \eta}), \qquad   h^{(1), \rm{tensor}}_{\hat{t}_E\eta,\rm{pg}}=\mathcal{O}(\eta e^{2 \eta}), \qquad   h^{(1), \rm{tensor}}_{\hat{t}_E\hat\hat\phi,\rm{pg}}=\mathcal{O}(\eta e^{ \eta}),\\
          &h^{(1), \rm{tensor}}_{\eta\eta,\rm{pg}} =\mathcal{O}(1), \qquad   h^{(1), \rm{tensor}}_{\eta\hat\phi,\rm{pg}}=\mathcal{O}(\eta e^{ \eta}), \qquad   h^{(1), \rm{tensor}}_{\hat\phi\hat\phi,\rm{pg}}= 0~,
        \end{split}
    \end{equation}
    for the tensor modes, and 
    \begin{equation}\label{eq:grow-rotation}
        \begin{split}
         &h^{(1), \rm{rot}}_{\hat{t}_E\hat{t}_E,\rm{pg}} = \mathcal{O}(e^{5 \eta}), \qquad   h^{(1), \rm{rot}}_{\hat{t}_E\eta,\rm{pg}}=\mathcal{O}( e^{5 \eta}), \qquad   h^{(1), \rm{rot}}_{\hat{t}_E\hat\phi,\rm{pg}}=\mathcal{O}( e^{4 \eta}),\\
          &h^{(1), \rm{rot}}_{\eta\eta,\rm{pg}} =\mathcal{O}(e^{3\eta}), \qquad   h^{(1), \rm{rot}}_{\eta\hat\phi,\rm{pg}}=\mathcal{O}(e^{ 4\eta}), \qquad   h^{(1), \rm{rot}}_{\hat\phi\hat\phi,\rm{pg}}= 0~,
        \end{split}
    \end{equation}
    for the rotational modes. Although both solutions grow rapidly at the boundary, $h^{(1)}_{\mu\nu,\rm{pg}}$ is normalisable in the near-horizon geometry for the tensor modes and non-normalisable for the rotational modes. Therefore, from the near-horizon point of view,  tensor modes comply with the assumptions laid out by perturbation theory. In stark contrast, the rotational mode satisfies the gauge condition (at linear order in $T$) only after including non-normalisable contributions to the eigenvector correction. This breaks the standard rules of perturbation theory, which will be reflected again in Sec.\,\ref{sec:grav-far} as we analyse the one-loop determinants in the far region. This is an indication that the correction to the eigenvalue \eqref{eq:nhg-rot-eigenval} should not be trusted.

\subsubsection{Absence of Kerr/CFT-like modes in the GPI}\label{sec:kerr-cft-no-go}

It is tempting to conclude the analysis of gravitational zero modes here, since tensor and rotational modes are the commonly discussed modes in the literature. However, in \eqref{eq:wn-def} we assumed that the modes were axisymmetric without justification. We turn now to analysing the most general scenario: constructing zero modes supported both along $\hat t_E$ and $\hat \phi$ complying with \eqref{eq:nhg-xi-eigenproblem}. 
This is motivated by the diffeomorphism first proposed in Kerr/CFT \cite{Guica:2008mu}, which was then adapted to BTZ in \cite{Balasubramanian:2009bg,Azeyanagi:2011zj}. In that construction, the diffeomorphisms are characterised by their angular support, and in a sense, are orthogonal to the tensor and rotational modes constructed above.

Rather than solving \eqref{eq:nhg-omega-eq}, it is simpler to proceed by noticing that, in three dimensions, the equations in \eqref{eq:nhg-xi-eigenproblem} are equivalent to \cite{Datta:2011za} 
\begin{equation}\label{eq:nhg-xi-first-order}
  \epsilon_\mu{}^{\nu\alpha} \bar{\nabla}_\nu \xi_\alpha = i\r{\lambda}{\ell}\xi_\mu~,
\end{equation}
with $\lambda = \pm 2$. Our convention for the Levi-Civita tensor in the ENHG background is $\epsilon_{\hat t_E \eta \hat\phi}=\sqrt{\bar g}$. To solve \eqref{eq:nhg-xi-first-order} in  the ENHG, we will do a Fourier decomposition in the $(\hat t_E,\hat \phi)$ variables for each component   
\begin{equation}\label{eq:nhg-xi-ansatz}
  \xi_\mu =\sum_{n,m\in\mathbb{Z}} e^{in\hat{t}_{\mr{E}}+im\hat{\phi}}({\xi}^{(\lambda)}_{nm})_\mu(\eta)~.
\end{equation}
With this, the radial component of \eqref{eq:nhg-xi-first-order} is algebraic, and we will use it to fix the radial component of $\xi$ as follows
\begin{equation}\label{eq:nhg-xi-radial}
  (\xi^{(\lambda)}_{nm})_\eta =\r{ \ell}{ \lambda r_0}\r{1}{\sinh\eta} \f(m\, ({\xi}^{(\lambda)}_{nm})_{\hat{t}_{\mr{E}}}-n \,({\xi}^{(\lambda)}_{nm})_{\hat{\phi}})~.
\end{equation}
For fixed $\lambda$, the  $(\hat t_E,\hat \phi)$ components of $\xi_\mu$ are then determined by a second-order ODE plus one constraint that relates the components; this follows the analysis of App.\,\ref{sec:gaugemodes} with appropriate modifications for the ENHG.  This gives rise to two linearly independent solutions: one that is polynomially regular at the horizon, $\eta=0$, and one that contains logarithmic and polynomially divergent terms. Focusing on the regular solution at the horizon, the solution reads
\begin{multline}\label{eq:nhg-xi-gen-1}
  \m{(\xi^{(-2)}_{nm})_{\hat{t}_{\mr{E}}} \\ (\xi^{(-2)}_{nm})_{\hat{\phi}}}= e^{i(n\hat{t}_{\mr{E}} + m\hat{\phi})} \tanh^{\abs{n}}\f(\r{\eta}{2}) \m{1 & 1 \\ 0 & -\r{2ir_0}{\ell}} \\
  \times \m{(1 - \abs{n} - \r{i\ell \sgn(n)m}{2r_0}) \,{}_2F_1(\r{i\ell \sgn(n)m}{2r_0}, \abs{n} - \r{i\ell \sgn(n)m}{2r_0}, 1+\abs{n},\tanh^2(\r{\eta}{2})) \\
  -(1 - \r{i\ell \sgn(n)m}{2r_0})\cosh^2\f(\r{\eta}{2}) \,{}_2F_1( \r{i\ell \sgn(n)m}{2r_0} -1, \abs{n} - \r{i\ell \sgn(n)m}{2r_0}-1, 1+\abs{n},\tanh^2(\r{\eta}{2}))}
\end{multline}
for $\lambda = -2$, and
\begin{multline}\label{eq:nhg-xi-gen-2}
  \m{(\xi^{(2)}_{nm})_{\hat{t}_{\mr{E}}} \\ (\xi^{(2)}_{nm})_{\hat{\phi}}}= e^{i(n\hat{t}_{\mr{E}} + m\hat{\phi})} \tanh^{\abs{n}}\f(\r{\eta}{2}) \m{1 & 1 \\ 0 & -\r{2ir_0}{\ell}} \\
  \times \m{(1 + \abs{n} + \r{i\ell \sgn(n)m}{2r_0})\cosh^2\f(\r{\eta}{2}) \,{}_2F_1(\r{i\ell \sgn(n)m}{2r_0}-1, \abs{n} - \r{i\ell \sgn(n)m}{2r_0}-1, 1+\abs{n},\tanh^2(\r{\eta}{2})) \\
  -(1 + \r{i\ell \sgn(n)m}{2r_0}) \,{}_2F_1(\r{i\ell \sgn(n)m}{2r_0}, \abs{n} - \r{i\ell \sgn(n)m}{2r_0}, 1+\abs{n},\tanh^2(\r{\eta}{2}))}
\end{multline}
for $\lambda = 2$. When $m=0$, the hypergeometric functions simplify to
\begin{equation}
    {}_2F_1(-1, -1+\abs{n}, 1+\abs{n},z) = 1 + \frac{1-|n|}{1+|n|}z ~,\qquad  {}_2F_1(0, \abs{n}, 1+\abs{n},z) = 1~.
\end{equation}
With this, it is straightforward to check that the tensor modes \eqref{eq:xi-tensor} and rotational modes \eqref{eq:xi-rot} arise for $m=0$ and $\lambda=-2$ and $\lambda=2$, respectively.  

As a final step, we still need to impose that the metric modes constructed from \eqref{eq:nhg-xi-gen-1} and \eqref{eq:nhg-xi-gen-2} are normalizable. For this, it is useful to record the behaviour of the hypergeometric functions near the boundary of the ENHG; taking $z=\tanh^2(\eta/2)$, in the limit $z\to 1$, we have \cite{abramowitz_handbook_1972}
\begin{equation}
\begin{aligned}
    {}_2F_1(i a-1,  \abs{n} - ia -1, 1+\abs{n},z) \underset{z\to 1}{=}&~ \frac{2\Gamma(1+|n|)}{\Gamma(2+|n|-ia)\Gamma(2+ia)} \\ &\quad + O(z-1)  + O(\log(1-z)(z-1)^3)~,\\  
    {}_2F_1(i a , \abs{n} - i a, 1+\abs{n},z) \underset{z\to 1}{=}&~\frac{\Gamma(1+|n|)}{\Gamma(1+|n|-ia)\Gamma(1+ia)}\\
   &\quad + O(z-1)^2 +O(\log(1-z) (z-1)) ~,
\end{aligned}
\end{equation}
where $a\in \R$.  The metric fluctuation constructed from \eqref{eq:nhg-xi-gen-1} and \eqref{eq:nhg-xi-gen-2},
\begin{equation}
    h_{\mu\nu} =\mc{L}_\xi \bar g_{\mu\nu}~, 
\end{equation}
will therefore behave near the boundary of the ENHG, at $\eta\to \infty$, as
\begin{equation}
\begin{aligned}\label{eq:h-bndy}
    h_{\hat t_E \hat t_E}\sim  O(e^{\eta})~, \quad
    h_{\hat \phi \hat \phi}\sim O(1)~,\quad 
    h_{\hat \phi \hat t_E}\sim O(1)~, \\
        h_{\hat t_E \eta}\sim O(1)  ~, \quad
    h_{\hat \phi \eta}\sim O(1)~,  \quad 
    h_{\eta \eta}\sim O(e^{-\eta})~,
\end{aligned}
\end{equation}
for $m\neq0$ and $\lambda =2$. It is straightforward, although tedious, to check that the norm of $h$ is divergent for $m\neq0$. The only way to eliminate this divergence is to set $m=0$.\footnote{It is instructive to compare \eqref{eq:h-bndy} with \eqref{eq:h-rot}. An important difference is that in \eqref{eq:h-bndy} we have $ h_{\hat \phi \hat \phi}\sim O(1)$, and this can be traced back to the divergence of the norm. The only way to turn off this component is by setting $m=0$.} With this, we confirm that the only normalizable zero modes in the ENHG of BTZ are tensor and rotational modes. 

It is worth discussing which modes we have excluded from the GPI and how they relate to Kerr/CFT.  Considering the case of $\lambda=2$, $n = 0$ and $m\neq0$, the diffeomorphism we constructed in \eqref{eq:nhg-xi-gen-2} behave as, up to normalisation,
\begin{equation}\label{eq:xi-kerr}
  \xi^{(2)}_{0m} \underset{\eta\rightarrow \infty}{\approx} e^{im\hat{\phi}}\f(i\partial_{\hat{\phi}}+m\partial_\eta)~.
\end{equation}
These are precisely the diffeomorphisms considered in \cite{Balasubramanian:2009bg,Azeyanagi:2011zj}, which are inspired by the original proposal of Kerr/CFT \cite{Guica:2008mu}. In those works, it was shown that these diffeomorphisms lead to an asymptotic symmetry group compatible with a Virasoro algebra containing a non-trivial central extension. It is interesting to note that an analysis of the classical phase space and asymptotic charges done there is insensitive to our normalisation condition of the metric fluctuations. What we have shown in this portion is that although some large diffeomorphisms might have interesting features at the (semi-)classical level, they do not survive the quantum realm. We see no place for diffeomorphism of the form \eqref{eq:xi-kerr} in the GPI of near-extremal BTZ. 

Finally, as mentioned below \eqref{eq:nhg-xi-radial}, these are not all possible solutions--each of the hypergeometrics in \eqref{eq:nhg-xi-gen-1} and \eqref{eq:nhg-xi-gen-2} can be replaced by a linearly independent solution, which amounts to substitution
\begin{equation}
  {}_2F_1(a,b,1+\abs{n},z)\quad \rightarrow\quad {}_2F_1(a,b,1+\abs{n},z)\log z + z^{-\abs{n}}\mc{F}_{a,b,n}(z)~,
\end{equation}
with $\mc{F}_{a,b,n}(z)$ analytic at $z=0$, and a change in the values of the relative normalisations of the entries in \eqref{eq:nhg-xi-gen-1} and \eqref{eq:nhg-xi-gen-2}. However, these solutions do not lead to an $h = \mc{L}_\xi g$ that is normalisable in a neighbourhood of $\eta=0$. Hence, we will not consider them as valid perturbations to include in the calculation of the one-loop determinant.

\subsection{Gauge zero modes}\label{sec:nhg-gauge}

Zero modes arising from gauge fields in three dimensions have not been discussed on the same footing as the metric perturbations. The culprit here is that it is simple to infer their contribution from the low-temperature behaviour of the CFT$_2$ partition function (the fermionic modes suffer from a similar curse). Here, we will analyse them solely from a near-horizon perspective and characterise their behaviour at zero and finite temperatures, first for $U(1)$ gauge fields and then for the non-abelian case. In addition to completing the discussion, the analysis of gauge modes will provide a natural point of comparison to the rotational modes. Finally, we recall that there are two pairs of Chern-Simons $(A,A')$ in our supergravity theory. We will focus on $A$, the sector which supports supersymmetry; the extension to $A'$ is straightforward.

Similar to the gravitational analysis in Sec.\,\ref{sec:nhg-grav}, we start by inspecting the path integral for the Chern-Simons fields around the ENHG background \eqref{Enhorizon} supported as well by the gauge field \eqref{eq:A-enhg}. The one-loop path integral \eqref{eq:cs-partition-function} is
\begin{equation}\label{eq:cs-1-loop-2}
    Z_{\mr{CS}}[\bar A, \bar g] = e^{i\theta[\bar{g},\bar{A}]}\sqrt{\r{\det(-(\bar{D}^\mu\bar{D}_\mu)_0)}{\abs{\det(\bar{\star}\bar{D}\big|_{\mr{T}})}}}~,
\end{equation}
where $\bar D_\mu$ is defined in \eqref{eq:def-D-A}. For the $U(1)$ theory, this derivative becomes a covariant derivative with respect to the background metric, $\bar D_\mu =\bar \nabla_\mu$. Hence, for this simple case, the background gauge field \eqref{eq:A-enhg} will not play a role. We will therefore have
\begin{equation}\label{eq:cs-u1}
    Z_{\mr{CS,U(1)}}[\bar A, \bar g] = e^{i\theta[\bar{g},\bar{A}]}\sqrt{\r{\det(-\bar\Box_0)}{\abs{\det(\bar{\star}\dd\big|_{\mr{T}})}}}~.
\end{equation}
The ghost determinant does not contain zero modes: all normalizable and periodic eigenfunctions of $\bar\Box_0$ have non-zero eigenvalue.
The determinant in the denominator will clearly have zero modes for closed one-forms that both respect the periodicity of the ENHG \eqref{eq:nhg-trivial-periodicity} and are normalizable. The general zero mode can be succinctly written as
\begin{equation} \label{eq:a-gauge-nhg}
  a^{\mr{gauge}}_n = \dd{\Phi_n}~,
\end{equation}
where $\Phi_n$ is a discrete family of solutions to $\bar{\Box}\Phi_n = 0$, given by
\begin{equation}\label{eq:scalar-gauge}
  \Phi_n = \r{e^{in\hat{t}_{\mr{E}}}}{2\pi}\sqrt{\r{\ell}{\abs{n}r_0}}\tanh^{\abs{n}}\f(\r{\eta}{2})~. \qquad |n|\ge1~.
\end{equation}
Notice that, similar to the gravitational zero modes, $\Phi_n$ are themselves not normalisable as scalar modes, but lead to normalisable one-forms $a^{\mr{gauge}}_n$, where  
\begin{equation}
     \langle a_{-m}^{\rm gauge} | a_n^{\rm gauge} \rangle = \ell\delta_{nm}~.
\end{equation}
In fact, the functions in \eqref{eq:scalar-gauge} are the only scalar modes satisfying $\bar{\Box} \Phi = 0$ for which $\dd{\Phi}$ is normalisable---solutions with non-trivial $\hat{\phi}$ dependence turn out to be non-normalizable. Note that the charge quantisation condition \eqref{eq:susy-charge-U(1)}, which distinguishes supersymmetric versus non-supersymmetric black holes, plays no role in $ a^{\mr{gauge}}_n $ in the Abelian case. This also means that there are zero modes for the primed sector, i.e., the path integral for the second Chern-Simons field $A'$. 

Generalising this analysis for the non-abelian case is straightforward. The zero modes are now given by
\begin{equation}\label{eq:zero-non-A}
    \begin{aligned}
        a^{\mr{gauge}}_n &=  g^{-1}\,\dd{\Phi^I_n} T_I \,g\\
        &= \bar D_\mu \le( g^{-1}\,\Phi^I_n T_I \,g\ri)\dd x^\mu~, 
    \end{aligned}
\end{equation}
where 
\begin{equation}\label{eq:g-path}
    g(x)={\rm P}\exp\le(\int_{\gamma(x)} \bar{A}\ri)~,
\end{equation}
and $\gamma(x)$ is any curve connecting the point of integration $x^\mu$ to some fixed reference point, and $\bar A$ is the background connection with support on the Cartan subalgebra of the gauge group. The scalar functions $\Phi^I_n$ satisfy
\begin{equation}
    \bar\Box\Phi^I_n= 0~.
\end{equation}
Requiring that $a^{\mr{gauge}}_n$ is normalizable and periodic in $\hat t_E$ reduces $\Phi^I_n$ to be of the form \eqref{eq:scalar-gauge}. However, $\bar A$ has angular support for non-zero charges, due to the group element $g$ in \eqref{eq:g-path}. Demanding therefore periodicity of \eqref{eq:zero-non-A} along $\hat \phi \sim \hat \phi +2\pi$ is non-trivial and further restricts the zero modes. To illustrate this, we consider the case of $SU(2)$, where the background gauge field is given by
\begin{equation}
     \bar A_{{\mf{su}(2)}} = GQ\, T_3\, \dd \hat \phi~,
\end{equation}
and the conditions for supersymmetry are given in \eqref{eq:susy-charge-SU(2)}. We therefore have two situations:

\paragraph{Supersymmetric black holes:} when $GQ\in\mathbb{Z}$, the group element $g$ changes by $\mathbb{1}$ or $-\mathbb{1}$ under $\hat\phi\sim \hat\phi +2\pi$, i.e., it changes by an element of the centre of the group. Therefore, $a^{\mr{gauge}}_n$ is periodic and normalizable in this case. The conclusion is that the ENHG of supersymmetric black holes will have three towers of gauge zero modes characterised by the functions $\Phi_n^I$, $I=1,2,3$ of the form \eqref{eq:scalar-gauge}. In the primed sector, we will only have one set of zero modes for ${\Phi'}_n^{I=3}$, which also takes the form as in \eqref{eq:zero-non-A} and \eqref{eq:scalar-gauge} with $\bar{A}$ replaced by $\bar{A'}$.

\paragraph{Non-supersymmetric black holes:} if $GQ\notin\mathbb{Z}$, $g$ is neither periodic nor anti-periodic when transported along $\hat\phi\sim \hat\phi +2\pi$. The only component of $a^{\mr{gauge}}_n$ that survives the periodicity condition along $\hat\phi \sim \hat\phi+2\pi$ is the one along $\Phi_n^{I=3}$. Therefore, we will have only one tower of gauge zero modes, where $\Phi_n^{I=1,2}=0$ and $\Phi_n^{I=3}$ is given by \eqref{eq:scalar-gauge}. The primed sector follows in parallel. In short, for non-supersymmetric black holes, increasing the group to be non-Abelian does not increase the number of zero modes, in contrast to the supersymmetric case. 

After having identified the zero modes in the ENHG, we now turn to how their eigenvalues are corrected by deforming the ENHG away from extremality.  From \eqref{eq:two-CS}, the  correction to the action of these modes due to turning on a small non-zero temperature is 
\begin{equation}\label{gaugeev-nhg}
  \delta\Lambda_n^{\mr{gauge}} = \int\dd[3]{x} \sqrt{\bar{g}}(a^{\mr{gauge}}_{-n})^\mu \delta(\bar\epsilon_\mu{}^{\nu\lambda}\bar D_\nu)(a^{\mr{gauge}}_n)_\lambda = 0~,
\end{equation}
since $\bar\epsilon_\mu{}^{\nu\lambda}\bar D_\nu$ does not respond to $T$ in the canonical ensemble; recall that the background connections \eqref{eq:A-enhg} are independent of temperature at fixed $Q$.
Hence, from the perspective of the ENHG and its deformations, the gauge zero modes contribute to the GPI at a subleading order in temperature. 

One can also revisit this analysis in an ensemble where $(\mu,\mu')$ are fixed rather than $(Q,Q')$. This requires a change in how we treat $(\bar A, \bar A')$ and, in turn, this will affect periodicities we impose on fluctuations \eqref{eq:zero-non-A}.  Still, even taking these modifications into account, the correction \eqref{gaugeev-nhg} still vanishes.

The conclusions we are reaching in this portion will turn out to be incompatible with the far region analysis, as will become apparent from the analysis of the modes in the full BTZ geometry in Sec.\,\ref{sec:fullbtz-gauge}. The culprit has the same origin as the mismatch for rotational modes mentioned below \eqref{eq:nhg-rot-eigenval}: there are normalizable solutions in the whole geometry that become non-normalizable in the ENHG.  The approach of this section is blind to non-normalizable contributions, reflecting once again that there are problems in the decoupling limit at the quantum level.

Similar to the gravitational case, for the gauge modes, we can flag an issue with the perturbative analysis by considering if gauge fixing conditions are satisfied to linear order in $T$. That is, we can determine what the gauge transformation is that ensures that the correction to the eigenmodes keeps the modes transverse gauge at linear order in $T$. Following \eqref{eq:split-h}, and for simplicity focusing on the $u(1)$ case, the eigenmode corrections can again be separated into a propagating part and a pure gauge part,
\begin{equation}
    A_\mu^{(1)}= \tilde A_\mu^{(1)} + A_{\mu,\rm{pg}}^{(1)} = \sum_m c_m A_{\mu,m}^{(0)} + \bar\nabla^{(0)}_\mu \Phi^{(1)}~,
\end{equation}
where we again used the fact that the propagating part of the eigenmode correction is normalisable. The gauge condition at linear order in $T$ is given by
\begin{equation}
    \bar\nabla^{(1)\mu}A^{(0)}_\mu + \bar\Box^{(0)}\Phi^{(1)}=0~.
\end{equation}
It is simple to solve for $\Phi^{(1)}$ such that the gauge condition is satisfied at linear order in $T$. At large $\eta$, we find that the pure gauge behaves as
\begin{equation}
    A^{(1)}_{\hat{t}_E,\rm{pg}}= \mathcal{O}(\eta), \qquad A^{(1)}_{\eta,\rm{pg}}= \mathcal{O}(1), \qquad A^{(1)}_{\hat\phi,\rm{pg}}= 0~,
\end{equation}
which leads to $A_{\mu,\rm{pg}}^{(1)}$ being non-normalisable in the near-horizon geometry. 
As for rotational modes, this again signals that perturbation theory in $T$ is breaking down for this class of modes.

\subsection{Fermionic zero modes}\label{sec:nhg-fermionic}

Finally, we turn to zero modes appearing in the fermionic operators controlling the spin-3/2 fields $\psi$ and $\psi'$.  The determinants in question are
\begin{equation}\label{eq:ferm-one-loop1-recap}
     Z_{3/2} [\bar g, \bar A]= \sqrt{\r{\det({\slashed{\bar D}}_{3/2}-\r{1}{2\ell})_{\rm T}}{\det({\slashed{\bar D}}_{1/2} - \r{3}{2\ell})^2}}~,\qquad Z'_{3/2} [\bar g, \bar A']= \sqrt{\r{\det({\slashed{\bar D'}}_{3/2}+\r{1}{2\ell})_{\rm T}}{\det({\slashed{\bar D'}}_{1/2} + \r{3}{2\ell})^2}} ~,
\end{equation}
with the differential operators $D$ and $D'$ defined in  \eqref{eq:def-covariant-1}. As done in the previous subsections, we will first search for zero modes in these determinants when $\bar g = g_{\rm ENHG}$ in \eqref{Enhorizon} and the connections $(\bar A,\bar A')$ given by  \eqref{eq:A-enhg}. For the spin-3/2 fields the zero modes arise as configuration where, for $\psi_\mu = e_\mu^a\psi_a$, we have
\begin{equation}\label{eq:defn-ferm-modes}
\begin{aligned}
    \psi_a= \bar{\cal D}_a \chi ~,  \qquad  \psi'_a= \bar{\cal D}'_a \chi' ~,
\end{aligned}
\end{equation}
with $\chi$ and $\chi'$ spin-1/2 fields. Imposing the gauge condition $\gamma^a \psi_a =0$ and $\gamma^a \psi'_a =0$ further sets
\begin{equation}\label{eq:dirac-chi}
    \slashed{\bar{D}} \chi = \frac{3}{2\ell}\chi~, \qquad\slashed{\bar{D'}} \chi' = -\frac{3}{2\ell}\chi'~.
\end{equation}
In the following we will first solve for the spinors $(\chi,\chi')$ satisfying \eqref{eq:dirac-chi} on the ENHG, and then construct $(\psi_a, \psi'_a)$ further imposing that their norm,
\begin{equation}\label{eq:norm-3/2}
    \langle \psi | \psi \rangle_{\rm 3/2} = \int \dd^3x\, \sqrt{\bar g} \,\psi_a^\dagger \psi^a~, 
\end{equation}
is finite. We will first start the analysis for ${\cal N}=(1,1)$, where the Chern-Simons fields will play no role; we will then generalise the construction to cases where the gauge fields are present. 

\paragraph{\texorpdfstring{${\cal N}=(1,1)$}{N=1} supergravity.}
We start by diagonalising the Dirac operator in the ENHG background \eqref{Enhorizon}, where we parametrise the eigenvalues as 
\begin{equation}\label{eq:nhg-dirac-eigen}
    \ell\bar{\slashed{D}} \chi = \f(i\lambda-\r{1}{2}) \chi~,
\end{equation}
where $\lambda$ will be allowed to take complex values.
Although this problem is broader than our aim of characterising zero modes, it will allow us to connect with other results in the literature with little effort. In relation to \eqref{eq:dirac-chi}, our interest will be in cases when $\lambda=-2i$ and $\lambda=i$. The operator $\bar{\slashed D}$ evaluated on \eqref{Enhorizon} reads
\begin{equation}\label{eq:nhg-dirac}
    \bar{\slashed{D}}_{\mr{ENHG}} = \r{2}{\ell}\f(\r{1}{\sinh\eta}\gamma^0\partial_{\hat{t}_{\mr{E}}} + \gamma^1 \partial_\eta) + \r{1}{r_0}\f(\gamma^2 + i\tanh\r{\eta}{2}\gamma^0)\partial_{\hat{\phi}} + \r{1}{\ell}\coth\eta\gamma^1 - \r{1}{2\ell}~.
\end{equation}
The strategy is to first solve for eigenvalues of the square of $\ell\bar{\slashed{D}}_{\mr{ENHG}} + 1/2$, which leads to a pair of hypergeometric equations for the components of the spinor. The relative normalisation of these solutions is then obtained by replacing them in the first-order equation (\ref{eq:nhg-dirac-eigen}). 
The resulting eigenmodes are
\begin{multline}\label{eq:nhg-dirac-mode-plus}
    \chi^{(+)}_{\lambda k m} =e^{i(k+\r{1}{2})\hat{t}_{\mr{E}}+im\hat{\phi}}(\cosh\r{\eta}{2})^{-\r{im\ell}{r_0}} \sinh^{k}\eta 
    \m{i\r{\lambda + m\ell/r_0}{2(k+1)}\sinh\r{\eta}{2}\,
     {}_2F_1(a,b; k+2; -\sinh^2\r{\eta}{2}) \\
  \cosh\r{\eta}{2}\, {}_2F_1(a ,b; k+1; -\sinh^2\r{\eta}{2})
    }~,
\end{multline}
and
\begin{multline}\label{eq:nhg-dirac-mode-minus}
    \chi^{(-)}_{\lambda k m} = e^{-i(k+\r{1}{2})\hat{t}_{\mr{E}}-im\hat{\phi}}(\cosh\r{\eta}{2})^{-\r{im\ell}{r_0}} \sinh^k\eta 
    \m{
    \cosh\r{\eta}{2} \,{}_2F_1(a, b; k+1; -\sinh^2\r{\eta}{2})
    \\
    i\r{\lambda + m\ell/r_0}{2(k+1)}
    \sinh\r{\eta}{2} \,{}_2F_1(a, b; k+2; -\sinh^2\r{\eta}{2})
    }~,
\end{multline}
where 
\begin{equation}
\begin{aligned}
a=    k + 1 -\r{im\ell}{2r_0} + \r{i\lambda}{2}~,\qquad b=      k + 1 -\r{im\ell}{2r_0} - \r{i\lambda}{2}~,
\end{aligned}
\end{equation}
and we have $k \in \mathbb{Z}_{\ge 0}$ and $m \in \mathbb{Z}$. Note that for $m=0$ they become just the Dirac eigenmodes in AdS$_2$, as derived in \cite{camporesi_eigenfunctions_1996} and reviewed in \cite{Sen2012b}. The modes with $\lambda \in \R$ 
satsify
\begin{equation}
  \ip*{\chi_{\lambda km}}{\chi_{\lambda'k'm'}}_{\mr{1/2}} = \delta(\lambda-\lambda')\delta_{kk'}\delta_{mm'}
\end{equation}
with respect to the norm
\begin{equation}\label{eq:norm-1/2}
     \langle \chi | \chi \rangle_{\rm 1/2} = \int\dd[3]{x}\sqrt{\bar{g}}\, \chi^\dagger\chi~.
\end{equation}

With this, we can start constructing zero modes for the gravitino using \eqref{eq:defn-ferm-modes}. Let us first consider constructing the gravitino zero modes of the unprimed field $\psi$. This is easily achieved by taking $\chi^{(\pm)}_{\lambda km}$ in \eqref{eq:nhg-dirac-mode-plus}-\eqref{eq:nhg-dirac-mode-minus} with $\lambda = -2i$, hence
\begin{equation}
    (\psi_{a})^{(\pm)}_{km} = \bar{\cal{D}}_a \chi^{(\pm)}_{(-2i)km}~.
\end{equation}
Locally, this defines a zero mode, but rather crucially, we need to further restrict $\chi$ such that the norm of $\psi^{(\pm)}_{km}$, defined in \eqref{eq:norm-3/2}, is finite. A convenient way to understand the mechanics behind this condition is to first bound the norm of $\psi$  by the norm of $\chi$; we find
\begin{equation}
\begin{aligned}
   \langle (\psi_a)_{km}^{(\pm)} | (\psi_a)_{km}^{(\pm)} \rangle_{\rm 3/2} \ge \,  \langle (\psi_2)_{km}^{(\pm)} | (\psi_2)_{km}^{(\pm)} \rangle_{\rm 3/2} &= \frac{1}{r_0^2} \langle \partial_{\hat\phi}\chi_{(-2i)km}^{(\pm)} | \partial_{\hat\phi}\chi_{(-2i)km}^{(\pm)} \rangle_{\rm 1/2}\\
   &= \frac{m^2}{r_0^2}\langle \chi_{(-2i)km}^{(\pm)} |\chi_{(-2i)km}^{(\pm)} \rangle_{\rm 1/2}~.
\end{aligned}  
\end{equation}
From our analysis above, it is simple to demonstrate that the norm of $\chi_{(-2i) km}$ diverges.
Thus, normalisability of $\psi_{km}^{(\pm)}$ forces us to choose $m=0$ as a necessary condition for the mode to be normalizable. One can further verify that $m=0$ is also sufficient by explicitly evaluating the norm. Hence, we find two families of zero modes,
\begin{equation}
    \psi_{k}^{(\pm)} =\bar{\mc{D}} \chi^{(\pm)}_{(-2i) k0}~, \qquad k \in \mathbb{Z}_{\ge 1}~.
\end{equation}
Note that the mode with $k=0$ will correspond to a Killing spinor, and hence $\psi$ vanishes by construction. Finally, we normalise the modes such that $\langle \psi_{k'}^{(\pm)} | \psi_{k}^{(\pm)} \rangle_{\rm 3/2} = \ell\delta_{k'k}$, leading to the explicit expressions
\begin{equation}\label{eq:ferm-zero-nh}
\begin{aligned}
    \psi_{k}^{(+)} &= (i\bar{e}^0+\bar{e}^1)\sqrt{\r{k}{k+1}}\r{e^{i(k+\r{1}{2})\hat{t}_{\mr{E}}}\tanh^k\r{\eta}{2}}{\pi\sqrt{\ell r_0}\sinh\eta\cosh\r{\eta}{2}}\m{0\\1}~,\\
    \psi_{k}^{(-)} &= (-i\bar{e}^0+\bar{e}^1)\sqrt{\r{k}{k+1}}\r{e^{-i(k+\r{1}{2})\hat{t}_{\mr{E}}}\tanh^k\r{\eta}{2}}{\pi\sqrt{\ell r_0}\sinh\eta\cosh\r{\eta}{2}}\m{1\\0}~.
\end{aligned}
\end{equation}
On the other hand, there are no normalisable zero modes in the primed sector of \eqref{eq:defn-ferm-modes}-\eqref{eq:dirac-chi}. In this case, we need to set $\lambda = i$, and tentatively, we would have
\begin{equation}
    {\psi'}_{km}^{(\pm)} =\bar{\mc{D}'} \chi^{(\pm)}_{(i) km}~. 
\end{equation}
The norm of $\psi'$ would now obey
\begin{equation}
\begin{aligned}
   \langle ({\psi'}_a)_{km}^{(\pm)} | ({\psi'}_a)_{km}^{(\pm)} \rangle_{\rm 3/2} \ge \,  \langle (\psi'_2)_{km}^{(\pm)} | (\psi'_2)_{km}^{(\pm)} \rangle_{\rm 1/2} &=\big\langle \big(\frac{1}{r_0} \partial_{\hat\phi} +\r{\gamma_2}{\ell} \big)\chi_{(i)km}^{(\pm)} | \big(\frac{1}{r_0} \partial_{\hat\phi} +\r{\gamma_2}{\ell} \big)\chi_{(i)km}^{(\pm)} \big\rangle_{\rm 1/2}\\
   &= \f(\r{m^2}{r_0^2} + \r{1}{\ell^2})\langle \chi_{(i)km}^{(\pm)} |\chi_{(i)km}^{(\pm)} \rangle_{\rm 1/2}~.
\end{aligned}  
\end{equation}
The norm of $\chi_{(i)km}^{(\pm)}$ is infinite, which implies that for any value of $m$ and $k$ the norm of ${\psi'}_{km}^{(\pm)}$ is infinite. Therefore, we do not have any zero modes for the determinant associated with $\psi'$. This is expected since the background did not support spinors in this sector, as argued around \eqref{eq:kilspin-eq-primed}.

Having identified that the only fermionic zero modes are those in \eqref{eq:ferm-zero-nh}, we now turn to how they respond under a temperature deformation of the ENHG \eqref{nhg-correction}. Following the logic leading to \eqref{eq:lambda3}, the correction to the eigenvalues coming from fermionic modes is simply given by
\begin{equation}\label{eq:nhg-psi-eigenval}
    \delta \lambda^{\rm ferm}_{k,\pm}= T\lambda^{(1)\mr{ferm}}_{k,\pm} = -\frac{i}{\langle \psi_k^{(\pm)}|\psi_k^{(\pm)}\rangle}\int\bar{\psi}_k^{(\pm)}\wedge(\delta\mc{D})\psi_{k}^{(\pm)} = (k+\r{1}{2})\r{\pi\ell}{r_0}T~.
\end{equation}

\paragraph{\texorpdfstring{${\cal N}=(2,2)$}{N=2} and  \texorpdfstring{${\cal N}=(4,4)$}{N=4}  supergravity.} Turning on now to cases where we have more supercharges, the analysis of zero modes needs to take into the account that the derivatives $\mc{D}$ and $\mc{D}'$ in \eqref{eq:fermionic-action} have contributions from the background gauge fields $(\bar A, \bar A')$, according to \eqref{eq:def-covariant} and \eqref{eq:def-covariant-1}. In the case of a non-zero background $U(1)$ field, relevant for ${\cal N}=(2,2)$ supergravity, the zero modes are now 
\begin{equation}\label{eq:zero-n=2}
  \psi^{(\pm)}_{\scaleto{\rm U(1)}{4pt} k} =  g_{\scaleto{\rm U(1)}{4pt}}(x) \psi^{(\pm)}_{k}~, \qquad g_{\scaleto{\rm U(1)}{4pt}}(x) :=\exp\le(-i\int_{\gamma(x)} \bar{A}\ri)~,
\end{equation}
where $\psi^{(\pm)}_{k}$ is given by \eqref{eq:ferm-zero-nh} and  $\gamma(x)$ is any curve connecting $x^\mu$ to some fixed reference point. Their (anti-)periodicity under $\phi \sim \phi + 2\pi$ requires that $GQ = \r{n}{2}$ with $n \in \Z$; therefore, only supersymmetric black holes have fermionic zero modes in the canonical ensemble. Recall that, in Lorentzian signature, the spinors in the ${\cal N}=(1,1)$ theory are Majorana, while for ${\cal N}=(2,2)$ we are treating them as Dirac. In that context, we have now doubled the number of zero modes. 

Similarily, when the gauge group is $SU(2)$,  we have
\begin{equation}\label{eq:zero-n=4}
  (\psi^i_{\scaleto{\rm SU(2)}{4pt}})^{ (\pm)}_{k} =  g_{\scaleto{\rm SU(2)}{4pt}}(x)\varepsilon^i\psi^{(\pm)}_{k}~,\qquad  g_{\scaleto{\rm SU(2)}{4pt}}(x) :={\rm P} \exp \left(-\int_{\gamma(x)} \bar{A}\right)~,
\end{equation}
which will be (anti-)periodic if $GQ=n$ and $n \in \Z$. Thus, the conditions for the existence of fermionic zero modes are the same as the conditions for the existence of a Killing spinor discussed in Sec.\,\ref{sec:bhs}. Here $\varepsilon^i$ is a complex constant, simply indicating that we now have more zero modes, and effectively double it relative to the ${\cal N}=(2,2)$ case.

Finally, we remark that the correction to the eigenvalues arising from turning on temperature in the ENHG is still given by \eqref{eq:nhg-psi-eigenval}. The reason is that $\bar A$ does not change under a temperature deformation. In these cases, we just have to keep track of how many additional modes there are relative to the ${\cal N}=(1,1)$ analysis. 

\subsection{\texorpdfstring{Log-$T$}{Log-T} corrections to the GPI}\label{sec:final-GPInear}

In this last portion, we now discuss how the zero modes make an imprint on the GPI from the perspective of the ENHG in the canonical ensemble. At zero temperature, the GPI around the ENHG takes the form
\begin{equation}\label{eq:ZNH}
    Z_{\rm ENHG}[T=0] \approx Z_{0}[\bar g_{\rm ENHG}, \bar A_{\rm ENHG}] Z_{\rm z.m.}Z_{\rm n.z.m.}~,
\end{equation}
where $Z_{0}$ is the tree-level contribution, and we have split the one-loop contributions \eqref{eq:one-loop} into two: contributions from zero modes, $Z_{\rm z.m.}$, and non-zero modes $Z_{\rm n.z.m.}$. In \eqref{eq:ZNH}, we are neglecting higher loop corrections and non-perturbative effects. As we saw, there is an infinite set of zero modes in the ENHG background, and the proposal of  \cite{Iliesiu:2022onk} is to introduce temperature as a regulator, by which we deform the background from $\bar g_{\rm ENHG}$ to $\bar g_{\rm ENHG}+T \bar g^{(1)}$. Every term in \eqref{eq:ZNH} will clearly be affected by this deformation, but the leading effect at low temperatures comes from the zero modes:
\begin{equation}
\begin{aligned}\label{eq:ZNH-lowT}
    \log  Z_{\rm ENHG}[T] &\approx \log Z_{\rm z.m.}[T] +\cdots\\
    & \approx \log \prod_n \left(T \, \Lambda_n^{(1)}\right)^{\pm 1/2} + \cdots~,
\end{aligned}
\end{equation}
where $\delta \Lambda_n = T \Lambda_n^{(1)}$ is the zero mode contribution to the correction in the action due to the deformation of the background, and the $\pm$ reflect whether the determinants are Grassmann or not. Here, we will combine the corrections we found for each of the zero modes discussed, both bosonic and fermionic, and quantify the effect they have on $Z_{\rm ENHG}[T]$. The two main cases to differentiate are between supersymmetric and non-supersymmetric backgrounds at $T=0$.  

It is important to stress that these manipulations are optimistic. We have shown that the standard assumptions of perturbation theory break down for the rotational and gauge modes, since the gauge-fixing conditions cannot be satisfied while requiring that the modes be normalizable. The purpose of this portion is to serve as a summary and point of comparison with the far region analysis in Sec.\,\ref{sec:far}.

\paragraph{Non-supersymmetric black holes.} At $T=0$, this case corresponds to the near-horizon geometry of an extremal charged black hole that does not support a Killing spinor. For this class of ENHG background, we only encountered bosonic zero modes: tensor and rotational modes arising from the graviton one-loop determinant, and gauge zero modes from the Chern-Simons fields. Only the tensor modes had a non-trivial temperature correction to their eigenvalues; rotational and gauge modes continued to be zero modes after adding temperature effects. Therefore, we have  
\begin{equation}
\begin{aligned}\label{eq:ZNH-non-BPS}
    \log  Z_{\rm ENHG}^{ \scaleto{\rm non-BPS}{4pt}}[T,Q] &\approx \log Z_{\rm z.m.}^{\scaleto{\rm tensor}{4pt}}[T]  +\cdots\\
    & \approx-\frac{1}{2} \log \prod_{|n|\geq2} \left( \r{\abs{n}\pi\ell^2 }{2r_0} T\right)  \cdots\\
    &\approx \r{3}{2}\log\left(\r{T}{T_{\rm break}}\frac{1}{S_0}\right) +\cdots,
\end{aligned}
\end{equation}
where in the second line we used \eqref{eq:nhg-sch-eigenval}. In the third line, we introduced the following quantities:
\begin{equation}
    S_0 = \frac{\pi r_0}{2G } ~, \qquad T_{\rm break} = \frac{6}{\pi^2 c \, \ell}~,\qquad c=\frac{3\ell}{2G}~,
\end{equation}
with $S_0$ the Bekenstein-Hawking entropy at $T=0$, and $T_{\rm break}$ which controls the linear temperature response of the Bekenstein-Hawking entropy. We also regulated the products in \eqref{eq:ZNH-non-BPS} by using zeta function regularisation, in particular,
\begin{equation}\label{eq:zeta-1}
  \prod_{n\ge 2} \r{\alpha}{n} = \r{1}{\alpha^{3/2}\sqrt{2\pi}}~.
\end{equation}

\paragraph{Supersymmetric black holes.} Focusing next on cases where the black hole supersymmetry, we now have, in addition to the bosonic modes, the fermionic modes to account for as well. Schematically, we will have 
\begin{equation}
\begin{aligned}\label{eq:ZNH-BPS}
    \log  Z_{\rm ENHG}^{ \scaleto{\rm BPS}{4pt}}[T,Q] &\approx \log Z_{\rm z.m.}^{\scaleto{\rm tensor}{4pt}}[T] +  \log Z_{\rm z.m.}^{\scaleto{\rm fermion}{4pt}}[T]\cdots~.
\end{aligned}
\end{equation}
Despite the presence of gauge zero modes in the Chern-Simons path integral, we have not included them in \eqref{eq:ZNH-BPS}; as we saw in \eqref{gaugeev-nhg}, their eigenvalues are not corrected from the ENHG perspective.

To quantify the fermionic contribution it is instructive to start with the simplest case of ${\cal N}=(1,1)$. The zero modes are given by \eqref{eq:ferm-zero-nh}, and the correction to the eigenvalue is \eqref{eq:nhg-psi-eigenval}. Therefore, the fermionic contribution is 
    \begin{equation}\label{eq:ZNH-fermion}
        \begin{aligned}
             \log Z_{\rm z.m.}^{\scaleto{\rm fermion}{4pt}}[T]&=  \r{1}{2}\sum_{k\geqslant1} (\log \delta\Lambda^{\mr{ferm}}_{k,+} + \log \delta\Lambda^{\mr{ferm}}_{k,-}) \\
             & = \sum_{k\geqslant1} (k+\r{1}{2})\r{\pi\ell^2T}{r_0} = -\log\left(\r{T}{T_{\rm break}}\frac{1}{2S_0}\right)~.
        \end{aligned}
    \end{equation}
Note that the factor of $1/2$ in the first line is halving the degree of freedom due to the Majorana condition on the spinors.    The infinite sum is evaluated by an appropriate zeta function regularisation, where we used 
    \begin{equation}\label{eq:zeta-2}
 \prod_{k\geqslant1}\r{\alpha}{2k+1} = \r{1}{\alpha}~.
\end{equation}
For the extended supergravities under consideration, the modifications to \eqref{eq:ZNH-fermion} are simple to record. In the cases that we have been tracking explicitly, we have
\begin{equation}
    \begin{aligned}
        {\cal N}=(2,2): \qquad  \log Z_{\rm z.m.}^{\scaleto{\rm fermion}{4pt}}[T]=  -2\log\left(\r{T}{T_{\rm break}}\frac{1}{2S_0}\right)~,\\
        {\cal N}=(4,4): \qquad   \log Z_{\rm z.m.}^{\scaleto{\rm fermion}{4pt}}[T]=  -4\log\left(\r{T}{T_{\rm break}}\frac{1}{2S_0}\right)~,
    \end{aligned}
\end{equation}
where we have incorporated in each case the appropriate number of fermionic zero modes based on the number of degrees of freedom as discussed around \eqref{eq:zero-n=2}-\eqref{eq:zero-n=4}.

\section{GPI at low-temperature: far-away analysis}\label{sec:far}

The NHG simplifies the analysis of quantum corrections because of its enhanced symmetry relative to the parent black hole, with the AdS$_2$ portion of the geometry playing a central role. This, however, comes at a price. Since the NHG is only an approximation of the physical spacetime near the horizon, it does not necessarily capture all of the aspects of the modes arising there. In this section, we will show explicitly the quantum aspects of the black hole that are missed by taking a near-horizon limit. 

We will focus on the GPI around the BTZ black hole, and use the locally AdS$_3$ structure of the spacetime to study the behaviour of eigenmodes in the one-loop determinant as one takes the black hole temperature to zero.  This includes a construction of the eigenmodes for the graviton, gauge and gravitino, which become zero modes in the limit. More explicitly, at leading order near zero temperature, the GPI we will analyse is
\begin{equation}\label{eq:Zbtz-1}
\begin{aligned}
    Z_{\rm BTZ}[T] &~ = ~ Z_0[T] Z_{\rm one-loop}[T]  \\
    &\underset{T\to0}{\approx} Z_0[T] Z_{\rm low}[T] + \cdots ~.
\end{aligned}
\end{equation}
Note that we will not be taking the low-temperature limit of the final expression for the one-loop determinant $ Z_{\rm one-loop}[T]$ as done in, e.g., \cite{HeyIli20,GhoMax19}. The analysis we will perform is an explicit construction of the modes in the path integral with the appropriate zero-temperature limit along the lines of \cite{KolMar24,Acito:2025hka}. 
We will quantify $Z_{\rm low}[T]$ and qualitatively compare it against $Z_{\rm z.m.}[T]$ in \eqref{eq:ZNH-lowT}.\footnote{In this section our results are naturally in the grand canonical ensemble, whereas in Sec.\,\ref{sec:near} the analysis is primarily in the canonical ensemble. As we do appropriate comparisons, the change of ensemble will be taken into account.} 
As we will find out, while for some of the zero modes in the NHG analysis (such as the tensor and gravitino) correctly matches with \eqref{eq:Zbtz-1}, for others (such as the rotational and gauge zero modes) the NHG misses parts of the eigenvalue and eigenmode response to the temperature, and this leads to an incorrect imprint of the these corrections in the GPI.

In contrast to Sec.\,\ref{sec:near}, in this section, the saddle point $\bar{g}$ of the GPI around which we calculate the one-loop corrections discussed in Sec.\,\ref{sec:one-loop} will not be the near-horizon approximation \eqref{Enhorizon}, but instead the Euclidean BTZ geometry obtained from \eqref{btzm1} by Wick rotation $t = -it_{\mr{E}}$. Explicitly, the line element is
\begin{equation}\label{eq:Euc-BTZ}
  \bar{g}_{\mu\nu}\dd x^\mu \dd x^\nu = f(r)\dd{t_{\mr{E}}}^2 + \r{\dd{r}^2}{f(r)} + r^2\f(\dd{\phi} + \r{ir_+r_-}{\ell r^2}\dd{t_{\mr{E}}})^2,\quad f(r) \coloneqq \r{(r^2-r_+^2)(r^2-r_-^2)}{\ell^2r^2}~.
\end{equation}
We will also use the complex coordinates $(w,\bar w)$ introduced in \eqref{eq:wwbar-euc}, and the spatial and thermal identifications are specified around \eqref{eq:thermal-cycle}.
This solution will be supported by a pair of connections $(\bar A,\bar A')$, and the quantity that we will be prominent in the coming discussion is the corresponding holonomy,
\begin{equation}\label{eq:btz-holonomies}
     g(x) = {\rm P}\,\exp\left({\int_{\gamma(x)} \bar{A}}\right)~, \quad  g'(x) = {\rm P}\,\exp\left({\int_{\gamma(x)} \bar{A}'}\right)~.
\end{equation}
The holonomy of $g(x)$ and $g(x)'$ around the spatial cycle measures the charge $Q$ and $Q'$, respectively; the holonomy along the thermal cycle is trivial, which fixes the charge in terms of the potential according to \eqref{eq:u(1)-potential}. 
Along the lines of \eqref{eq:rpm-btz}-\eqref{eq:low-t-mass}, the near-extremal limit is defined as $r_\pm \rightarrow r_0 > 0$, or equivalently
\begin{equation}
  T_L = \frac{r_+-r_-}{2 \pi \ell^2} \rightarrow 0~,\qquad T_R = \frac{r_++r_-}{2 \pi \ell^2}\rightarrow \frac{r_0}{\pi\ell^2}~.
\end{equation}
To compare with \cite{HeyIli20}, we note that the near-extremal limit there holds fixed the chemical potential $\alpha$ fixed, which in our notation translates to 
\begin{equation}
    2\pi i \alpha = \beta \mu~. 
\end{equation}
Hence, as we take the extremal limit, $\beta\to\infty$, the appropriate limit used in \cite{HeyIli20} requires $\mu\to0$. This is a {\it near-BPS} limit due to \eqref{eq:u(1)-potential}: we are taking $Q\to 0$, which is compatible with the solution supporting a Killing spinor. This near-BPS limit can be generalised according to \cite{Larsen:2021wnu}, where the general condition to reach a BPS black hole is that
\begin{equation}\label{eq:near-BPS}
   \mu - k (1 +\ell \Omega) \to 0 ~,
\end{equation}
where $k \in \mathbb{Z}$ or $\mathbb{Z}/2$ depending on the gauge group; see \eqref{eq:susy-charge-U(1)} and  \eqref{eq:susy-charge-SU(2)}.

Around this solution, we will consider the eigenvalue equations
\begin{equation}
  \bar{\Delta} \varphi_n = \lambda_n \varphi_n~,
\end{equation}
for the various quadratic operators appearing in the formulas for one-loop determinants in Sec.\,\ref{sec:one-loop}. In the grand canonical ensemble, we will require that the fields are subject to the boundary conditions
\begin{equation}\label{eq:bc-fields}
    \varphi_n(w+2\pi,\bar w +2\pi)=\varphi_n(w,\bar w)~,\qquad     \varphi_n(w+2\pi\tau,\bar w +2\pi\bar\tau)=\varphi_n(w,\bar w)~,  
\end{equation}
that is, requiring that fields are single-valued along the spatial and thermal cycle, and appropriate modifications if fields are fermionic. We will focus on eigenmodes for which $\lambda_n \rightarrow 0$ in the near-extremal limit: these are the modes that contribute to $Z_{\rm low}[T]$. The analysis in this portion follows the approach taken in \cite{KolMar24}, which uses techniques developed in \cite{Datta:2011za,castro_tweaking_2017}, where a class of gravitational modes was constructed; in particular, those connected to the tensor modes in Sec.\,\ref{sec:nhg-grav}. Here we will complete the discussion of gravitational modes, and include as well a complete analysis of modes arising from the quantum fluctuations of the gauge fields and gravitino.  Several of the technical aspects of the construction are delegated to App.\,\ref{sec:app-full-btz}.

\subsection{Graviton modes}\label{sec:grav-far}

In this section, we analyse transverse-traceless (TT) perturbations of the BTZ background that become zero modes in the extremal limit. These modes govern the behaviour of the graviton one-loop determinant appearing in \eqref{eq:eh-partition-function-reduced}. The relevant quadratic operator acting on TT perturbations is
\begin{equation}\label{eq:graviton-quadratic}
    \bar{\Delta}_{\rm grav}\big|_{\rm TT}
    = -\frac{1}{4}\left(\bar{\Box}_2+\frac{2}{\ell^2}\right).
\end{equation}
To determine the spectrum of this operator, it is useful to decompose more explicitly the conditions behind it. One rewriting of \eqref{eq:graviton-quadratic} is as follows
\begin{equation}\label{eq:ads3-spin2-second-order-mt}
\begin{aligned}
    (\ell^2\bar\Box+2)\mathsf{h}_{\mu\nu}= (\lambda^2-1)\mathsf{h}_{\mu\nu}~,
    \\ 
    \bar\nabla^\mu \mathsf{h}_{\mu\nu}=0~,
    \quad 
    \mathsf{h}^\mu{}_\mu=0 ~,
\end{aligned}
\end{equation}
where $(1-\lambda^2)/4$ labels the eigenvalues of \eqref{eq:graviton-quadratic}, and we have stated the transverse and traceless conditions explicitly. However, in three dimensions, there is a simpler way to repackage \eqref{eq:ads3-spin2-second-order-mt} in terms of a first-order equation \cite{vasiliev_lagrangian_1997,Datta:2011za}, given by
\begin{equation}\label{eq:ads3-spin2-first-order-mt}
    \bar\epsilon_{\mu}{}^{\lambda\sigma}\bar\nabla_\lambda \mathsf{h}_{\sigma\nu}
    =  \frac{i\lambda}{\ell}\,\mathsf{h}_{\mu\nu}~.
\end{equation}
In App.\,\ref{sec:app-btz-graviton}, we obtain the general solution to these equations in the BTZ geometry and analyse its eigenvalues under \eqref{eq:graviton-quadratic}. Among all metric perturbations that are transverse and traceless, we find that only two families of modes develop vanishing eigenvalues as the black hole approaches extremality, that is 
\begin{equation}\label{eq:condition-lambda}
    \lambda(T)\underset{T\to 0}{\rightarrow} \pm 1~.
\end{equation}
These are the modes that contribute to $Z_{\rm low}[T]$ from the gravitational. We will call them $\mathsf{T}$-modes and $\mathsf{R}$-modes due to their close relation with the tensor and rotational modes in the near-horizon region. 

\paragraph{$\mathsf{T}$-modes:} This class of metric fluctuations were originally found in \cite{KolMar24}, and they take the form
\begin{multline}\label{eq:schwarzian-btz}
    \mathsf{h}^{\mathsf{T}}_n = N^{\mathsf{T}}_n e^{\r{2\pi i n}{\beta}t_E} r^2 \f(\r{r^2-r_+^2}{r^2-r_-^2})^{\r{\abs{n}}{2}-2}\f(\r{r_+^2-r_-^2}{r^2-r^2_-})^{4-\r{\abs{n}}{2}\r{r_+-r_-}{r_+}} \\
    \times \f(\dd{r} - \sgn(n)\r{r f(r)}{r_++r_-}(\ell\dd{\phi}-i\dd{t_E}))^2~,
\end{multline}
with $n$ an integer and $N^{\mathsf{T}}_n$ an overall normalisation. They also comply with the periodicity \eqref{eq:bc-fields}. Their eigenvalue with respect to the operator in the graviton one-loop determinant \eqref{eq:graviton-quadratic} is
\begin{equation}\label{eq:btz-sch-eigenval}
\left(\bar{\Delta}_{\mr{grav}}|_{\rm{TT}}\right)\mathsf{h}_n^{\mathsf{T}} = \r{\abs{n}T}{2\ell^2T_R}\f(1-\r{\abs{n}T}{2T_R})\mathsf{h}_n^{\mathsf{T}}~.
\end{equation}
From this expression, it is clear that in the extremal limit, where $T\to 0$ and $T_R$ is finite, the eigenvalue of the modes approaches zero. 

For the metric fluctuation $ \mathsf{h}^{\mathsf{T}}_n$ to contribute to the path integral $Z_{\rm BTZ}[T]$, we still need to impose two additional conditions: the modes should be {\it normalizable} and comply with appropriate {\it boundary conditions}. Imposing normalizability restricts $n$ to   
\begin{equation}\label{eq:norm-cond-T}
    2 \leqslant \abs{n} < 3\r{T_R}{T}~.
\end{equation}
Within this range, we will then have
\begin{equation}\label{normten}
    \ip*{\mathsf{h}^{\mathsf{T}}_{-m}}{\mathsf{h}^{\mathsf{T}}_{n}}
    = \ell^2\delta_{nm}~,
\end{equation}
where we have chosen the normalisation constant in \eqref{eq:schwarzian-btz} to be 
\begin{equation}\label{eq:btz-sch-norm-const}
   N^{\mathsf{T}}_n =\frac{\ell^2}{(r_+^2-r_-^2)^2} \sqrt{\r{\Gamma(2+\abs{n}\r{r_-}{r_+})}{8\pi^2r_+  \Gamma(\abs{n}-1)\Gamma(3-\abs{n}(1-\r{r_-}{r_+}))}}~.
\end{equation}
Our second condition is based on boundary conditions. For this, we inspect the behaviour of the modes at the asymptotic boundary of BTZ. In the limit $r\to\infty$, the tensor modes behave as
\begin{equation}
\lim_{r\to\infty}\mathsf h^{\mathsf{T}}_n
= N^{\mathsf{T}}_n e^{\frac{2\pi i n}{\beta}t_E}
\left(\frac{r_+^2-r_-^2}{r^2}\right)^{3-\frac{|n|}{2}\frac{r_+-r_-}{r_+}}
\left(\dd r-\sgn(n)\frac{r^3}{r_++r_-}(\ell\,\dd\phi-i\,\dd t_E)\right)^2 .
\end{equation}
From this expression, one finds that the boundary metric components scale as
\begin{equation}
(\mathsf{h}^{\mathsf{T}}_n)_{ij}\sim r^{\,|n|\frac{T}{T_R}} ~.
\end{equation}
Thus, these components grow at large $r$ for all $n$ even if we impose \eqref{eq:norm-cond-T}. In the context of the standard Brown-Henneaux boundary conditions used in AdS$_3$/CFT$_2$, the sources for the metric, that is, the temperature and angular potential, are carried by the boundary metric; in the notation used in this section, this means that source terms are of order
\begin{equation}\label{eq:bc-Dir}
    \bar g_{ij}\Big|_{\rm sources} \sim O(r^2)~.
\end{equation}
Requiring that our quantum fluctuations $\mathsf{h}^{\mathsf{T}}_n$ do not alter the boundary sources leads to the more stringent upper bound on $|n|$, leading to the condition
\begin{equation}\label{eq:final-cond-T-modes}
2\leq |n|\leq \frac{r_+}{r_+-r_-} = \frac{T_R}{T}~.
\end{equation}

It is interesting to note that these modes $\mathsf{h}^{\mathsf{T}}_n$ are rather peculiar. First, how many contribute to the GPI is a function of temperature, which turns them from being rather scarce at finite $T$ to suddenly overpowering the low-temperature behaviour of  $Z_{\rm BTZ}[T]$. Second, even after imposing \eqref{eq:final-cond-T-modes}, it is somewhat strange that the modes grow so quickly near the boundary. For example, in the construction of a classical phase space, one would expect to have $\mathsf{h}_{ij}\sim O(1)$. As we will discuss below, the conditions imposed here are compatible with other methods to quantify $Z_{\rm low}[T]$, hence it suffices to impose \eqref{eq:final-cond-T-modes}.  

\paragraph{$\mathsf{R}$-modes:} The second class of modes that comply with \eqref{eq:condition-lambda} and \eqref{eq:bc-fields} are given by
\begin{multline}\label{eq:r-modes}
    \mathsf{h}^{\mathsf{R}}_n = N^{\mathsf{R}}_n e^{\r{2\pi i n}{\beta}t_E}\r{r_+-r_-}{\ell^4}\f(\r{r^2-r_+^2}{r^2-r_-^2})^{\r{\abs{n}}{2}}\f(\r{r_+^2-r_-^2}{r^2-r_-^2})^{-\r{r_+-r_-}{r_+}\r{\abs{n}}{2}}
    \f(\r{r_++r_-}{r f(r)}\sgn(n)\dd{r} - (\ell \dd\phi - i\dd{t_E})) \\
    \times\left[ \r{\abs{n}(r_+-r_-)(r^2+r_+r_-)-r_+(2r^2-r_+^2-r_-^2)}{r_+ r f(r)}\dd{r} - 2(r_++r_-)\sgn(n)(\ell\dd{\phi}+i\dd{t_E}) \right.\\
    \left. - \r{\abs{n}(r_+-r_-)(r^2+r_+r_-)+r_+(2r^2-r_+^2-r_-^2)}{r_+(r_++r_-)}\sgn(n)(\ell\dd{\phi}-i\dd{t_E})
    \right]~,
\end{multline}
with $n$ an integer and $N^{\mathsf{R}}_n$ an overall normalisation. Their eigenvalue with respect to \eqref{eq:graviton-quadratic} is
\begin{equation}\label{eq:btz-rot-eigenval}
    \left(\bar{\Delta}_{\mr{grav}}|_{\rm TT}\right)\mathsf{h}^{\mathsf{R}}_n
    = -\r{\abs{n}T}{2\ell^2T_R}\f(1+\r{\abs{n}T}{2T_R})\mathsf{h}^{\mathsf{R}}_n~.
\end{equation}
It is important to notice that, in contrast to \eqref{eq:btz-sch-eigenval}, the eigenvalues of these modes are negative.\footnote{ The negativity in \eqref{eq:btz-rot-eigenval} could be used at this stage as a reason to dismiss the modes since the Gaussian integral is not convergent. Still, there are a plethora of examples where negative eigenvalues appear in GPIs, and a common approach is to perform suitable deformations of the contour of integration in $Z_{\rm grav}$ to make it convergent. See, for example, \cite{Marolf:2022ntb,Liu:2023jvm,blacker_quantum_2025} for recent discussions in this regard. } Similar to the $\mathsf{T}$-modes, the first condition is to require that the modes are normalizable, which enforces
\begin{equation}
    1 \leqslant \abs{n} < \r{T_R}{T}~.
\end{equation}
Given this restriction, the norm is then given by
\begin{equation}
    \ip*{\mathsf{h}^{\mathsf{R}}_{-m}}{\mathsf{h}^{\mathsf{R}}_n}
    = \ell^2\delta_{nm}~,
\end{equation}
where the normalisation factor in \eqref{eq:r-modes} is given by
\begin{equation}
  N^{\mathsf{R}}_n = \sqrt{\r{l^4r_- (\abs{n}+1)\Gamma(1+\abs{n}\r{r_-}{r_+})}{8\pi^2 r_+\f(2r_+ + (r_+-r_-)n^2 + (3r_+-r_-)\abs{n})\Gamma(\abs{n})\Gamma(1-\abs{n}(1-\r{r_-}{r_+}))}}~.
\end{equation}
The second criterion on the modes comes from the boundary conditions they respect at the asymptotic boundary of BTZ. For the $\mathsf{R}$-modes, the large-$r$ limit gives
\begin{multline}
    \lim_{r\to \infty}\mathsf{h}^{\mathsf{R}}_n = N^{\mathsf{R}}_n e^{\r{2\pi i n}{\beta}t_E}\r{r_+-r_-}{\ell^4}\f(\r{r_+^2-r_-^2}{r^2})^{-\r{r_+-r_-}{r_+}\r{\abs{n}}{2}}
    \f(\r{r_++r_-}{r^3}\dd{r} - \sgn(n)(\ell \dd\bar w)) \\
    \times\left[ \r{\abs{n}(r_+-r_-)-2r_+}{r_+ r}\dd{r} - 2(r_++r_-)\sgn(n)(\ell\dd w) - r^2\r{\abs{n}(r_+-r_-)+2 r_+}{r_+(r_++r_-)}\sgn(n)(\ell \dd\bar w)
    \right]~,
\end{multline}
where $w, \bar{w}$ are defined in \eqref{eq:wwbar-euc}.
The asymptotic scaling of the individual components is
\begin{equation}
\begin{aligned}\label{eq:far-r-modes}
(\mathsf h^{\mathsf{R}}_n)_{rr}&\sim r^{|n|\frac{r_+-r_-}{r_+}-4}, 
&\qquad 
(\mathsf h^{\mathsf{R}}_n)_{r\bar w}&\sim r^{|n|\frac{r_+-r_-}{r_+}-1},
&\qquad 
(\mathsf h^{\mathsf{R}}_n)_{rw}&=\mathcal O(1),\\[4pt]
(\mathsf h^{\mathsf{R}}_n)_{\bar w\bar w}&\sim r^{|n|\frac{r_+-r_-}{r_+}+2},
&\qquad 
(\mathsf h^{\mathsf{R}}_n)_{w\bar w}&\sim r^{|n|\frac{r_+-r_-}{r_+}} .
\end{aligned}
\end{equation}
The rapidly growing $(\bar w\bar w)$-component of the fluctuation violates the boundary condition stated in \eqref{eq:bc-Dir}, i.e., that fluctuations do not change the temperature and angular momentum when working in the standard AdS$_3$/CFT$_2$ setup. Therefore, $\mathsf{R}$-mode will never contribute in this scenario, which is compatible with the previous analysis of $Z_{\rm BTZ}[T]$ in \cite{KapLaw24,GhoMax19}. However, AdS$_3$ admits a whole plethora of boundary conditions that deviate significantly from the standard Brown-Henneaux conditions; see, for example, \cite{Grumiller:2016pqb}. One particular choice that seems fitting to the behaviour in \eqref{eq:far-r-modes} are the CSS boundary conditions \cite{Compere:2013bya}. CSS allows for the boundary metric to fluctuate and, in particular, grow as $\bar g_{\bar w\bar w}=\mathcal O(r^2)$. This modification would allow the incorporation of $\mathsf{h}^{\mathsf{R}}_n$ in the path integral and hence contribute to $Z_{\rm low}[T]$.

\paragraph{Comparison with the NHG.} The next step is to compare the $\mathsf{T}$ and $\mathsf{R}$ modes in this section, obtained at finite temperature on BTZ, with the tensor and rotational modes of Sec.\,\ref{sec:nhg-grav}, derived from the ENHG. Starting from our modes at finite temperature, we will apply the transformations \eqref{eq:decouple-coords} to \eqref{eq:schwarzian-btz} to expand the modes in \eqref{eq:schwarzian-btz} and \eqref{eq:r-modes}; this gives
\begin{equation}\label{eq:expand-eigenmodes}
  \mathsf{h}^{\mathsf{T}/\mathsf{R}}_n = h^{\mr{tensor/rot}}_n + T\delta \mathsf{h}^{\mathsf{T}/\mathsf{R}} + \mc{O}(T^2)~,
\end{equation}
where $h^{\mr{tensor/rot}}_n$ are given by \eqref{eq:nhg-sch} and \eqref{eq:h-rot}. Here we see that at $T=0$ the modes in the far region correctly match with the modes in the near region. We can do the same procedure on the eigenvalues in \eqref{eq:btz-sch-eigenval} and \eqref{eq:btz-rot-eigenval}, where it is simple to see that
\begin{equation}
  \mathsf{\lambda}_n^{\mathsf{T}/\mathsf{R}} =  0 +  \delta\lambda_n^{\mathsf{T}/\mathsf{R}} + \mc{O}(T^2)~,
\end{equation}
hence confirming that at $T=0$, both approaches agree.

The interesting comparison between the near and far regions comes from the leading order correction in temperature. Starting with the $\mathsf{T}$-modes, we find from \eqref{eq:btz-sch-eigenval} that the leading-order correction to the action  is
\begin{equation}\label{evcorr}
        \delta\lambda_n^{\mathsf{T}} = \r{\abs{n}\pi }{2r_0} T~, 
\end{equation}
where we used  $T_R= \frac{r_0}{\pi \ell^2}$ at extremality.
Comparing with \eqref{eq:nhg-sch-eigenval}, we see that the eigenvalue correction of the $\mathsf{T}$-modes agrees with the near-horizon analysis: we happily report that $ \delta\lambda_n^{\mathsf{T}}=  T \lambda_n^{(1)\rm tensor}$.  Next, we can also inspect the correction to the eigenmodes in \eqref{eq:expand-eigenmodes}.
The precise form of $\delta h^{\mathsf{T}}$ is not very illuminating, but it is instructive to inspect their asymptotic behaviours. We find  
\begin{equation}
  \begin{gathered}
    \delta \mathsf{h}^{\mathsf{T}}_{\hat{t}_{\mr{E}}\hat{t}_{\mr{E}}} = \mc{O}(\eta)~,\quad
    \delta \mathsf{h}^{\mathsf{T}}_{\hat{t}_{\mr{E}}\eta} = \mc{O}(\eta e^{-\eta})~,\quad
    \delta\mathsf{h}^{\mathsf{T}}_{\eta\eta} = \mc{O}(\eta e^{-2\eta})~, \\
    \delta \mathsf{h}^{\mathsf{T}}_{\hat{t}_{\mr{E}}\hat{\phi}} = \mc{O}(1)~,\quad
    \delta \mathsf{h}^{\mathsf{T}}_{\eta \hat{\phi}} = \mc{O}(e^{-\eta})~,\quad
    \delta \mathsf{h}^{\mathsf{T}}_{\hat{\phi}\hat{\phi}} = 0~.
  \end{gathered}
\end{equation}
It is simple to verify that the leading-order correction in $T$ of the eigenmodes remains well-behaved and normalisable in the NHG. 

Next, let us turn to the $\mathsf{R}$-modes. The correction to the eigenvalue from \eqref{eq:btz-rot-eigenval} reads
\begin{equation}\label{eq:xyz123}
            \delta\lambda_n^{\mathsf{R}} = -\r{\abs{n}\pi }{2r_0} T~.
\end{equation}
In contrast to $\mathsf{T}$-modes, for the $\mathsf{R}$-modes there is a disagreement between $\delta\lambda_n^{\mathsf{R}}$ 
and the correction for rotational modes in \eqref{eq:nhg-rot-eigenval}, where we found that $\lambda^{(1)\rm rot}_n=0$. The source of the discrepancy originates from the corrections to eigenmodes $\delta h^{\mathsf{R}}$. 
We can easily read off this correction from \eqref{eq:r-modes}: near the boundary of the ENHG, this correction behaves as 
\begin{equation}
  \begin{gathered}
    \delta \mathsf{h}^{\mathsf{R}}_{\hat{t}_{\mr{E}}\hat{t}_{\mr{E}}} = \mc{O}(e^{2\eta})~,\quad
    \delta \mathsf{h}^{\mathsf{R}}_{\hat{t}_{\mr{E}}\eta} = \mc{O}(1)~,\quad
    \delta \mathsf{h}^{\mathsf{R}}_{\eta\eta} = \mc{O}(1)~, \\
    \delta \mathsf{h}^{\mathsf{R}}_{\hat{t}_{\mr{E}}\hat{\phi}} = \mc{O}(e^\eta)~,\quad
    \delta \mathsf{h}^{\mathsf{R}}_{\eta \hat{\phi}} = \mc{O}(\eta e^{-\eta})~,\quad
    \delta \mathsf{h}^{\mathsf{R}}_{\hat{\phi}\hat{\phi}} = \mc{O}(1)~,
  \end{gathered}
\end{equation}
and is therefore non-normalisable in the ENHG. This indicates that perturbation theory breaks down for these modes, and therefore, \eqref{eq:nhg-rot-eigenval} should not be expected to give the correction to the eigenvalues of rotational modes. Recall that \eqref{eq:nhg-rot-eigenval} assumes that the second term on the right-hand side of the general expression \eqref{eq:lambda2} cancels the third term, which is appropriate under the assumptions that the perturbation in $T$ is adiabatic.  This breakdown arises because we have
\begin{equation}
   \delta \mathsf{h}^{\mathsf{R}} = \delta\mathsf{h}^{\infty} + \delta h^{\rm rot}~,
\end{equation}
where $\delta\mathsf{h}^{\infty}$ is the non-normalizable piece of the eigenmode and $\delta h^{\rm rot}$ is the normalizable contribution. In particular, $\delta h^{\rm rot}$ is constructed as a superposition of normalizable eigenmodes of  $\bar{\Delta}^{(0)}_{\mr{grav}}$. The correction we evaluated in  \eqref{eq:nhg-rot-eigenval} arises from assuming that the corrected eigenmode is just $\delta h^{\rm rot}$, which leads to \eqref{eq:lambda3}.
What the far region analysis tells us is that we need to account for the contribution arising from $\delta\mathsf{h}^{\infty}$ to the total correction of the eigenvalue.
Using \eqref{eq:lambda2} and the explicit expression for $\delta \mathsf{h}^{\mathsf{R}}$, we have that the correction to the eigenvalue arising from the non-normalizable mode is
\begin{equation}
\begin{aligned}\label{eq:rot-far-correction}
     \delta\lambda_n^{\infty}&= T\frac{\ip*{h^{\mr{rot}}_{-n}}{\bar{\Delta}^{(0)}_{\mr{grav}} \delta\mathsf{h}^{\mathsf{R}}_n}}{\ip*{h^{\mr{rot}}_{-n}}{h_n^{\rm rot}}} \\
     &= T\frac{\ip*{h^{\mr{rot}}_{-n}}{\bar{\Delta}^{(0)}_{\mr{grav}} \delta\mathsf{h}^{\infty}_n}}{\ip*{h^{\mr{rot}}_{-n}}{h_n^{\rm rot}}}  =  -\r{\abs{n}\pi }{2r_0} T~,
\end{aligned}
\end{equation}
which agrees with \eqref{eq:xyz123}, that is $ \delta\lambda_n^{\mathsf{R}}  = \delta\lambda_n^{\infty} $. The appearance of $\delta\mathsf{h}^{\infty}$ was something we suspected from \eqref{eq:grow-rotation}: we couldn't satisfy the gauge conditions while keeping modes normalizable. Although $\delta\mathsf{h}^{\infty}$ is not pure gauge, we consider the analysis surrounding  \eqref{eq:grow-rotation} indicative of the mismatch between the near-horizon and BTZ analysis. 
We conclude that the near-horizon analysis misses contributions from modes that appear non-normalisable within the NHG but become normalisable in the full geometry. Consequently, the decoupling limit fails for the $\mathsf{R}$-modes, and the full BTZ geometry is required to obtain the correct eigenvalue correction.

Finally, it is interesting to remark that the analysis in this section confirms our findings in Sec.\,\ref{sec:kerr-cft-no-go}. $\mathsf{T}$ and $\mathsf{R}$ are the only modes whose eigenvalues tend to zero as $T\to0$ and comply with appropriate regularity conditions. This is compatible with the statement that there are no other large diffeomorphisms present in the NHG that contribute to the GPI.

\paragraph{Ghosts.} We now analyse the spectrum of the ghost determinant appearing in 
\eqref{eq:eh-partition-function-reduced} to assess if there are contributions to $Z_{\rm low}[T]$ coming from that sector. The goal is to identify transverse vector modes whose eigenvalues under the quadratic operator
\begin{equation}
    \left(\bar\Box_1 - \frac{2}{\ell^2}\right)~,
\end{equation}
vanish in the extremal limit. Similar to the treatment of the graviton, we solve for 
\begin{equation}\label{eq:ads3-spin1-eqmaintext}
    (\ell^2\bar\Box -2)\mathsf{a}_{\mu} = ( \lambda^2 - 4)\mathsf{a}_{\mu}~,\quad
    \bar\nabla^\mu \mathsf{a}_\mu = 0~,
\end{equation}
in the BTZ background by rewriting the second-order
equation as the equivalent first-order equation
\eqref{eq:ads3-spin1-firsteq}. The details are in App.\,\ref{sec:gaugemodes}. Our analysis shows that no normalisable vector modes exist such that $\lambda^2 = 4$ at $T=0$. Consequently, in agreement with the near-horizon analysis, we conclude that the ghost determinant does
not contain any modes whose eigenvalue approaches zero as we lower the temperature.

\subsection{Gauge modes}\label{sec:fullbtz-gauge}

In this portion, we analyse the GPI of the Chern-Simons fields $(A, A')$, with particular emphasis on the modes contributing to $Z_{\rm low}[T]$. A useful reference that guides this analysis is \cite{Porrati:2019knx}. We will denote this set of eigenmodes as $\mathsf{G}$-modes. As argued in Sec.\,\ref{sec:gauge-oneloop}, after the gauge-fixing procedure, the one-loop determinant for the Chern-Simons gauge field is expressed in terms of the operator \eqref{eq:cs-partition-function},
\begin{equation}\label{eq:btz-gauge-operators}
    (\bars{\star}\bar{D}\mathsf{a})_\mu = \r{1}{2}\epsilon_\mu{}^{\nu\lambda}(\bar{D}_\nu \mathsf{a}_\lambda)~,
\end{equation}
acting on Lie algebra-valued one-forms satisfying the gauge condition $\bar{D}^\mu a_\mu=0$. 

Let us first consider the $\mathcal{N}= (2,2)$ case, where the gauge group is $U(1) \times U(1)$, and the operator simply becomes $\bars{\star}\bar{\nabla}$. The details of the spectrum of this operator are App.\,\ref{sec:gaugemodes}. The $\mathsf{G}$-modes are given by
\begin{equation}\label{eq:fullbtz-gauge-modes}
  \mathsf{a}^{\mathsf{G}}_n = N^{\mathsf{G}}_{n}e^{\r{2\pi in}{\beta}t_E}
  \f(\r{r^2-r_+^2}{r^2-r_-^2})^{\r{\abs{n}}{2}}
  \f(\r{r_+^2-r_-^2}{r^2-r_-^2})^{-\r{r_+-r_-}{r_+}\r{\abs{n}}{2}}
  \f[\r{r_++r_-}{r f(r)}\dd{r} - \sgn(n)(\ell\dd{\phi}-i\dd{t_E})]~,
\end{equation}
where $n$ is an integer and $N^{\mathsf{G}}_{n}$ is an overall normalisation, and comply with \eqref{eq:bc-fields}. Furthermore, we have 
\begin{equation}\label{eq:fullbtz-gauge-eigenval}
  \bar\epsilon_\mu{}^{\nu\rho}\bar\nabla_\nu (\mathsf{a}_n^{\mathsf{G}})_\rho = \r{i\abs{n}T}{\ell T_R}(\mathsf{a}_n^{\mathsf{G}})_\mu~.
\end{equation}
Hence, these eigenmodes have a vanishing eigenvalue at extremality.\footnote{Notice that the eigenvalue in \eqref{eq:fullbtz-gauge-eigenval} is imaginary, not real. We attribute this to the framing anomaly of Chern-Simons, and it should be compared to the fact that rotational modes have a negative eigenvalue.} It is important to stress that there are two copies of these eigenmodes: one for the path integral of $A$ and a separate twin tower of modes for $A'$. Following the same outline as for gravitational fluctuations, the next step is to determine when the modes are normalizable and inspect the boundary conditions they satisfy. The modes are normalisable for
\begin{equation}\label{eq:norm-g-mode}
  1 \leqslant \abs{n} < \r{T_R}{T}~.
\end{equation}
Under this range of $n$, we set the norm of the eigenmodes to be
\begin{equation}
  \ip{\mathsf{a}^{\mathsf{G}}_{-m}}{\mathsf{a}^{\mathsf{G}}_n} = \ell \delta_{nm}~,
\end{equation}
with the normalisation constant 
\begin{equation}
 N^{\mathsf{G}}_{n}= \sqrt{\r{(r_+-r_-)^2\abs{n}\Gamma(1+\abs{n}\r{r_-}{r_+})}{4\pi^2\ell^3 r_+ \Gamma(1+\abs{n})\Gamma(1-\abs{n}(1-\r{r_-}{r_+}))}}~.
\end{equation}
Near the asymptotic boundary of BTZ, these eigenmodes have the following radial falloffs, 
\begin{equation}\label{eq:fullbtz-gauge-large r}
  (\mathsf{a}^{\mathsf{G}}_n)_{\bar w} \sim r^{|n|\frac{r_+-r_-}{r_+}}, \qquad (\mathsf{a}^{\mathsf{G}}_n)_{r} \sim r^{|n|\frac{r_+-r_-}{r_+}-3}, \qquad (\mathsf{a}^{\mathsf{G}}_n)_{w} = 0 ~,
\end{equation}
where $w, \bar{w}$ are defined in \eqref{eq:wwbar-euc}. It is interesting to contrast this behaviour against \eqref{eq:aw-potential}-\eqref{eq:aphi-charge}, where we set up the notion of sources and charges for the gauge fields.   It is at least satisfying to see that for the modes in the path integral for $A$, these modes will be compatible in a grand canonical ensemble where the source is held fixed. One aspect that is still strange, but aligns with the gravitational fluctuations, is that the scaling in $r$ of the mode along the components that capture charges, is stronger than their classical counterpart; classically, we would expect $(\mathsf{a}^{\mathsf{G}}_n)_{\bar w} \sim O(r^0)$ when the charge is supported by $A$.  In contrast, for the path integral for $A'$, these eigenmodes clearly tamper with the source, which aligns now along the $\bar w$ component of the connection. Therefore, in the grand canonical ensemble, the modes in the $A'$ sector do not contribute to the GPI.  If we worked in the canonical ensemble, where $(Q,Q')$ are fixed, the roles would be reversed: the $\mathsf{G}$-modes in the $A$ sector would not contribute to the GPI, while they would appear in $A'$.

\paragraph{Non-abelian case.}
The generalisation to the non-abelian case is analogous to that performed in Sec.\,\ref{sec:nhg-gauge}, where the effect we need to take into account is that the differential operator \eqref{eq:btz-gauge-operators} depends on the background gauge field $\bar{A}_\mu$, and respectively for the primed sector. Therefore, it is important to take into account the behaviour of the potential (and charges) of the black hole as we take a low-temperature limit.

Similar to the analysis in the NHG in \eqref{eq:zero-non-A}, the candidates for our modes are of the form
\begin{equation}\label{eq:btz-nonabelian-modes}
\mathsf{a}^{\mathsf{G}(I)}_\mu  =  g^{-1} \mathsf{a}_\mu\, T_I \, g~,
\end{equation}
where $T_I$ are the generators of the Lie algebra $\mathfrak{g}$ and the holonomy $g(x)$ is given in \eqref{eq:btz-holonomies}.
The components  $\mathsf{a}_\mu$ are eigenmodes of the operator $\bar{\star}\bar{\nabla}$. Note that imposing the periodicity conditions \eqref{eq:bc-fields} on $\mathsf{a}^{\mathsf{G}(I)}_\mu $ leads to twisted boundary conditions of $\mathsf{a}_\mu$ along the spatial direction. 

To make the construction of $\mathsf{a}^{\mathsf{G}}_\mu $ more explicit, we will focus on the case of $SU(2)$. For this choice of gauge group the background connection is given by \eqref{eq:gauge-fields-bh}.
The three conditions on the set of modes we require are: first, normalisability under the inner product
\begin{equation}
  \ip*{\mathsf{a}}{\tilde{\mathsf{a}}} = \int \dd[3]{x}\sqrt{\bar{g}} \Tr(\mathsf{a}^\mu\tilde{\mathsf{a}}_\mu)~,
\end{equation}
second, the corresponding eigenvalue goes to zero in the near-extremal limit,
and third, the periodicity under both the spatial and thermal identifications.
One solution to these three conditions is
\begin{equation}\label{eq:gauge-nonab-zeromodes3}
    \mathsf{a}^{\mathsf{G}(3)}_{n} = g^{-1} \mathsf{a}_{\lambda_n n 0}\, T_3\, g~, \quad  \lambda_n = \abs{n}\frac{T}{T_R}~,\quad n =\pm 1, \pm2, \ldots~,
\end{equation}
with $\mathsf{a}^{(3)}_{\lambda_n n 0}$ given by \eqref{eq:sol-gauge-gen}, which is the same profile that leads to \eqref{eq:fullbtz-gauge-modes}, and
\begin{equation}
\bar\epsilon_\mu{}^{\nu\rho}\bar D_\nu (\mathsf{a}_n^{\mathsf{G}})_\rho = \r{i\abs{n}T}{\ell T_R}(\mathsf{a}_n^{\mathsf{G}})_\mu~.
\end{equation}
Another set that complies with periodicity and normalizability consists of  
\begin{equation}
\begin{split}
   \mathsf{a}^{(+)}_{nm} &= g^{-1}\mathsf{a}_{\tilde{\lambda}_{nm} n (m - GQ)}T_+ \, g~,\quad n =\pm 1, \pm2, \ldots~, \\
    \mathsf{a}^{(-)}_{nm} &= g^{-1} \mathsf{a}_{\tilde{\lambda}_{nm} (-n) (-m + GQ)}T_-\, g~,\quad n =\pm 1, \pm2, \ldots~,
\end{split}
\end{equation}
where $m\in\Z$, 
\begin{equation}\label{eq:gauged-spin1-lambdan}
  \tilde{\lambda}_{nm}  = \abs{n}\frac{T}{T_R} + \frac{i\ell\sgn(n)(m-GQ)}{r_+}~,
\end{equation}
and $ T^\pm = \frac{1}{\sqrt{2}}(T^1 \pm iT^2)$. However, as they stand, the eigenvalue in \eqref{eq:gauged-spin1-lambdan} is non-zero for $T=0$, unless we impose an extra condition: the near-BPS condition. That is, if we take $GQ\in \Z$, which is equivalent to \eqref{eq:near-BPS}, we can set $m=GQ$, and obtain a set of modes that comply with all three conditions:
\begin{equation}\label{eq:gauged-spin1-lambdan-1}
   \mathsf{a}^{{\mathsf{G}}(\pm)}_{n} = \mathsf{a}^{(\pm)}_{n(GQ)}~.
\end{equation}
Hence, there are three towers of zero modes in the near-BPS limit, and only one in the near-extremal non-BPS limit. In the primed sector, where the background connection is $\bar A'$ in \eqref{eq:gauge-fields-bh}, we will only have one set of zero modes.

\paragraph{Comparison with NHG.} We will focus on the $U(1)\times U(1)$ case; still, the conclusion extends to the non-Abelian case. In Sec.\,\ref{sec:nhg-gauge}, we found that the leading temperature correction to the gauge eigenvalue vanishes, both in the canonical and grand canonical ensemble. This contradicts \eqref{eq:fullbtz-gauge-eigenval}, where the gauge modes possess a non-zero eigenvalue under \( \bar\star \bar\nabla \) if we do the analysis on BTZ at finite temperature and then take $T\to 0$. Concretely, we find that
\begin{equation}
     \mathsf{\lambda}_n^{\mathsf{G}} =  0 +  \delta\lambda_n^{\mathsf{G}} + \mc{O}(T^2)~,
\end{equation}
with
\begin{equation}\label{eq:corrected-G-mode}
     \delta\lambda_n^{\mathsf{G}} =   \frac{i\ell\pi |n|}{r_0}\, T~.
\end{equation}
In \eqref{gaugeev-nhg} we had $\delta\lambda_n^{\rm gauge}=0$.
This tension is accounted for by a more careful analysis of the $\mathsf{G}$-modes. Using the explicit form of the modes \eqref{eq:fullbtz-gauge-modes}, we expand them at small temperature, which gives up to an overall normalisation
\begin{equation}
\mathsf{a}^{\mathsf{G}}_n = a^{\mathrm{gauge}}_n + T\,\delta \mathsf{a}^{\mathsf{G}}_n + \mathcal{O}(T^2)\, .
\end{equation}
Here, $a^{\mathrm{gauge}}_n$ is given by \eqref{eq:a-gauge-nhg}-\eqref{eq:scalar-gauge} and $\delta a^{\mathsf{G}}_n$ is the first order correction. This correction grows toward the boundary as 
\begin{equation}
(\delta \mathsf{a}^{\mathsf{G}}_n)_{\hat t_E}=\mathcal{O}(\eta)~, \qquad
(\delta \mathsf{a}^{\mathsf{G}}_n)_{\eta}=\mathcal{O}(\eta e^{-\eta})~, \qquad
(\delta \mathsf{a}^{\mathsf{G}}_n)_{\hat\phi}=\mathcal{O}(1)\, . 
\end{equation}
Consequently, these corrections are non-normalisable in the ENHG. Therefore, just as for the $\mathsf{R}$-modes, when thinking of the corrected eigenvalue as a perturbation, the second term of \eqref{eq:lambda2} cannot be neglected. Evaluating the non-normalizable contribution to the eigenvalue gives
\begin{equation}
\frac{\left\langle a^{\mathrm{gauge}}_{-n}\,|\,(\bar\star\bar \nabla)^{(0)}\, T \delta \mathsf{a}^{\mathsf{G}}_n \right\rangle}{\langle a^{\mathrm{gauge}}_{-n}| a^{\mathrm{gauge}}_{n}\rangle}
= \frac{i\ell\pi |n|}{r_0}\, T .
\end{equation}
This exactly captures \eqref{eq:corrected-G-mode}. Unfortunately, starting from the ENHG, there is no systematic way to account for a non-normalizable mode such as $\delta \mathsf{a}^{\mathsf{G}}_n$. This effect is similar to our findings for rotational modes in \eqref{eq:rot-far-correction}.  We see that there is an impediment in capturing the GPI of the black hole from the near-horizon perspective.

\paragraph{Ghosts.} The quadratic operator relevant for the ghost one-loop determinant is 
\begin{equation}
 -(\bar{D}_\mu \bar{D}^\mu)_0~,  
\end{equation}
which reduces to the scalar Laplacian ($\bar\Box_0$) in the $\mathcal{N}=(2,2)$ case, where the gauge group is $U(1)\times U(1)$. We analyse the spectrum of the scalar Laplacian in App.\,\ref{sec:scalarmodes} and find that there are no normalisable scalar modes with eigenvalues approaching zero near extremality.
The eigenmodes in the non-abelian case take the form $g^{-1}\phi g$ where $\phi$ is an eigenmode of the scalar Laplacian $\Box_0$. Since there are no normalisable eigenmodes for the scalar Laplacian, there will be no normalisable eigenmodes for the ghost operator in the non-Abelian case as well.

\subsection{Fermionic modes}\label{sec:fermi-far}

Finally, we turn to the fermionic contributions to $Z_{\rm low}[T]$. These arise from the spin-3/2 fields $\psi$ and $\psi'$ in our supergravity theory.  The gauge-fixed one-loop fermionic partition function \eqref{eq:ferm-one-loop-N=1} is expressed in terms of determinants of operators
\begin{equation}
  (\bars{\star}\bar{\mc{D}}\bm{\psi})_\mu = \r{1}{2}\bar{\epsilon}_\mu{}^{\nu\lambda}\bar{\mc{D}}_\nu \bm{\psi}_\lambda~,
\end{equation}
acting on spin-$\r{3}{2}$ gravitino fields $\bm{\psi}$ satisfying the gauge condition $\gamma^a \bm{\psi}_a = 0$. The ghost contribution is describe by a spin-$\r{1}{2}$ Dirac ghost fields $\bm{\chi}$ with differential operator
\begin{equation}
 \gamma^a\bar{\mc{D}}_a \bm{\chi}~.
\end{equation}
 Similar operators hold for the primed sector.

We will first focus on the case $\mc{N} = (1,1)$, the simplest supergravity under consideration. In this case, the ghost operators for the unprimed and primed sectors are given by
\begin{equation}
  \gamma^a\bar{\mc{D}}_a \bm{\chi}= \left(\gamma^a\bar D_a -\frac{3}{2\ell}\right)\bm{\chi}, \qquad   \gamma^a\bar{\mc{D}'}_a \bm{\chi}'= \left(\gamma^a\bar D_a +\frac{3}{2\ell}\right)\bm{\chi}'~,
\end{equation}
where $D_a$ is the covariant derivative defined in \eqref{eq:def-covariant-1} with the gauge field set to zero.  In App.\,\ref{spin12-btz-app} we show that the Dirac operator $\slashed{D} \coloneqq \gamma^aD_a$ does not have modes with eigenvalues approaching $\pm\r{3}{2\ell}$ near extremality, and therefore the ghost operators do not have modes with eigenvalues approaching zero in that limit.
On the other hand, the gravitino operator $\bars{\star}\bar{\mc{D}}$ has two families of modes $\bm{\psi}^+_k$ and $\bm{\psi}^-_k$ described in \eqref{eq:ads3-psi-zm-plus} and \eqref{eq:ads3-psi-zm-minus}, with
\begin{equation}\label{eq:eigen-btz-spin3/2}
  -i\epsilon_\mu{}^{\nu\lambda}\mc{D}_\nu(\bm{\psi}^{(\pm)}_k)_\lambda = \r{T}{\ell T_R} (k+\r{1}{2}) (\bm{\psi}^{(\pm)}_k)_\mu~,
\end{equation}
which are periodic and normalisable for
\begin{equation}
  1 \leqslant k < \frac{1}{2}\left(\r{T_R}{T} - 1\right)~.
\end{equation}
If we take the zero temperature limit, the eigenvalue behaves as
\begin{equation}
     \mathsf{\lambda}_n^{\bm{\psi}} =  0 +  \delta\lambda_n^{\bm{\psi}} + \mc{O}(T^2)~,
\end{equation}
with
\begin{equation}\label{eq:corrected-psi-mode}
     \delta\lambda_n^{\bm{\psi}} =  (k+\r{1}{2}) \r{\pi \ell}{r_0} T ~.
\end{equation}
This agrees with the corrections to the eigenvalues calculated in the NHG \eqref{eq:nhg-psi-eigenval}, and hence leads to the same coefficient of the $\log T$ correction to entropy. For the eigenmodes, we find
\begin{equation}
\bm{\psi}^{(\pm)}_k = {\psi}^{(\pm)}_k + T\,\delta \bm{\psi}^{\pm}_k + \mathcal{O}(T^2)~,
\end{equation}
where ${\psi}^{(\pm)}_k $ is given by \eqref{eq:ferm-zero-nh} and $\delta \bm{\psi}^{\pm}_k$ is normalizable in the ENHG.  
Similar analysis for the operator $\bar{\star}\bar{\mc{D}}'$ shows that the path integral of $\psi'$ does not have analogous modes, and hence does not contribute to the $\log T$ corrections. 

\paragraph{Charged black holes.}
Similarly to the case of the gauge modes in Sec.\,\ref{sec:fullbtz-gauge}, when considering one-loop corrections around saddles with a non-zero charge, we have to consider its impact on the differential operators whose modes we are looking for. In the case of fermions, the operators change both in the Abelian and non-Abelian cases. 
The construction of the eventually-zero fermionic modes goes as follows:
\begin{itemize}
\item \textbf{$U(1)$ gauge group:} We are looking for gravitino modes which are normalisable under the Euclidean spinor inner product
\begin{equation}
  \ip*{\bm{\psi}}{\tilde{\bm{\psi}}} = \int \dd[3]{x}\sqrt{\bar{g}}\, \bm{\psi}^\mu\tilde{\bm{\psi}}_\mu~,
\end{equation}
and satisfy
\begin{equation}
\bar{\epsilon}_\mu{}^{\nu\lambda}\bar{\mc{D}}_{\nu}\bm{\psi}_\lambda = \bar{\epsilon}_\mu{}^{\nu\lambda}\bar{D}_{\nu}\bm{\psi}_\lambda + \frac{1}{2\ell}\bm{\psi}_\mu = \frac{i}{\ell}\f(\lambda + \frac{1}{2})\bm{\psi}_\mu~,
\end{equation}
with the eigenvalue $\lambda$ going to $-\frac{1}{2}$ in the near-extremal limit. One can check that the space of solutions to these two conditions is spanned by
\begin{equation}
  \bm{\psi}^{\scaleto{\rm U(1)}{4pt},\pm}_{k} = g(x)^{-1} \bm{\psi}^{\pm}_{(\lambda^{\scaleto{\rm U(1)}{4pt}}_\pm)\,k\,\mp(m-GQ)}~,\quad
  \lambda^{\scaleto{\rm U(1)}{4pt}}_\pm = -\frac{1}{2} + (k+\frac{1}{2})\frac{T}{T_R} \pm \frac{i\ell(m-GQ)}{r_+}~,
\end{equation}
where $g(x)$ are the elements of the Lie group \eqref{eq:btz-holonomies} acting in the fundamental $U(1)$ representation and $\bm{\psi}^\pm_{\lambda km}$ are the eignmodes of the uncharged version of the operator $\bar{\star}\bar{D}$ derived in App.\,\ref{spin32-btz-app}. 
Since the total holonomy of $\bar{A}$ around the spatial circle is $iGQ$, the condition for $\bm{\psi}^{\scaleto{\rm U(1)}{4pt},\pm}_{k}$ to be periodic or anti-periodic is $m \in \Z$ or  $m \in \Z+\frac{1}{2}$, respectively. The modes will have the property of $\lambda\to-\frac{1}{2}$ as $T\to 0$ provided we also set $m=GQ$. This is only consistent when the black hole supports a Killing spinor, i.e., $GQ$ complies with \eqref{sec:bhs}. Therefore, this class of modes are only supported in the near-BPS limit. 

\item \textbf{$SU(2)$ gauge group:} Since the background gauge field \eqref{eq:gauge-fields-bh} is supported only along $T_3$, the action of the charged differential operators acting on spinors will be diagonalised by modes proportional to the two eigenvectors of $T_3$, namely
  \begin{equation}
    (\varepsilon_{1})^i \coloneqq \m{1\\0}~,\quad (\varepsilon_{2})^i \coloneqq \m{0 \\ 1}~.
  \end{equation}
  Thus we have two families of eventually-zero modes
  \begin{equation}
    \begin{split}
      (\bm{\psi}^{\scaleto{\rm SU(2)}{4pt},\pm}_{k,1})^i_\mu &= (g^{-1})^i{}_{j} \varepsilon^j_1 \bm{\psi}^\pm_{(\lambda^{\scaleto{\rm SU(2)}{4pt}}_\pm)\, k\,\mp(m - \frac{GQ}{2})} ~,\\
      (\bm{\psi}^{\scaleto{\rm SU(2)}{4pt},\pm}_{k,2})^i_\mu &= (g^{-1})^i{}_{j} \varepsilon^j_2 \bm{\psi}^\pm_{(\lambda^{\scaleto{\rm SU(2)}{4pt}}_\mp)\, k\,\pm(m - \frac{GQ}{2})}~,
    \end{split}
  \end{equation}
  where
  \begin{equation}
    \lambda^{\scaleto{\rm SU(2)}{4pt}}_\pm = -\frac{1}{2} + (k+\frac{1}{2})\frac{T}{T_R} \pm \frac{i\ell(m-\frac{GQ}{2})}{r_+}~.
  \end{equation}
Here $m$ is allowed to be either an integer or a half-integer, depending on whether the fermion is periodic or antiperiodic along the spatial cycle. Analogous to the $U(1)$ case above, modes will have $\lambda\to-\frac{1}{2}$ at zero temperature provided $GQ$ complies with \eqref{eq:susy-charge-SU(2)} and $m=GQ/2$. Hence, the modes are only present in the near-BPS limit, and there is a total of four of them.  These modes are not present in the primed sector. 
\end{itemize}

\subsection{\texorpdfstring{Log-$T$}{Log-T} corrections to the GPI} \label{sec:final-GPIfar}

To conclude this section, we will piece together all the contributions to $Z_{\rm low}[T]$ in the grand canonical ensemble. There are two axes to take into consideration in this discussion. We first have to take into account the boundary condition imposed on the black hole: either Dirichlet boundary conditions, compatible with AdS$_3$/CFT$_2$, or a looser condition, such as the CSS boundary conditions that are more compatible with the dual being a warped CFT$_2$. The second consideration regards the properties of the black hole background: whether the extremal black hole is supersymmetric or non-supersymmetric. 

Similar to the discussion in Sec.\,\ref{sec:final-GPInear}, the behaviour of the low temperature GPI of BTZ will be of the form
\begin{equation}
\begin{aligned}\label{eq:ZlowT}
    \log  Z_{\rm BTZ}[T] &\approx \log Z_{\rm low}[T] + \log Z_0[T]+\cdots\\
    & \approx \log \prod_n \left(\delta\Lambda_n\right)^{\pm 1/2} + \log Z_0[T]+ \cdots~,
\end{aligned}
\end{equation}
where the contributions in this product are the eigenmodes with vanishing eigenvalue at $T=0$, and the $\pm$ in the exponent is a reminder that some modes are Grassmann (anti-commuting) versus bosonic (commuting). As in Sec.\,\ref{sec:final-GPInear}, the infinite products will be regulated using zeta function regularisation, where the relevant identities are \eqref{eq:zeta-1} and \eqref{eq:zeta-2}. Collecting all the species of fields we had, we write
\begin{equation}\label{eq:Zlow-final}
    \log Z_{\rm low}[T] =\left(a_{\mathsf{T}}+ a_{\mathsf{R}}+ a_{\mathsf{G}}+a_{\mathsf{G}'} + a_{\bm \psi}+a_{\bm \psi'} \right)\log T~,
\end{equation}
where $a_{\mathsf{T}}$ and $a_{\mathsf{R}}$ are the contributions from the gravitational sector in Sec.\,\ref{sec:grav-far}; $a_{\mathsf{G}}$ is the contribution from the Chern-Simons sector in Sec.\,\ref{sec:fullbtz-gauge}, and $a_{\bm \psi}$ the contribution from the gravitini fields in Sec.\ref{sec:fermi-far}, which we divided intro unprimed and primed sector. We consistently have $a_{\bm \psi'}=0$ for all near-extremal black holes due to the chirality of the spinors, and $a_{\mathsf{G}'}=0$ due to our choice of boundary conditions. The values we found are summarised in Table\,\ref{tab:results-far}. 

\renewcommand{\arraystretch}{1.3}
\begin{table}[!ht]
    \centering
    \begin{tabular}{|c|ccc|ccc|c|c|}
    \hline
         & \multicolumn{3}{c|}{near-BPS} & \multicolumn{3}{c|}{near-ext}  & {\centering near-ext} & \multirow{3}{*}{$\log Z_{\rm z.m}[T]$} \\
\cline{2-8}
 &       \multicolumn{3}{c|}{DBC} & \multicolumn{3}{c|}{DBC}  & {\centering MBC}  &\\ 
\cline{1-8}
        ${\cal N}$ &\multicolumn{1}{c|}{$(1,1)$}  & \multicolumn{1}{c|}{$(2,2)$} & $(4,4)$ & \multicolumn{1}{c|}{$(0,0)$} & \multicolumn{1}{c|}{$(2,2)$} & $(4,4)$  & $(0,0)$ & \\
        \hline
        \hline
        $a_{\mathsf{T}}$ & \multicolumn{1}{c|}{$\frac{3}{2}$}  & \multicolumn{1}{c|}{$\frac{3}{2}$} & $\frac{3}{2}$ & \multicolumn{1}{c|}{$\frac{3}{2}$} & \multicolumn{1}{c|}{$\frac{3}{2}$} & $\frac{3}{2}$ & $\frac{3}{2}$ & \checkmark\\
        \hline
        $a_{\mathsf{R}}$& \multicolumn{1}{c|}{$0$}  & \multicolumn{1}{c|}{$0$} & $0$ & \multicolumn{1}{c|}{$0$} & \multicolumn{1}{c|}{$0$} & $0$ & $\frac{1}{2}$ & \ding{55}\\
        \hline
        $a_{\mathsf{G}}$ & \multicolumn{1}{c|}{$0$}  & \multicolumn{1}{c|}{$\frac{1}{2}$}  & $\frac{3}{2}$ & \multicolumn{1}{c|}{$0$}  & \multicolumn{1}{c|}{$\frac{1}{2}$}  & $\frac{1}{2}$ & \multicolumn{1}{c|}{$0$} & \ding{55} \\
        \hline 
        $a_{\bm\psi}$ & \multicolumn{1}{c|}{$-1$} & \multicolumn{1}{c|}{$-2$} & $-4$ & \multicolumn{1}{c|}{$0$} & \multicolumn{1}{c|}{$0$} & 0 & 0 & \checkmark\\
        \hline
    \end{tabular}
    \caption{Contributions to $\log Z_{\rm low}[T]$ for near-BPS and near-extremal black holes. We have included the different classes of supersymmetric theories and different boundary conditions. In the second row, ``DBC'' denotes Dirichlet boundary conditions, which should be contrasted with the dual CFT$_2$; ``MBC'' denotes modified boundary conditions, which can be more relaxed, such as CSS boundary conditions. In the last column, we compare our results against the counterpart analysis from the ENHG in Sec.\,\ref{sec:final-GPInear}, where \checkmark means the answers agree, and \ding{55} means there is a disagreement. }
    \label{tab:results-far}
\end{table}

In comparison to previous studies of $Z_{\rm BTZ}[T]$, without supersymmetry and with DBC, such as \cite{KapLaw24,KolMar24,Acito:2025hka,GhoMax19}, our results here agree. Our result in the grand canonical ensemble for the near-extremal black hole with MBC is compatible with the grand canonical partition function of a WCFT reported in \cite{Aggarwal:2022xfd}. In the case of ${\cal N}=(4,4)$ theory and near-BPS black holes, the result in our Table\,\ref{tab:results-far} agrees with \cite{HeyIli20}, where both expressions in the grand canonical ensemble have a term $-\log T$ when $\mu\to 0$ and $2\pi i\alpha = \beta \mu$ is fixed.  The case of ${\cal N}=(1,1)$ has not been discussed explicitly for BTZ before in this ensemble. However, this case can be compared with the results in \cite{Stanford:2017thb} for the super-Schwarzian theory, where we find agreement again in the near-BPS black hole. 

\section{Discussion}\label{sec:discussion}

In this work, we have revisited the low-temperature behaviour of the gravitational path integral of black holes in AdS$_3$ supergravity, with particular emphasis on the relation between near-horizon and asymptotic analyses. Our results highlight a number of conceptual and technical subtleties that refine the standard expectations regarding how the gravitational path integral behaves in the far and near regions.

A central outcome of our analysis is the failure of a naive decoupling between the near-horizon region and the full asymptotically AdS$_3$ geometry at the quantum level, which we summarised in Table\,\ref{tab:results-far}. While the classical contribution to the path integral agrees in both descriptions, we have shown that the one-loop contributions can differ significantly. In particular, certain fluctuations that are non-normalisable in the Euclidean near-horizon geometry become normalisable when embedded in the  BTZ background. This mismatch implies that the near-horizon computation is not, in general, sufficient to capture the full set of quantum effects governing the low-temperature regime. From a broader perspective, this suggests that similar limitations may arise in higher-dimensional rotating black holes, where near-horizon analyses are often assumed to encode the relevant infrared physics. 

Our analysis includes both the BPS and non-BPS black holes. It is expected that their loop corrections have different low-temperature behaviours, and we are in compliance with that expectation. For example, near-BPS black holes have contributions from fermionic modes which are absent in the generic near-extremal setup. It is worth highlighting that the differences we observe between $Z_{\rm z.m}[T]$ and $Z_{\rm low}[T]$ are neither attributed to nor fixed by supersymmetry.

The problematic modes are the rotational and gauge modes in the NHG, which do not match the $\mathsf{R}/\mathsf{G}$-modes in the BTZ background. A peculiar property of the $\mathsf{R}$- and $\mathsf{G}$-modes is that their eigenvalues are negative and imaginary, respectively, which can raise concerns about the convergence of the path integral. One property is that these modes are complex yet satisfy the KS criteria \cite{Kontsevich:2021dmb}; hence, they should contribute to the GPI upon appropriate contour rotation. For the $\mathsf{G}$-modes, we can attest that they match the dual SCFT$_2$ results and hence the deformation of the contour seems appropriate. For $\mathsf{R}$-modes, it would be interesting to understand this effect since it is present in higher-dimensional black holes as well \cite{blacker_quantum_2025}. The analysis of the forthcoming work in \cite{DDMT} should also serve as a lamppost in resolving these issues. 

A separate question we addressed is the role of asymptotic symmetries in the near-horizon region within the GPI, in particular those appearing in Kerr/CFT-inspired analyses. As we have discussed, while such symmetries can be defined at the level of the classical phase space, they do not, in general, give rise to additional contributions to the gravitational path integral under the boundary conditions considered here. This reinforces the idea that the presence of an asymptotic symmetry algebra does not automatically imply the existence of corresponding quantum degrees of freedom contributing to the partition function. It would be interesting to investigate this further for other near-horizon geometries: is there a background for which the asymptotic symmetry group advocated in Kerr/CFT contributes to the GPI?

Overall, our results point towards a more nuanced picture of infrared dynamics in AdS$_3$ gravity, in which near-horizon data alone is insufficient to characterise the quantum theory. Instead, a consistent treatment requires combining the near-horizon analysis with the global structure of the spacetime and a precise specification of boundary conditions. Extending this perspective to other settings—such as higher-dimensional rotating black holes or theories with additional fields—may provide further insight into the role of infrared regions in gravitational path integrals.  One approach along this line would be the work in \cite{ArnBon24}; it would also be interesting to connect with the infrared limits studied in \cite{CasMan25}.

\section*{Acknowledgments}
We thank Matt Blacker, Bob Knighton, Maciej Kolanowski,  James Lucietti, Robinson Mancilla, and Joaquin Turiaci for helpful discussions. In particular, we thank Nabamita Banerjee and Muktajyoti Saha for pointing out a key correction to the rotational modes in our first version. We also thank the organisers and participants of the 2026 CERN Winter School on Supergravity, Strings and Gauge Theory, where portions of these results were presented. This work has been partially supported by STFC consolidated grant ST/X000664/1. AB is supported by the Cambridge Trust International Scholarship. DJ is supported by the Simons Collaboration for Celestial Holography. 

\appendix

\section{Conventions}\label{app:conventions}
In this Appendix, we list some of the conventions used in the main text.   We first list a few identities involving spinors in three dimensions in App.\,\ref{app:spinors}, and in App.\,\ref{app:ortho-basis} we list our choices of orthonormal basis for the near-horizon and black hole backgrounds.

\subsection{Spinors}\label{app:spinors}
The spinor degrees of freedom are Grassmann variables, and they satisfy
\begin{equation}
   \quad \bar{\psi}\chi = \bar{\chi}\psi~.
\end{equation}
Also, using the identity $\gamma^0\gamma^a\gamma^0 = (\gamma^a)^T$, for gamma matrices, we have
\begin{equation}\label{gamma-identities}
    \ol{\gamma^a \psi} \coloneq (\gamma^a \psi)^T \gamma^0 = -\bar{\psi}\gamma^a~.
\end{equation}
The derivative operators $\mc{D}$ and $\mc{D}'$ satisfy 
\begin{equation}
\begin{split}
    \dd{\f(\bar{\psi}\wedge \chi)} &= \ol{\mc{D}\psi}\wedge\chi + (-1)^k \bar{\psi}\wedge \mc{D}\chi~,\\
    \dd{\f(\bar{\psi}\wedge \chi)} &= \ol{\mc{D}^{'}\psi}\wedge\chi + (-1)^k \bar{\psi}\wedge \mc{D}^{'}\chi~,
\end{split}
\end{equation}
where $\dd$ is the exterior derivative.
Here we used (\ref{gamma-identities}) and the antisymmetry of the gauge field, when represented as a matrix in the fundamental representation. Moreover,
\begin{equation}
\begin{split}
\mc{D}^{2} &= \r{1}{4}\mc{R}^{ab}\gamma_{ab} + F + \r{1}{8\ell}\f(\bar{\psi}\wedge\gamma^a\psi+\bar{\psi}'\wedge\gamma^a\psi')\gamma_a + \r{1}{4\ell^2}e^a\wedge e^b \gamma_{ab}~,\\
\mc{D}^{'2} &= \r{1}{4}\mc{R}^{ab}\gamma_{ab} + F' - \r{1}{8\ell}\f(\bar{\psi}\wedge\gamma^a\psi+\bar{\psi}'\wedge\gamma^a\psi')\gamma_a + \r{1}{4\ell^2}e^a\wedge e^b \gamma_{ab}~,
\end{split}
\end{equation}
 which in particular implies that both $\mc{D}^{2} = 0$ and $\mc{D}^{'2} = 0$ vanish on bosonic solutions to the classical equations of motion.

 Furthermore, the spin connection is not treated as independent degrees of freedom: its equations of motion are imposed at all times, and they fix its torsion to be 
\begin{equation}\label{torsion}
    \dd{e^a} + \omega^a{}_b\wedge e^b = -\r{1}{4}\f(\bar{\psi}\wedge\gamma^a\psi+\bar{\psi}'\wedge\gamma^a\psi')~.
\end{equation}
In particular, the torsion vanishes for bosonic solutions.

\subsection{Orthonormal basis}\label{app:ortho-basis}
In this portion we collect our conventions for the orthonormal frames $e^{\mu}_a$ and spin connections $\omega^a{}_b$ used throughout the text in Euclidean signature. The basic definitions are
\begin{equation}
    g_{\mu\nu} = e_{\mu}^a\, e_{\nu}^b\, \delta_{ab} ~,\quad 
    \dd{e^a} + \omega^a{}_b\wedge e^b=0~. 
\end{equation}
Our orientation for the Levi-Civita tensor is consistently $\epsilon_{012}=1$ in the orthonormal frame.
\paragraph{Euclidean BTZ metric \eqref{eq:Euc-BTZ}.}
\begin{equation}\label{btzframe}
\begin{aligned}
    &e^0 = \frac{\sqrt{(r^2-r_+^2)(r^2-r_-^2)}}{\ell r}\dd{t_{\mr{E}}}~,\\
    &e^1 = \frac{\ell r}{\sqrt{(r^2-r_+^2)(r^2-r_-^2)}}\dd{r}~,\\
    &e^2 = r\f(\dd{\phi} - \r{r_+r_-}{\ell r^2}\dd{t_{\mr{E}}})~.
\end{aligned}
\end{equation}
In order to go back to Lorentzian signature, we use $(e^0)_{\rm Lor} = -i(e^0)_{\rm Euc}$. The corresponding spin-connections are given by
\begin{equation}
    \begin{aligned}
    &\omega_{01} = \r{r_+r_-}{\ell r}\dd{\phi} - \r{r}{\ell^2}\dd{t}~,\\
    &\omega_{12} = -\r{\sqrt{(r^2-r_+^2)(r^2-r_-^2)}}{\ell r}\dd{\phi}~,\\
    &\omega_{20} = -\r{r_+r_-}{r\sqrt{(r^2-r_+^2)(r^2-r_-^2)}}\dd{r}~.
    \end{aligned}
\end{equation}

\paragraph{Euclidean near-horizon metric \eqref{Enhorizon}.}
\begin{equation}\label{nhframe}
    e^0 = \r{\ell}{2}\sinh\eta\dd{\hat{t}_{\mr{E}}}~,\quad e^1 = \r{\ell}{2}\dd{\eta}~,\quad e^2 = r_0\f(\dd{\hat{\phi}} - \r{i\ell}{r_0}\sinh^2\r{\eta}{2}\dd{\hat{t}_{\mr{E}}})~.
\end{equation}
With respect to this frame, the spin-connections are given by
\begin{equation}\label{nhg-connection}
    \omega_{01} = \cosh^2\r{\eta}{2}\dd{\hat t_E} - \r{ir_0}{\ell}\dd{\hat \phi}~,\quad
    \omega_{12} = \r{i}{2}\sinh\eta \dd{\hat t_E}~,\quad
    \omega_{20} = \r{i}{2}\dd{\eta}~.
\end{equation}
\paragraph{Euclidean near-horizon near-extremal metric \eqref{ebtz-expansion}-\eqref{nhg-correction}.}
\begin{equation}
\begin{aligned}\label{nhnextframe}
    & e^0 = \left(\frac{\ell}{2}  \sinh \eta -\frac{\pi  \ell^3 T \sinh ^4\left(\frac{\eta }{2}\right) (\cosh \eta +2) \text{csch}\eta }{2 r_0}\right) \dd\hat{t}_{\mr{E}}~, \\
    & e^1 = \left(\frac{\ell}{2}+ \frac{\pi  \ell^3 T (\cosh \eta +2) \tanh ^2\left(\frac{\eta }{2}\right)}{8 r_0}\right) \dd\eta~,\\
    & e^2 = \left(r_0+ \frac{1}{2} \pi  \ell^2 T \cosh \eta \right)\dd\hat{\phi} + \left(-\frac{i \ell}{2}(\cosh \eta -1)+ \frac{i \pi  \ell^3 T \sinh ^2\left(\frac{\eta }{2}\right) (\cosh \eta +3)}{4 r_0}\right)\dd\hat{t}_{\mr{E}}~.
\end{aligned}
\end{equation}
The spin-connections are given by
\begin{equation}
    \begin{split}
        &\omega_{01} = \left(\cosh ^2\left(\frac{\eta }{2}\right)-\frac{\pi  \ell^2 T \sinh ^2\left(\frac{\eta }{2}\right) (\cosh \eta +3)}{4 r_0}\right) \dd\hat t_E+ \left(-\frac{i r_0}{\ell}+ \frac{1}{2} i \pi  \ell T \cosh \eta\right) \dd\hat \phi~, \\
        &\omega_{12} = \left(\frac{i}{2}  \sinh \eta -\frac{i \pi  \ell^2 T \left(9 \sinh \left(\frac{3 \eta }{2}\right)+\sinh \left(\frac{5 \eta }{2}\right)\right)\text{sech}\left(\frac{\eta }{2}\right)}{32 r_0}\right)\dd\hat t_E-\pi \ell T \sinh \eta~ \dd\hat\phi ~,\\
    &\omega_{02} = \left(-\frac{i}{2}+\frac{i \pi  \ell^2 T (6 \cosh \eta +3 \cosh (2 \eta )+7)}{16 r_0 (\cosh \eta +1)}\right)\dd\eta~.
    \end{split}
\end{equation}

\section{Eigenmodes in Euclidean BTZ}\label{sec:app-full-btz}

Here we summarise the derivation of the results for the zero modes in the Euclidean BTZ background. Following \cite{Datta:2011za,castro_tweaking_2017}, we express locally AdS$_3$ spaces using the coordinate system
\begin{equation}\label{eq:ads3-glob-coord}
    g = \ell^2(\dd{\xi}^2 + \sinh^2\xi\dd{T_E}^2 + \cosh^2\xi\dd{\Phi}^2)~,
\end{equation}
which is related to the Euclidean BTZ by
\begin{equation}\label{eq:ads3-btz-trans}
    \tanh^2\xi = \r{r^2-r_+^2}{r^2-r_-^2},\quad T_E = \r{1}{\ell^2}(r_+ t_E - ir_-\ell \phi),\quad \Phi = \r{1}{\ell^2}(r_+\ell\phi + ir_- t_E)~.
\end{equation}
We will also, at various points, use the coordinates
\begin{equation}
    z \coloneqq \tanh^2\xi
\end{equation}
and
\begin{equation}\label{eq:ads3-lr-coords}
    x_L = -iT_E+\Phi~,\quad x_R = -iT_E-\Phi~.
\end{equation}
We will work with the following frame and spin connections for the metric \eqref{eq:ads3-glob-coord}:
\begin{equation}\label{eq:ads3-glob-frame}
\begin{gathered}
    e^0 = \ell\sinh\xi \dd{T_E}~,\quad e^1 = \ell\dd{\xi}~,\quad e^2 = \ell\cosh\xi \dd{\Phi}~, \\
    \omega_{01} = \frac{1}{\ell}\coth\xi e^0~,\quad \omega_{12} = -\frac{1}{\ell}\tanh\xi e^2~,\quad \omega_{20} = 0~.
\end{gathered}
\end{equation}

In \cite{vasiliev_lagrangian_1997,Datta:2011za} it was shown that on this background, the  equations for the spin-$s$ symmetric, transverse and traceless massive field with eigenvalue $\lambda^2$,
\begin{equation}\label{eq:ads3-spins-second-order}
    (\ell^2\Box - \lambda^2 + (s+1)) \mathsf{T}_{\mu_1\cdots\mu_s} = 0~,\quad
    g^{\mu_1\mu_2}\mathsf{T}_{\mu_1\mu_2\ldots\mu_s} = 0~,\quad
    \nabla^{\mu_1}\mathsf{T}_{\mu_1\mu_2\ldots\mu_s} = 0~,
\end{equation}
are equivalent to the first-order equation
\begin{equation}\label{eq:ads3-spins-first-order}
    \epsilon_{\mu_1}{}^{\nu\lambda}\nabla_\nu \mathsf{T}_{\lambda\mu_2\cdots\mu_s} = \r{i\lambda}{\ell^2}\mathsf{T}_{\mu_1\mu_2\cdots\mu_s}
\end{equation}
for one of the choices of the sign of $\lambda$.

This leads to the following strategy for solving for modes satisfying \eqref{eq:ads3-spins-second-order}: first one solves for the $T_E,\Phi$ components by expressing the first equation in \eqref{eq:ads3-spins-second-order} in terms of the scalar Laplacian and using \eqref{eq:ads3-spins-first-order} to replace the remaining components (those partially in the $\xi$ direction) in terms of derivatives of the $T_E,\Phi$ components. The resulting equation turns out to be easily diagonalisable by changing coordinates from the $T_E,\Phi$ to $x_L,x_R$, and can then be solved using hypergeometrics. Having solved for the $x_L,x_R$ components, up to relative normalisation constants, one then uses \eqref{eq:ads3-spins-first-order} to extract from them the components along $\xi$. Finally, making sure that all the components satisfy \eqref{eq:ads3-spins-first-order} in the $z\ar 0$ limit fixes the relative normalisation constants.

\subsection{Spin-0 (scalar) eigenmodes}\label{sec:scalarmodes}
Let us begin by solving for scalar eigenmodes which satisfy 
\begin{equation}\label{eq:ads3-scalar-eigeneq}
  (\ell^2\Box - \lambda^2+1)\mathsf{f} = 0~,
\end{equation}
where scalar Laplacian in the metric \eqref{eq:ads3-glob-coord} is given by
\begin{equation}\label{eq:ads3-scalar-lap}
    \Box \mathsf{f} \coloneqq \r{1}{\sqrt{g}}\partial_\mu \f(\sqrt{g}g^{\mu\nu}\partial_\nu \mathsf{f}) = \r{\partial_\xi\f(\sinh(2\xi)\partial_\xi \mathsf{f})}{\ell^2\sinh(2\xi)} + \r{\partial_{T_E}^2 \mathsf{f}}{\ell^2\sinh^2\xi} + \r{\partial_\Phi^2 \mathsf{f}}{\ell^2\cosh^2\xi}~,
\end{equation}
and investigating their properties with respect to the norm given by the inner product
\begin{equation}\label{eq:ads3-scalar-norm}
  \ip*{\mathsf{f}}{\mathsf{f}'} = \int\dd[3]{x} \sqrt{g}\,\mathsf{f}\, \mathsf{f}'~.
\end{equation}
Due to symmetries of the metric, the solutions can be decomposed into
\begin{equation}\label{eq:ads3-scalar-ansatz}
    \mathsf{f} = \sum_{n,l} e^{i(nT_E+l\Phi)}R_{\lambda nm}(\xi)=\sum_{n,m}e^{\r{2\pi in}{\beta}t_E}e^{im(\phi+i\Omega t_E)}R_{\lambda nm}(\xi)~,
\end{equation}
where
\begin{equation}\label{eq:btz-lmn}
    l = \r{\ell m + ir_- n}{r_+}~,
\end{equation}
and $(t_E,\phi)$ are related to $(T_E,\Phi)$ via transformations given in \eqref{eq:ads3-btz-trans}. Demanding the modes to be periodic under $(t_E,\phi)\sim(t_E,\phi+2\pi)$ and $(t_E,\phi)\sim(t_E+\beta,\phi-i\beta\Omega)$ sets $n,m\in\Z$. This ansatz reduces \eqref{eq:ads3-scalar-eigeneq} into a hypergeometric-type equation for the radial profile $R_{\lambda nm}$. The solution can be written as a linear combination of the two linearly independent profiles
\begin{equation}\label{eq:ads3-scalar-gen-sol}
  R^{(1)}_{\lambda nm} = \r{(\tanh\xi)^{\abs{n}}}{(\cosh\xi)^{\lambda+1}} \,{}_2F_1\f(\alpha^+_{nm}(\lambda),\beta^+_{nm}(\lambda),1+ \abs{n};\tanh^2\xi)
\end{equation}
and
\begin{equation}\label{eq:ads3-scalar-gen-sol-log}
   R^{(2)}_{\lambda nm} =
   \begin{cases}
     \r{(\tanh\xi)^{-\abs{n}}}{(\cosh\xi)^{\lambda+1}}\,{}_2F_1\f(\alpha^-_{nm}(\lambda),\beta^-_{nm}(\lambda),1- \abs{n};\tanh^2\xi)  & \text{for } n \notin \Z \\
     R^{(1)}_{\lambda nm}(\xi)\log(\tanh^2\xi) + (\tanh\xi)^{-\abs{n}}\mc{F}_{\lambda nm}(\tanh^2\xi) & \text{for } n \in \Z
   \end{cases}~,
 \end{equation}
where
\begin{equation}\label{eq:scalar-alpha-beta}
    \alpha^\pm_{nm}(\lambda) = \r{1}{2}\f(\lambda+1\pm\r{\abs{n}T}{T_R} \pm \r{i\ell \sgn(n)m}{r_+})~,\quad \beta^\pm_{nm}(\lambda) = \r{1}{2}\f(\lambda+1\pm\r{\abs{n}T}{T_L} \mp \r{i\ell \sgn(n)m}{r_+})~,
\end{equation}
and $\mc{F}_{\lambda nm}(z)$ are functions analytic at $z=0$, whose precise form will not be important for us (see eqs. 15.5.16-21 in \cite{abramowitz_handbook_1972} for details).
Moreover, the solutions \eqref{eq:ads3-scalar-gen-sol-log} are not normalisable in the neighbourhood of $z=\xi = 0$ (corresponding to $r=r_+$). Thus the scalar modes normalisable near the horizon are
\begin{equation}
  \mathsf{f}_{\lambda nm} \coloneqq e^{\r{2\pi in}{\beta}t_E}e^{im(\phi+i\Omega t_E)}R^{(1)}_{\lambda nm}
\end{equation}
and their asymptotic behaviour in the $\xi \to \infty$ limit is given by
\begin{equation}\label{eq:ads3-scalar-asymp}
  R^{(1)}_{\lambda nm} \sim 
  \begin{cases}
    C_{nm}(\lambda)e^{(\lambda-1)\xi} + C_{nm}(-\lambda)e^{(-\lambda-1)\xi} & \text{for } \lambda \notin \Z \\
    C_{nm}(\abs{\lambda})e^{(\abs{\lambda}-1)\xi} + C_{nm}(-\abs{\lambda})\xi e^{(-\abs{\lambda}-1)\xi} & \text{for } \lambda \in \Z
  \end{cases}~,
\end{equation}
where
\begin{equation}\label{cs}
  C_{nm}(\lambda) \coloneqq \r{\Gamma(1+\abs{n})}{2^{\lambda-1}\Gamma(\alpha^+_{nm}(\lambda))\Gamma(\beta^+_{nm}(\lambda))}\times
  \
  \begin{cases}
    \Gamma(\lambda) & \text{for } \lambda \notin \Z \\
    2(-1)^{\lambda} & \text{for } \lambda \in \Z
  \end{cases}
  ~.
\end{equation}
Hence, these modes are non-normalisable with respect to \eqref{eq:ads3-scalar-norm} for $\Re \lambda \ne 0$. The solutions with $\Re \lambda = 0$ belong to the continuous spectrum of the scalar Laplacian.

\subsection{Spin-1 (gauge) eigenmodes}\label{sec:gaugemodes} 
As argued in Sec.\,\ref{sec:gauge-oneloop}, after the gauge-fixing procedure, the one-loop determinant for the Chern-Simons gauge field in \eqref{eq:cs-partition-function} is expressed in terms of the operator
\begin{equation}
    (\bar\star\bar D|_T \mathsf{a})_\mu = \r{1}{2}\epsilon_\mu{}^{\nu\lambda}\bar D_\nu \mathsf{a}_\lambda~,
\end{equation}
acting on the perturbations of the Chern-Simons field (which are Lie algebra valued one-forms)  satisfying $\bar D^\mu \mathsf{a}_\mu =0$. Before considering the zero modes of these operators, let us consider the operator $(\bar\star\bar D|_T)^2 = -\bar D^\mu \bar D_\mu - \frac{2}{\ell^2}$ acting on spin-1 fields satisfying the transverse condition. We further restrict to the $U(1)$ case first, where the relevant operator is $-\bar \Box_1- \frac{2}{\ell^2}$, whose spectra we can investigate using the general strategy described at the beginning of this section. The modifications necessary to obtain the solutions in the case of non-abelian theories are discussed in Sec.\,\ref{sec:fullbtz-gauge}. We are dropping bars over the differential operators in the next parts to avoid cluttering. 

We  consider the eigenvalue problem \eqref{eq:ads3-spins-second-order} for $s=1$, 
\begin{equation}\label{eq:ads3-spin1-eq}
    (\ell^2\Box - \lambda^2 + 2)\mathsf{a}_{\mu} = 0~,\quad
    \nabla^\mu \mathsf{a}_\mu = 0~,
\end{equation}
which is equivalent to the first-order equation,
\begin{equation}\label{eq:ads3-spin1-firsteq}
  \epsilon_{\mu}{}^{\nu\rho}\nabla_\nu \mathsf{a}_\rho = \r{i\lambda}{\ell}\mathsf{a}_\mu~.
\end{equation}
 We will consider the normalisability of the solutions to these equations under the inner product
\begin{equation}\label{eq:ads3-vector-ip}
  \ip*{\mathsf{a}}{\mathsf{a}'} \coloneqq \int\dd[3]{x} \sqrt{g}\, \mathsf{a}^\mu \mathsf{a}'_\mu~.
\end{equation}
Changing the basis from $T_E$,$\Phi$ to $x_L$, $x_R$ defined in \eqref{eq:ads3-lr-coords},
\begin{equation}
    \m{i\mathsf{a}_{T_E}\\\mathsf{a}_\Phi} = \m{1 & 1 \\ 1 & -1}\m{\mathsf{a}_L\\ \mathsf{a}_R}~,
\end{equation}
and using \eqref{eq:ads3-spin1-firsteq} to express $\mathsf{a}_\xi$ in terms of $\mathsf{a}_L$ and $\mathsf{a}_R$, \eqref{eq:ads3-spin1-eq} reduces to the scalar eigenvalue equations
\begin{equation}
    (\ell^2\Box-(\lambda+1)^2+1)\mathsf{a}_L = 0,\quad
    (\ell^2\Box-(\lambda-1)^2+1)\mathsf{a}_R = 0~.
\end{equation}
The solutions to the above equations take the form
\begin{equation}
\begin{split}\label{eq:sol-gauge-gen}
   \mathsf{a}_L = \sum_{n,m \in \mathbb{Z}}e^{\r{2\pi in}{\beta}t_E}e^{im(\phi+i\Omega t_E)}\f(c_L^{(\lambda nm)} R^{(1)}_{(\lambda+1)nm}(\xi)+ b_L^{(\lambda nm)} R^{(2)}_{(\lambda+1)nm}(\xi))~,\\
   \mathsf{a}_R = \sum_{n,m \in \mathbb{Z}}e^{\r{2\pi in}{\beta}t_E}e^{im(\phi+i\Omega t_E)}\f(c_R^{(\lambda nm)} R^{(1)}_{(\lambda-1)nm}(\xi)+ b_R^{(\lambda nm)} R^{(2)}_{(\lambda-1)nm}(\xi))~,
\end{split}
\end{equation}
where $R^{(i)}_{\lambda nm}$ are given in \eqref{eq:ads3-scalar-gen-sol} and \eqref{eq:ads3-scalar-gen-sol-log}. One can show that in order to obtain vector modes normalisable around $\xi=0$, we need to exclude the radial profiles singular at this point, i.e. set $b_L^{(\lambda nm)}=b_R^{(\lambda nm)}=0$. The first order equation \eqref{eq:ads3-spin1-firsteq} then leads to a linear equation in $c_L^{(\lambda nm)}$ and $c_R^{(\lambda nm)}$. The result is that the solutions normalisable at the horizon are combinations of modes $\mathsf{a}_{\lambda nm}$ with
\begin{equation}
\begin{aligned}
     \m{(\mathsf{a}_{\lambda nm})_{L} \\ (\mathsf{a}_{\lambda nm})_{R}}
  &\coloneqq e^{\r{2\pi in}{\beta}t_E}e^{im(\phi+i\Omega t_E)}\m{e_L^{(\lambda nm)}R^{(1)}_{(\lambda+1)nm} \\ e_R^{(\lambda nm)}R^{(1)}_{(\lambda-1)nm}}~, \\
  \m{e_L^{(\lambda nm)} \\ e_R^{(\lambda nm)}} &\coloneqq \m{\lambda + \abs{n}\r{T}{T_L} - \r{i\ell \sgn(n)m}{r_+} \\ -\lambda + \abs{n}\r{T}{T_R} + \r{i\ell \sgn(n)m}{r_+}}~,
\end{aligned}
\end{equation}
and $(\mathsf{a}_{\lambda nm})_\xi$ is uniquely determined in terms of these components by \eqref{eq:ads3-spin1-firsteq}.
Next, we investigate how the norm \eqref{eq:ads3-vector-ip} of these modes behaves at $\xi \to \infty$. Using the asymptotic behaviour of the radial profiles \eqref{eq:ads3-scalar-asymp}, for $\lambda,\lambda'\notin \Z$, we have
\begin{multline}\label{b26}
    \ell^2\sqrt{g} (\mathsf{a}_{\lambda'(-n)(-m)})^\mu (\mathsf{a}_{\lambda nm})_{\mu}= -2e^{(\lambda nm)}_Le^{(\lambda' nm)}_RC_{nm}(\lambda+1)C_{nm}(-\lambda'+1)e^{(\lambda-\lambda')\xi} \\ + (\lambda \leftrightarrow \lambda') + \mc{O}(e^{2(\max(\abs{\Re\lambda},\abs{\Re\lambda'})-2)\xi})~,
\end{multline}
where  $C_{mn}$ are defined in \eqref{cs}.  When one or both of the values $\lambda,\lambda'$ are integers, some of the terms in \eqref{b26} get multiplied by a polynomial in $\xi$ (either constant or linear). This, however, doesn't affect the integrability of the whole function. From the second line of \eqref{b26}, it is clear that these modes are non-normalisable for $|\rm{Re} \lambda|\geq 2$ and become normalisble for $\abs{\Re\lambda} < 2$ when either $e^{(\lambda nm)}_L = 0$ or $e^{(\lambda nm)}_R = 0$. Restricting to the subset of modes for which $\lambda \to 0$ as $T_L\to 0$ requires setting $m = 0$ and $\lambda = \abs{n}\r{T}{T_R}$, which makes $e_R^{(\lambda nm)}$ vanish. Hence, these modes have support only along $x_L$. The explicit expression for these modes and their properties are given in Sec.\,\ref{sec:fullbtz-gauge}. Apart from the discrete modes, we obtain a continuous spectrum of normalisable modes for $\Re \lambda=0$.

\subsection{Spin-2 (graviton) eigenmodes}\label{sec:app-btz-graviton}
Next, we solve for the spin-2 eigenmodes with eigenvalue equation obtained by setting $s=2$ in \eqref{eq:ads3-spins-second-order}
\begin{equation}\label{eq:ads3-spin2-second-order}
    (\ell^2\Box - \lambda^2 + 3)\mathsf{h}_{\mu\nu} = 0~,\quad \nabla^\mu \mathsf{h}_{\mu\nu} = 0,\quad \mathsf{h}^\mu{}_\mu = 0~,
\end{equation}
equivalent to
\begin{equation}\label{eq:ads3-spin2-first-order}
    \epsilon_\mu{}^{\lambda\sigma} \nabla_{\lambda} \mathsf{h}_{\sigma\nu} = \r{i\lambda}{\ell} \mathsf{h}_{\mu\nu}~,
\end{equation}
and consider the normalisability of the solutions with respect to
\begin{equation}\label{eq:ads3-graviton-ip}
  \ip*{\mathsf{h}}{\mathsf{h}'} \coloneqq \int\dd[3]{x}\sqrt{g}\,\mathsf{h}^{\mu\nu}\mathsf{h}'_{\mu\nu}~.
\end{equation}
Using above \eqref{eq:ads3-spin2-first-order}, one can show \cite{Datta:2011za} that in the coordinates \eqref{eq:ads3-glob-coord} the Laplace-Beltrami operator $\Box_2$ is related to the scalar Laplacian \eqref{eq:ads3-scalar-lap} in the following way:
\begin{equation}
    \ell^2\Box_2 \m{-\mathsf{h}_{T_E T_E}\\i\mathsf{h}_{T_E \Phi}\\\mathsf{h}_{\Phi\Phi}} = \f(\ell^2\Box - \m{4 & 4\lambda & 2 \\ 2\lambda & 6 & 2\lambda \\ 2 & 4\lambda & 4})\m{-\mathsf{h}_{T_E T_E}\\i\mathsf{h}_{T_E \Phi}\\\mathsf{h}_{\Phi\Phi}}~.
\end{equation}
Diagonalising the mass matrix above amounts to changing coordinates to \eqref{eq:ads3-lr-coords},
\begin{equation}\label{eq:lr-to-tp}
    \m{-\mathsf{h}_{T_E T_E}\\i\mathsf{h}_{T_E \Phi}\\\mathsf{h}_{\Phi\Phi}} = V\m{\mathsf{h}_{LL}\\\mathsf{h}_{LR}\\\mathsf{h}_{RR}}~,\quad
    V = \m{1 & 2 & 1 \\ 1 & 0 & -1 \\ 1 & -2 & 1}~.
\end{equation}
In the diagonal basis, the eigenmode equations  \eqref{eq:ads3-spin2-second-order} acting on the $T_E$, $\Phi$ components are equivalent to
\begin{equation}\label{eq:ads3-spin2-diagonalised}
    (\ell^2\Box - (\lambda+2)^2+1)\mathsf{h}_{LL} = 0~, \quad
    (\ell^2\Box - \lambda^2+1)\mathsf{h}_{LR} = 0~, \quad
    (\ell^2\Box - (\lambda-2)^2+1)\mathsf{h}_{RR} = 0~,
\end{equation}
and can be solved using the results from the previous Sec.\,\ref{sec:scalarmodes} to obtain modes \cite{Datta:2011za,castro_tweaking_2017}
\begin{equation}\label{eq:ads3-spin2-gen-sol}
\begin{split}
   \mathsf{h}_{LL} &= \sum_{n,m \in \mathbb{Z}}e^{\r{2\pi in}{\beta}t_E}e^{im(\phi+i\Omega t_E)}\f(c_{LL}^{(\lambda nm)} R^{(1)}_{(\lambda+2)nm}+ b_{LL}^{(\lambda nm)} R^{(2)}_{(\lambda+2)nm})~,\\
   \mathsf{h}_{LR} &= \sum_{n,m \in \mathbb{Z}}e^{\r{2\pi in}{\beta}t_E}e^{im(\phi+i\Omega t_E)}\f(c_{LR}^{(\lambda nm)} R^{(1)}_{\lambda nm}+ b_{LR}^{(\lambda nm)} R^{(2)}_{\lambda nm})~,\\
   \mathsf{h}_{RR} &= \sum_{n,m \in \mathbb{Z}}e^{\r{2\pi in}{\beta}t_E}e^{im(\phi+i\Omega t_E)}\f(c_{RR}^{(\lambda nm)} R^{(1)}_{(\lambda-2)nm}+ b_{RR}^{(\lambda nm)} R^{(2)}_{(\lambda-2)nm})~.\\
\end{split}
\end{equation}
Similarly as in the spin-1 case, one can show that allowing for contributions from the modes $R^{(2)}_{\lambda nm}$ leads to spin-2 modes non-normalisable around $\xi=0$, so we have to set $b^{(\lambda nm)}_{ij} = 0$. Requiring that \eqref{eq:ads3-spin2-first-order} is satisfied in the $\xi\ar 0$ limit (i.e. at the horizon) imposes two linear equations on the remaining coefficients $c^{(\cdots)}_{ij}$ and the final solution normalisable around $\xi=0$ can be written as a combination of modes $\mathsf{h}_{\lambda nm}$ where \cite{castro_tweaking_2017}:
\begin{equation}\label{eq:ads3-rel-norms}
\m{(\mathsf{h}_{\lambda nm})_{LL} \\ (\mathsf{h}_{\lambda nm})_{LR} \\ (\mathsf{h}_{\lambda nm})_{RR}} \coloneqq e^{\r{2\pi in}{\beta}t_E}e^{im(\phi+i\Omega t_E)}\m{e_{LL}^{(\lambda nm)}R^{(1)}_{(\lambda+2) nm} \\ e_{LR}^{(\lambda nm)}R^{(1)}_{\lambda nm} \\ e_{RR}^{(\lambda nm)}R^{(1)}_{(\lambda-2) nm}}~,
\end{equation}
\begin{equation}
\m{e_{LL}^{(\lambda nm)} \\ e_{LR}^{(\lambda nm)} \\ e_{RR}^{(\lambda nm)}} \coloneqq 
\m{
  \f(\lambda-1+\abs{n}\r{T}{T_L} - \r{i\ell\sgn(n)m}{r_+})\f(\lambda+1+\abs{n}\r{T}{T_L} - \r{i\ell\sgn(n)m}{r_+}) \\
    -\f(\lambda-1+\abs{n}\r{T}{T_L} - \r{i\ell\sgn(n)m}{r_+})\f(\lambda+1-\abs{n}\r{T}{T_R} - \r{i\ell\sgn(n)m}{r_+}) \\
    \f(\lambda-1-\abs{n}\r{T}{T_R} - \r{i\ell\sgn(n)m}{r_+})\f(\lambda+1-\abs{n}\r{T}{T_R} - \r{i\ell\sgn(n)m}{r_+})
}~.
\end{equation}
The remaining components ($\mathsf{h}_{\xi T_{\mr{E}\Phi}}$, $\mathsf{h}_{\xi \Phi}$ and $\mathsf{h}_{\xi\xi}$) are then extracted using \eqref{eq:ads3-spin2-first-order} and the tracelessness condition.

To find the set of normalisable eigenmodes, we use the asymptotic behaviour of the radial profiles \eqref{eq:ads3-scalar-asymp} to estimate the asymptotic ($\xi \to \infty$) behaviour of the integrand in \eqref{eq:ads3-graviton-ip}. In the case where $\lambda,\lambda' \notin \Z$ we get
\begin{multline}\label{eq:btz-grav-norm-asym}
    \ell^2\sqrt{g} (\mathsf{h}_{\lambda'(-n)(-m)})_{\mu\nu}(\mathsf{h}_{\lambda nm})^{\mu\nu} = 16e_{LL}^{(\lambda'nm)}e_{RR}^{(\lambda nm)}C_{nm}(\lambda'+2)C_{nm}(-\lambda+2)e^{(\lambda'-\lambda)\xi} \\ + (\lambda \leftrightarrow \lambda')
    + \mc{O}(e^{2\f(\max(\abs{\Re\lambda},\abs{\Re\lambda'})-2)\xi})~,
\end{multline}
where the coefficients $C_{nm}$ are defined in \eqref{cs}. We used \eqref{eq:ads3-spin2-second-order} and \eqref{eq:ads3-spin2-first-order} to get the asymptotic behaviour of ($\mathsf{h}_{\xi\mu}$) components. When one or both of the values $\lambda,\lambda'$ are integers, some of the terms in \eqref{eq:btz-grav-norm-asym} get multiplied by a polynomial in $\xi$ (either constant or linear). This, however, doesn't affect the integrability of the whole function. Thus, as for the spin-1 case, for $\abs{\Re\lambda} < 2$, we obtain a set of discrete normalisable eigenmodes when either $e^{(\lambda nm)}_{LL} = 0$ or $e^{(\lambda nm)}_{RR} = 0$. Setting $e^{(\lambda nm)}_{RR} = 0$ results in modes that become zero modes of $\bar{\Delta}_{\mr{grav}}$ \eqref{eq:graviton-quadratic} in the limit $T_L \ar 0$ ($r_- \ar r_+$), while the modes $e^{(\lambda nm)}_{LL} = 0$ become zero modes in the $T_R \ar 0$ ($r_- \ar -r_+$) limit -- as everywhere else in this paper we restrict to the first case without loss of generality. Thus we have two families of ``eventually zero'' modes: \textbf{$\mathsf{T}$-modes}, with $\lambda = -1 + \abs{n}\r{T}{T_R}$, and \textbf{$\mathsf{R}$-modes}, with $\lambda = 1 + \abs{n}\r{T}{T_R}$. We further discuss their properties in Sec.\,\ref{sec:far}.

In addition, we have continuous spectrum when $\Re\lambda = 0$ -- indeed, from \eqref{eq:ads3-spin2-second-order} we get
\begin{equation}
\begin{split}
    \ip*{\mathsf{h}_{\lambda' nm}}{\mathsf{h}_{\lambda nm}} &= \r{1}{\lambda^2-\lambda^{\prime 2}}\int_M \f((\mathsf{h}_{\lambda'nm})_{\mu\nu}\Box (\mathsf{h}_{\lambda nm})^{\mu\nu} - (\mathsf{h}_{\lambda nm})_{\mu\nu}\Box (\mathsf{h}_{\lambda'nm})^{\mu\nu})\sqrt{g_{M}}\dd[3]{x} \\
    &= \r{1}{\lambda^2-\lambda^{\prime 2}}\int_{\partial M} n^\sigma \f((\mathsf{h}_{\lambda'nm})_{\mu\nu}\nabla_\sigma (\mathsf{h}_{\lambda nm})^{\mu\nu} - (\mathsf{h}_{\lambda nm})_{\mu\nu}\nabla_\sigma (\mathsf{h}_{\lambda'nm})^{\mu\nu})\sqrt{g_{\partial M}}\dd[2]{x} \\
    &= \r{2\pi\ell \beta}{\lambda^2-\lambda^{\prime 2}}\eval{\left[ \f((\mathsf{h}_{\lambda'nm})_{\mu\nu}\partial_\xi (\mathsf{h}_{\lambda nm})^{\mu\nu} - (\mathsf{h}_{\lambda nm})_{\mu\nu}\partial_\xi (\mathsf{h}_{\lambda'nm})^{\mu\nu})\sinh\xi\cosh\xi \right]}_{\xi=0}^{\xi=+\infty} \\
    &= 32\pi\ell\beta 
    \lim_{\xi\ar\infty} \left[\r{e_{LL}^{(\lambda'nm)}e_{RR}^{(\lambda nm)}C_{nm}(\lambda'+2)C_{nm}(-\lambda+2)e^{(\lambda-\lambda')\xi} - (\lambda \leftrightarrow \lambda')}{\lambda-\lambda'} \right.\\
    &\qquad\left.+ \mc{O}\f(e^{2\f(\max(\abs{\Re\lambda},\abs{\Re\lambda'})-2)\xi})
    \vphantom{\r{C_-^{(\lambda')}}{\lambda'}}\right]~,
\end{split}
\end{equation}
and hence
\begin{equation}
    \ip*{\mathsf{h}'}{\mathsf{h}} = \mathcal{C}\delta(\abs{\lambda-\lambda'})\quad \text{for} \quad \Re\lambda=\Re\lambda'=0~,
\end{equation}
with
\begin{equation}
    \mathcal{C} = 64\pi\ell\beta e_{LL}^{(\lambda nm)}e_{RR}^{(\lambda nm)}C_{nm}(\lambda+2)C_{nm}(-\lambda+2)~.
\end{equation}
\subsection{Spin-\texorpdfstring{$\r{1}{2}$}{1/2} (fermion) eigenmodes}\label{spin12-btz-app}
In this subsection, we solve for Dirac eigenmodes satisfying
\begin{equation}\label{eq:ads3-dirac-eigenproblem}
  \ell\slashed{D}\chi = s \chi
\end{equation}
on the background \eqref{eq:ads3-glob-frame}. This becomes more tractable if one performes the substitution $\chi \ar \hat{\chi}$ defined by \cite{das_black_1999,becar_dirac_2013}
\begin{equation}\label{eq:ads3-spinor-rescaling}
  \chi \eqqcolon \exp(\r{\xi}{2}\gamma^1)\hat{\chi}~,
\end{equation}
with which \eqref{eq:ads3-dirac-eigenproblem} becomes equivalent to
\begin{multline}\label{eq:ads3-spin12-first-order}
  \partial_\xi\f(\sqrt{\sinh(2\xi)}\hat{\chi}) =\\ \f(\gamma^2(i\coth\xi \partial_{T_E} + \tanh\xi\partial_\Phi) - i\gamma^0(i\partial_{T_E} + \partial_\Phi) + \gamma^1(s-\r{1}{2}))\sqrt{\sinh(2\xi)}\hat{\chi}~,
\end{multline}
where we used the frames defined in \eqref{eq:ads3-glob-frame}. Squaring the above equation gives the following second-order equation
\begin{equation}\label{eq:ads3-spinor-eq-squared}
  \f(\r{\partial_\xi(\sinh(2\xi)\partial_\xi)}{\sinh(2\xi)} - \r{(i\partial_{T_E} - \r{\gamma^2}{2})^2}{\sinh^2\xi} + \r{(\partial_\Phi - \r{\gamma^2}{2})^2}{\cosh^2\xi} + 1 - (s-\r{1}{2})^2)\m{\hat{\chi}_1\\\hat{\chi}_2} = 0~.
\end{equation}
The isometries of BTZ allow us to use the ansatz
\begin{equation}
  \hat{\chi}_r = \hat{\chi}^+_r + \hat{\chi}^-_r~,\quad
  \hat{\chi}^\pm_r = \sum_{k=0}^\infty\sum_{m\in \Z/2}e^{\pm\r{2\pi i}{\beta}(k+\r{1}{2})t_E}e^{\pm im(\phi+i\Omega t_E)} R^\pm_{skmr}(\xi)~,
\end{equation}
where $r$ labels the two components of the Dirac spinor. Using the above ansatz, we obtain a second-order radial equations whose solution take the form
\begin{equation}
  \hat{\chi}^\pm_r = \sum_{k=0}^\infty\sum_{m\in \Z/2}e^{\pm\r{2\pi i}{\beta}(k+\r{1}{2})t_E}e^{\pm im(\phi+i\Omega t_E)} \f(c^{(skm\pm)}_rR^{\pm(1)}_{skmr}(\xi) + b^{(skm\pm)}_rR^{\pm(2)}_{skmr}(\xi))~,
\end{equation}
with
\begin{equation}
  \begin{split}
  R^{+(i)}_{skm1}(\xi) &= R^{-(i)}_{skm2}(\xi) = R^{(i)}_{(s-\r{1}{2})(k+1)(m+\r{i}{2\ell}(r_+-r_-))}~,\quad \\
  R^{-(i)}_{skm1}(\xi) &= R^{+(i)}_{skm2}(\xi) = R^{(i)}_{(s-\r{1}{2})k(m-\r{i}{2\ell}(r_+-r_-))}~,
  \end{split}
\end{equation}
where $R^{(i)}_{\lambda nm}$ are the radial profiles \eqref{eq:ads3-scalar-gen-sol} and \eqref{eq:ads3-scalar-gen-sol-log}.
Requiring normalisability around $\xi=0$ forces us to exclude the contributions singular at $\xi=0$, i.e. set $b^{skm\pm}_r = 0$. To solve for the remaining constants $c^{(snm)}_i$, we plug these solutions into the first-order equation \eqref{eq:ads3-spin12-first-order}. The result is
\begin{equation}
  \r{c^{(skm+)}_1}{c^{(skm+)}_2} =  \r{c^{(skm-)}_2}{c^{(skm-)}_1} = \r{s - \r{1}{2} - \r{T}{T_L}(k+\r{1}{2}) + \r{i\sgn(n)m}{r_+}}{2(k+1)}~.
\end{equation}
Finally, the solution for the Dirac spinor $\hat{\chi}$ is a linear combination of the following modes
\begin{equation}\label{eq:spin12-hat-sol-plus}
  \hat{\chi}^{+}_{skm} = e^{\r{2\pi i}{\beta}(k+\r{1}{2})t_E} e^{im(\phi + i\Omega t_E)}\r{(\tanh\xi)^{k}}{(\cosh\xi)^{s+\r{1}{2}}} \m{(s-\r{1}{2}-\beta_{skm})\tanh\xi~ {}_2F_1(\alpha_{skm}, \beta_{skm}+1, k+2,z) \\ (k+1)~ {}_2F_1(\alpha_{skm},\beta_{skm},k+1,z)}~,
\end{equation}
and
\begin{equation}\label{eq:spin12-hat-sol-minus}
  \hat{\chi}^{-}_{skm} = e^{-\r{2\pi i}{\beta}(k+\r{1}{2})t_E} e^{-im(\phi + i\Omega t_E)}\r{(\tanh\xi)^{k}}{(\cosh\xi)^{s+\r{1}{2}}} \m{(k+1)~ {}_2F_1(\alpha_{skm},\beta_{skm},k+1,z) \\ (s-\r{1}{2}-\beta_{skm})\tanh\xi ~ {}_2F_1(\alpha_{skm}, \beta_{skm}+1, k+2,z)}~,
\end{equation}
where in both $k \ge 0$, and
\begin{equation}
  2\alpha_{skm} \coloneqq s+\r{1}{2} + \r{T}{T_R}(k + \r{1}{2}) + \r{i\ell m}{r_+} ~,\quad
  2\beta_{skm} \coloneqq s-\r{1}{2} + \r{T}{T_L}(k + \r{1}{2}) - \r{i\ell m}{r_+}~,\quad
\end{equation}
which satisfy $\alpha_{skm} + \beta_{skm} = s + k + \r{1}{2}$. Let us now analyse normalisability of the actual Dirac eigenmodes (solutions of \eqref{eq:ads3-dirac-eigenproblem}), which are related to the modes $\hat{\chi}^{\pm}_{skm}$ by \eqref{eq:ads3-spinor-rescaling}.
The norm is defined as
\begin{equation}
    \langle\chi|\chi'\rangle \coloneqq \int \dd[3]{x}\,\sqrt{g}\, \chi^{\dagger} \chi'~.
\end{equation}
The product of $\chi$'s in the integrand takes the following form when expressed in terms of the spinor $\hat \chi$
\begin{equation}\label{eq:spin12-chi-squared}
    (\chi^\pm_{skm})^\dagger(\chi^\pm_{skm}) = \r{1}{2}\f(e^\xi \langle{(\hat{\chi}^\pm_{skm})_1 + (\hat{\chi}^\pm_{skm})_2}\rangle + e^{-\xi}\langle{(\hat{\chi}^\pm_{skm})_1 - (\hat{\chi}^\pm_{skm})_2}\rangle)~.
\end{equation}
Next, we use \eqref{eq:ads3-scalar-asymp} to show that
\begin{multline}
  e^{-i(\ldots)}(\hat{\chi}^+_{skm})_1 = e^{i(\ldots)}(\hat{\chi}^-_{skm})_2 
  = \r{s-\r{1}{2}-\beta_{skm}}{s-\r{1}{2}} D^{(1)}_{skm}e^{(s-\r{1}{2})\xi} - \r{1}{2}D^{(2)}_{skm}e^{(-s+\r{1}{2})\xi} \\
  + \mc{O}(e^{(s-\r{1}{2})\xi}) + \mc{O}(e^{(-s+\r{1}{2})\xi}) ~,
\end{multline}
\begin{multline}
  e^{-i(\ldots)}(\hat{\chi}^+_{skm})_2 = e^{i(\ldots)}(\hat{\chi}^-_{skm})_1
  = \r{\beta_{skm}}{s-\r{1}{2}}D^{(1)}_{skm}e^{(s-\r{1}{2})\xi} + \r{1}{2}D^{(2)}_{skm}e^{(-s+\r{1}{2})\xi} \\
  + \mc{O}(e^{(s-\r{1}{2})\xi}) + \mc{O}(e^{(-s+\r{1}{2})\xi}) ~,
\end{multline}
where
\begin{equation}\label{eq:ads3-dirac-constants}
  D^{(1)}_{skm} = \r{\Gamma(k+2)\Gamma(s+\r{1}{2})}{2^{s-\r{1}{2}}\Gamma(\alpha_{skm})\Gamma(\beta_{skm}+1)}~,\quad
  D^{(2)}_{skm} = \r{\Gamma(k+2)\Gamma(-s+\r{1}{2})}{2^{-s-\r{1}{2}}\Gamma(\alpha_{skm}-s+\r{1}{2})\Gamma(\beta_{skm}-s+\r{1}{2})}~.
\end{equation}
Thus, from \eqref{eq:spin12-chi-squared} we have
\begin{equation}\label{eq:spin12-norm-dens}
    (\chi^\pm_{skm})^\dagger(\chi^\pm_{skm})
    = 2\abs{D^{(1)}_{skm}}^2e^{(2\Re s - 2)\xi}
    + 2\abs{D^{(2)}_{skm}}^2e^{(-2\Re s - 2)\xi} + \mc{O}(e^{(2\abs{\Re s}-4)\xi})~.
\end{equation}
Including the factors of $\sqrt{g}$ in the norm, it is easy to check that these modes are not normalisable for any value of $s$ with $\Re s \ne 0$. Similarily, one can show that
\begin{equation}\label{eq:spin12-norm-dens-gamma1}
    (\chi^\pm_{skm})^\dagger\gamma^1(\chi^\pm_{skm})
    = 2\abs{D^{(1)}_{skm}}^2e^{(2\Re s - 2)\xi}
    - 2\abs{D^{(2)}_{skm}}^2e^{(-2\Re s - 2)\xi} + \mc{O}(e^{(2\abs{\Re s}-4)\xi})~.
\end{equation}
This equation will be useful to comment on the normalisability of spin-$\frac{3}{2}$ eigenmodes.

\subsection{Spin-\texorpdfstring{$\r{3}{2}$}{3/2} (gravitino) eigenmodes}\label{spin32-btz-app}
Following the discussion in Sec.\,\ref{sec:gravitini-oneloop}, we are looking for eigenmodes $\bm{\psi}$ of the operators $\star\mc{D}$ and $\star\mc{D}'$, whose eigenvalue approaches zero in the extremal limit, satisfying the gauge condition $\gamma^a\bm{\psi}_a=0$. Similarily to App.\,\ref{sec:gaugemodes}, we will consider the case when the background gauge fields $A$, $A'$ are zero, and the modifications necessary to obtain the modes of the charged operators are described in Sec.\,\ref{sec:fermi-far}. Starting with the eigenvalue equation of the form
\begin{equation}\label{eq:ads3-spin32-first-order1}
\epsilon_\mu{}^{\nu\lambda}D_\nu\bm{\psi}_\lambda = \r{i\lambda}{\ell}\bm{\psi}_\mu~,
\end{equation}
where $D_\nu$ is the covariant derivative defined in \eqref{eq:def-covariant-1} with $A=A'=0$, we can obtain eigenvalue equations for the operators $\star \mathcal{D}$ and $\star \mathcal{D}'$ by noting that 
\begin{equation}
\epsilon_\mu{}^{\nu\lambda}D_\nu\bm{\psi}_\lambda = \r{i\lambda}{\ell}\bm{\psi}_\mu \iff \epsilon_\mu{}^{\nu\lambda}(D_\nu - \r{\alpha}{\ell}\gamma_\nu)\bm{\psi}_\lambda = \r{i(\lambda+\alpha)}{\ell}\bm{\psi}_\mu~,\quad
  \text{for any } \alpha \in \C~,
\end{equation}
where we used $\gamma^{[a}\gamma^{b]} = i\epsilon^{abc}\gamma_c$ and the gauge condition ($\gamma^a\bm{\psi}_a=0$). Therefore, setting $\alpha = \pm \frac{1}{2}$, we obtain the eigenvalue equations 
\begin{equation}\label{eq:ads3-spin32-first-order}
 \begin{split}
  \epsilon_\mu{}^{\nu\lambda}\mc{D}_\nu\bm{\psi}_\lambda &= \r{i(\lambda+1/2)}{\ell}\bm{\psi}_\mu~,\\
  \epsilon_\mu{}^{\nu\lambda}\mc{D}'_\nu\bm{\psi}_\lambda &= \r{i(\lambda-1/2)}{\ell}\bm{\psi}_\mu~,
 \end{split}   
 \end{equation}
where $\mathcal{D}$ and $\mathcal{D'}$ are defined in \eqref{eq:def-covariant}. Therefore, solving for eigenmodes of $\mathcal{D}$ and $\mathcal{D}'$ is equivalent to solving \eqref{eq:ads3-spin32-first-order1}. Also note that the solutions of  \eqref{eq:ads3-spin32-first-order1}, together with the gauge condition $\gamma^a\bm{\psi}_a=0$ implies that $D^\mu \bm{\psi}_\mu = 0$. Indeed, this is true for any constant $\alpha$,
\begin{multline}
  \r{i(\lambda + \alpha)}{\ell}D^\mu \bm{\psi}_\mu = \r{i(\lambda+\alpha)}{\ell}(D^\mu - \r{\alpha}{\ell}\gamma^\mu) \bm{\psi}_\mu =  (D^\mu - \r{\alpha}{\ell}\gamma^\mu)\epsilon_\mu{}^{\nu\lambda}(D_\nu - \r{\alpha}{\ell}\gamma_\nu)\bm{\psi}_\lambda \\
= \r{1}{2}\epsilon^{\mu\nu\lambda}[D_{\mu},D_{\nu}]\bm{\psi}_\lambda = \r{1}{2}\epsilon^{\mu\nu\lambda}\f[-R^\sigma{}_{\lambda\mu\nu}\bm{\psi}_\sigma + \r{1}{4}R_{\mu\nu\sigma\rho}\gamma^\sigma\gamma^\rho \bm{\psi}_\lambda - \r{1}{2}\Gamma^\sigma{}_{\lambda [\mu}(\omega_{ab})_{\nu]}\gamma^a\gamma^b \bm{\psi}_\sigma] = 0~,
\end{multline}
where in addition to the previous identities we used $\gamma^{[\mu}\gamma^\nu\gamma^{\lambda]} = i\epsilon^{\mu\nu\lambda}$.

One can approach solving \eqref{eq:ads3-spin32-first-order1} in two ways -- one, analogous to the method used in the previous subsection for fields with integer spin, is to square the first-order operator to arrive at
\begin{equation}\label{eq:ads3-spin32-2nd-order}
    \f(\ell^2\slashed{D}^2_{\r{1}{2}} - 2\m{1 & \lambda \\ \lambda & 1} - \lambda^2 + 1)\m{i\bm{\psi}_{T_E}\\ \bm{\psi}_{\Phi}} = 0~,
\end{equation}
where we used $\gamma^a \bm\psi_a =0$ to express $\bm \psi_\xi$ in terms of $T_E$ and $\Phi$ components. We can again diagonalise the above equation by going to $x_L$ and $x_R$ coordinates and solve for all the components of $i\bm{\psi}_{T_E} \pm \bm{\psi}_\Phi$ separately. As in the cases of spin-1 and spin-2, the relative normalisation of these modes can be determined by plugging them back into the first-order equation.

In this subsection, we will, however, elaborate on a simpler way to solve \eqref{eq:ads3-spin32-first-order1}. Note that for fields satisfying $\gamma^a\bm{\psi}_a=0$ in three dimensions we have 
\begin{equation}
  \slashed{D}_{\r{3}{2}} \bm{\psi}_\mu \coloneqq \gamma^\nu D_\nu \bm{\psi}_\mu = i\epsilon_{\mu}{}^{\nu\lambda} D_\nu \bm{\psi}_\lambda= \gamma_{\mu}^{\nu\lambda}D_{\nu}\psi_{\lambda}~.
\end{equation}
The second equality can be proven using $\gamma^\mu\bm{\psi}_\mu = D^\mu\bm{\psi}_\mu = 0$ as follows:
  \begin{equation}
    \begin{split}
      i\epsilon^{\mu\nu\lambda}D_\nu \bm{\psi}_\lambda &= \gamma^{[\mu}\gamma^\nu \gamma^{\lambda]}D_\nu \bm{\psi}_\lambda \\
      &= (\gamma^\mu\gamma^\nu\gamma^\lambda - g^{\mu\nu}\gamma^\lambda - g^{\nu\lambda}\gamma^\mu + g^{\mu\lambda}\gamma^\nu) D_\nu  \bm{\psi}_\lambda \\
      &= \gamma^\nu D_\nu  \bm{\psi}^\mu~,
    \end{split}
  \end{equation}
where the first two terms in the second line vanish because $\gamma^\lambda\bm{\psi}_\lambda = 0$ and because $\gamma$s can be commuted over spinor covariant derivatives, and the third term vanishes because $D^\lambda \bm{\psi}_\lambda = 0$. We use this identity to 
express \eqref{eq:ads3-spin32-first-order1} as $\slashed{D}_{\r{3}{2}}\bm{\psi}_\mu = -\r{\lambda}{\ell}\bm{\psi}_\mu$. Furthermore, using the explicit form of spin-connections, $\slashed D_{\frac{3}{2}}$ acting on $T_E$ and $\Phi$ can be further simplified as follows
\begin{equation}
  \slashed{D}_{\r{3}{2}} \bm{\psi}_{T_E} = \slashed{D}_{\r{1}{2}}\bm{\psi}_{T_E} - \Gamma^\lambda{}_{T_E \mu} e^\mu_a\gamma^a \bm{\psi}_\lambda
  = \slashed{D}_{\r{1}{2}}\bm{\psi}_{T_E} + \cosh\xi ( \gamma^0 \bm{\psi}_1 - \gamma^1 \bm{\psi}_0)
  = \slashed{D}_{\r{1}{2}}\bm{\psi}_{T_E} - i\bm{\psi}_\Phi~.
\end{equation}
Similarly one can show that $\slashed{D}_{\r{3}{2}}\bm{\psi}_\Phi  = \slashed{D}_{\r{1}{2}}\bm{\psi}_\Phi + i\bm{\psi}_{T_E}$. Hence the $\mu = T_E$ and $\mu = \Phi$ components of \eqref{eq:ads3-spin32-first-order1} are equivalent to
\begin{equation}\label{eq:ads3-spin32-diag}
  \f(\ell\slashed{D}_{\r{1}{2}} + \m{0&1 \\ 1&0} + \lambda)\m{i\bm{\psi}_{T_E}\\\bm{\psi}_\Phi} = 0
  \iff \f(\ell\slashed{D}_{\r{1}{2}} + \lambda \pm 1)(i\bm{\psi}_{T_E} \pm \bm{\psi}_\Phi) = 0~,
\end{equation}
which indeed squares to \eqref{eq:ads3-spin32-2nd-order}.

Let us define
\begin{equation}\label{eq:psi-lr}
  \bm{\psi}_L = \r{i\bm{\psi}_{T_E} + \bm{\psi}_\Phi}{2}~,\quad
  \bm{\psi}_R = \r{i\bm{\psi}_{T_E} - \bm{\psi}_\Phi}{2}~.
\end{equation}
We just showed that the $\mu = T_E,\Phi$ components of \eqref{eq:ads3-spin32-first-order} are satisfied if and only if $\bm{\psi}_L$ and $\bm{\psi}_R$ are eigenmodes of the Dirac operator with eigenvalues $-(\lambda+1)/\ell$ and $-(\lambda-1)/\ell$ respectively. Using the ansatz
\begin{equation}
  \bm \psi_L = \bm \psi_L^+ + \bm \psi_L^-~,\quad
  \bm \psi_L^\pm = \sum_{k=0}^\infty\sum_{m\in \Z/2}c_{skm}e^{\pm\r{2\pi i}{\beta}(k+\r{1}{2})t_E}e^{\pm im(\phi+i\Omega t_E)} (\bm{\psi}^{\pm}_{s km})_L
\end{equation}
for $\bm \psi_L$ and a similar one $\bm \psi_R$ and using \eqref{eq:spin12-hat-sol-plus} and \eqref{eq:spin12-hat-sol-minus}, we find
\begin{equation}
  (\bm{\psi}^{\pm}_{\lambda km})_L = \tilde{e}_L^{(\lambda km)} \chi^\pm_{(-\lambda-1)km},\quad
  (\bm{\psi}^{\pm}_{\lambda km})_R = \tilde{e}_R^{(\lambda km)} \chi^\pm_{(-\lambda+1)km}~.
\end{equation}
Using the gauge condition, the remaining component, $\bm \psi_\xi$, is given in terms of $\bm\psi_{T_E}$ and $\bm\psi_\Phi$: 
\begin{equation}\label{eq:ads3-psixi}
  \bm{\psi}_\xi = \r{i\gamma^2}{\sinh\xi}\bm{\psi}_{T_E} - \r{i\gamma^0}{\cosh\xi}\bm{\psi}_\Phi~.
\end{equation}
Similar to the spin-$1$ case, the relative normalisation of these modes is not arbitrary -- it can be fixed by examining the leading order behaviour of 
of \eqref{eq:ads3-spin32-first-order} near $\xi = 0$. One finds that
\begin{equation}\label{eq:spin32-rel-prop}
\begin{aligned}
    \tilde{e}_L^{(\lambda nm)} &= \f(\lambda - \r{1}{2} + \r{T}{T_L}(k+\r{1}{2}) - \r{i\ell m}{r_+})~,\\
    \tilde{e}_R^{(\lambda nm)} &= -\f(\lambda + \r{1}{2} - \r{T}{T_R}(k+\r{1}{2}) - \r{i\ell m}{r_+})~.
\end{aligned}
\end{equation}
The norm density of the field can be simplified as
\begin{equation}\label{eq:spin32-norm-dens}
  \begin{split}
    \sqrt{g}\bm{\psi}^{\dagger}_\mu \bm{\psi}^{\mu}
    &= \ell\sinh\xi\cosh\xi\f(\r{\bm{\psi}^{\dagger}_{T_E} \bm{\psi}_{T_E}}{\sinh^2\xi} + \r{\bm{\psi}^{\dagger}_{\Phi} \bm{\psi}_{\Phi}}{\cosh^2\xi} + \bm{\psi}^{\dagger}_\xi \bm{\psi}_\xi) \\
    &= 2\ell\f(\coth\xi \bm{\psi}^{\dagger}_{T_E} \bm{\psi}_{T_E} + \tanh\xi \bm{\psi}^{\dagger}_{\Phi} \bm{\psi}_{\Phi} + \Re(i\bm{\psi}^{\dagger}_\Phi \gamma^1 \bm{\psi}_{T_E})) \\
    &= 2\ell\f(2\coth(2\xi) (\bm{\psi}^{\dagger}_L \bm{\psi}_L + \bm{\psi}^{\dagger}_R \bm{\psi}_R) + \r{4\Re (\bm{\psi}^{\dagger}_L \bm{\psi}_R)}{\sinh(2\xi)} + \bm{\psi}^{\dagger}_L \gamma^1 \bm{\psi}_L - \bm{\psi}^{\dagger}_R \gamma^1 \bm{\psi}_R)~, \\ 
  \end{split}
\end{equation}
where in the second equality we used $\gamma^a\bm{\psi}_a = 0$ and in the third -- the relation \eqref{eq:psi-lr}. Using it, and the asymptotic behaviour of the Dirac modes \eqref{eq:spin12-norm-dens}, \eqref{eq:spin12-norm-dens-gamma1}, one can extract the normalisable spectrum of operators \eqref{eq:ads3-spin32-first-order}. In particular, if one is interested in the zero modes of $\star\mc{D}$, according to \eqref{eq:ads3-spin32-first-order} one should look for modes $\bm{\psi}$ for which $\lambda$ approaches $-\r{1}{2}$ as one takes $T \ar 0$. From the discussion above, we know that these are constructed from the spin-$\r{1}{2}$ Dirac operator eigenmodes \eqref{eq:ads3-dirac-eigenproblem}, $\bm{\psi}_L$ and $\bm{\psi}_R$ with eigenvalues $s_L$ and $s_R$ approaching $-\r{1}{2}$ and $\r{3}{2}$, respectively. However, \eqref{eq:spin32-norm-dens} together with \eqref{eq:spin12-norm-dens} implies that this is not possible for $\bm{\psi}_R \ne 0$, since the Hermitian square of any Dirac eigenmode with eigenvalue close to $\r{3}{2\ell}$ grows too fast. Hence we have to set $\bm{\psi}_R = 0$, and choose $\lambda$ so that $\tilde{e}^{(\lambda nm)}_R = 0$. Thus, the only possible choice is 
\begin{equation}
  \lambda = -\r{1}{2} + \r{T}{T_R}(k+\r{1}{2})~,\quad m = 0~.
\end{equation}
Using this value of $\lambda$ in \eqref{eq:ads3-spin32-first-order}, we find that the modes that become zero modes at extremality obey 
\begin{equation}\label{lk}
  -i\epsilon_\mu{}^{\nu\lambda}\mc{D}_\nu\bm{\psi}_\lambda = \r{T}{\ell T_R} (k+\r{1}{2})\bm{\psi}_\mu~.
\end{equation}
So the zero modes appear in the unprimed sector only, as expected, and these modes are labelled by just one parameter $k$ ($\lambda$ is fixed in terms of $k$ via \eqref{lk} and $m=0$). Their explicit form in coordinates \eqref{eq:ads3-glob-coord} and frame \eqref{eq:ads3-glob-frame} is
\begin{equation}\label{eq:ads3-psi-zm-plus}
   (\bm{\psi}^{+}_{k})_{T_E} = (\bm{\psi}^{+}_{k})_\Phi = e^{\r{2\pi in}{\beta}t_E}(\tanh\xi)^k (\cosh\xi)^{\r{T}{T_R}(k+\r{1}{2})} \m{\cosh\r{\xi}{2}\tanh\xi - \sinh\r{\xi}{2} \\ \sinh\r{\xi}{2}\tanh\xi - \cosh\r{\xi}{2}}~,
\end{equation}
and
\begin{equation}\label{eq:ads3-psi-zm-minus}
   (\bm{\psi}^{-}_{k})_{T_E} = (\bm{\psi}^{-}_{k})_\Phi = e^{-\r{2\pi in}{\beta}t_E}(\tanh\xi)^k (\cosh\xi)^{\r{T}{T_R}(k+\r{1}{2})}\m{\sinh\r{\xi}{2}\tanh\xi - \cosh\r{\xi}{2} \\ \cosh\r{\xi}{2}\tanh\xi - \sinh\r{\xi}{2}}~,
 \end{equation}
with $\bm{\psi}_\xi$ given by \eqref{eq:ads3-psixi}.

\bibliography{all}{}
\bibliographystyle{ytphys}

\end{document}